\documentclass[sn-mathphys-num]{sn-jnl}% Math and Physical Sciences Numbered Reference Style 
\usepackage{graphicx}%
\usepackage{multirow}%
\usepackage{amsmath,amssymb,amsfonts}%
\usepackage{amsthm}%
\usepackage{mathrsfs}%
\usepackage[title]{appendix}%
\usepackage{xcolor}%
\usepackage{textcomp}%
\usepackage{manyfoot}%
\usepackage{booktabs}%
\usepackage{algorithm}%
\usepackage{algorithmicx}%
\usepackage{algpseudocode}%
\usepackage{listings}%
\usepackage{fullpage}
\usepackage{siunitx}
\usepackage{bm}
\UseRawInputEncoding
\usepackage{lineno}

\newcommand{\hii}{\mbox{H{\small II}}}

\newcommand{\micron}{\mbox{$\mathrm{\mu m}$}}

\newcommand{\arcsec}{\ensuremath{''}}

\newcommand{\changes}[1]{#1}

\usepackage{enumitem}
\usepackage{appendix}
\usepackage{hyperref}
\usepackage{amsmath,amstext}

\defcitealias{Draine2021}{D21}

\graphicspath{ {./figures/} } 

\begin{document}

% \linenumbers

% \title[Article Title]{JWST Captures Growth of Aromatic Hydrocarbon Dust Particles in the Extremely Metal-poor Galaxy Sextans A}
\title[Article Title]{Growth of Aromatic Hydrocarbon Dust Particles in the Extremely Metal-poor Galaxy Sextans A}


\author*[1]{\fnm{Elizabeth J.} \sur{Tarantino}}\email{etarantino@stsci.edu}

\author[1]{\fnm{Julia} \sur{Roman-Duval}}

\author[2]{\fnm{Karin M.} \sur{Sandstrom}}

\author[3]{\fnm{Cory M.} \sur{Whitcomb}}

\author[3]{\fnm{J.-D. T.} \sur{Smith}}

\author[4]{\fnm{Bruce T.} \sur{Draine}}

\author[1]{\fnm{Martha L.} \sur{Boyer}}

\author[5]{\fnm{J\'er\'emy} \sur{Chastenet}}

\author[6, 7]{\fnm{Ryan} \sur{Chown}}

\author[8]{\fnm{Christopher J. R.} \sur{Clark}}

\author[1]{\fnm{Meriem} \sur{Elyajouri}}

\author[1]{\fnm{Karl D.} \sur{Gordon}}

\author[9]{\fnm{Brandon S.} \sur{Hensley}}

\author[10]{\fnm{Thomas S.-Y.} \sur{Lai}}

\author[11, 1, 12]{\fnm{Christina W.} \sur{Lindberg}}

\author[1, 13]{\fnm{Kristen B. W.} \sur{McQuinn}}

\author[1]{\fnm{Max J. B.} \sur{Newman}}

\author[14]{\fnm{O. Grace} \sur{Telford}}

\author[15]{\fnm{Dries} \sur{Van De Putte}}

\author[16]{\fnm{Benjamin F.} \sur{Williams}}

\affil[1]{Space Telescope Science Institute, 3700 San Martin Drive, Baltimore, MD 21218, USA}

\affil[2]{Department of Astronomy \& Astrophysics, University of California, San Diego, La Jolla, CA 92093, USA}

\affil[3]{Ritter Astrophysical Research Center, Department of Physics \& Astronomy, University of Toledo, Toledo, OH 43606, USA}

\affil[4]{Department of Astrophysical Sciences, Princeton University, Princeton, NJ 08544, USA}

\affil[5]{Sterrenkundig Observatorium, Universiteit Gent, Krijgslaan 281 S9, B-9000 Gent, Belgium}

\affil[6]{Faculty of Computer Science \& Technology, Algoma University, Sault Ste. Marie, ON P6A 2G4, Canada}

\affil[7]{Department of Astronomy, The Ohio State University, 140 West 18th Avenue, Columbus, OH 43210, USA}

\affil[8]{AURA for the European Space Agency, Space Telescope Science Institute, 3700 San Martin Drive, Baltimore, MD 21218, USA}

\affil[9]{Jet Propulsion Laboratory, California Institute of Technology, 4800 Oak Grove Drive, Pasadena, CA 91109, USA}

\affil[10]{IPAC, California Institute of Technology, 1200 East California Boulevard, Pasadena, CA 91125, USA}

\affil[11]{Center for Astrophysics $\vert$ Harvard \& Smithsonian, 60 Garden St., Cambridge, MA 02138, USA}

\affil[12]{The William H. Miller III Department of Physics \& Astronomy, Bloomberg Center for Physics and Astronomy, Johns Hopkins University, 3400 N. Charles Street, Baltimore, MD 21218, USA}

\affil[13]{Department of Physics and Astronomy, Rutgers the State University of New Jersey, 136 Frelinghuysen Rd., Piscataway, NJ, 08854, USA}

\affil[14]{Department of Physics and Astronomy, University of Utah, 270 S 1400 E, Salt Lake City, UT 84112, USA}  

\affil[15]{Department of Physics \& Astronomy, The University of Western Ontario, London ON N6A 3K7, Canada}

\affil[16]{Department of Astronomy, University of Washington, Box 351580, U.W., Seattle, WA 98195-1580, USA}

%==================================%%
% Sample for unstructured abstract %%
%==================================%%

\abstract{
\unboldmath
The mid-infrared spectrum of star-forming, high metallicity galaxies is dominated by emission features from aromatic and aliphatic bonds in small carbonaceous dust grains, often referred to as polycyclic aromatic hydrocarbons (PAHs). In metal-poor galaxies, the abundance of PAHs relative to the total dust sharply declines, but the origin of this deficit is unknown. We present JWST observations that detect and resolve emission from PAHs in the 7\% Solar metallicity galaxy Sextans A, representing the lowest metallicity detection of PAH emission to date. In contrast to higher metallicity galaxies, the clumps of PAH emission are compact (0.5-1.5\arcsec or 3-10 pc), which explains why PAH emission evaded detection by lower resolution instruments. Ratios between the 3.3, 7.7, and 11.3~\micron\ PAH features indicate that the PAH grains in Sextans A are small and neutral, with no evidence of significant processing from radiation fields. These results favor inhibited grain growth over enhanced destruction as the origin of the low PAH abundance in Sextans A. The compact clumps of PAH emission are likely active sites of in-situ PAH growth within a dense, well-shielded phase of the interstellar medium. Our results show that PAHs can form and survive in extremely metal-poor environments common early in the evolution of the Universe.
}

% ================================%%
% Sample for structured abstract %%
% ================================%%

\maketitle

\section{Introduction}\label{sec:intro}

The mid-infrared (MIR) spectrum of interstellar material is dominated by broad emission features, at 3.3, 3.4, 5.3, 5.6, 6.3, 7.7, 8.6, 11.3, 12.6, and 17.3 \micron , that are attributed to vibrational modes of carbonaceous nanoparticles often referred to as polycyclic aromatic hydrocarbons (PAHs; \cite{Knacke1977, Allamandola1985, Tielens2008, Li2020}). Because PAH features are excited by ultraviolet (UV) photons, they are bright in regions powered by young massive stars, accounting for 5-20\% of the total IR luminosity of high metallicity galaxies \cite{Smith2007}. As a result, PAH emission is used as a tracer of star formation rate \citep[SFR;][]{Shipley2016, Ronayne2024} and molecular gas \citep{Cortzen2019, ShivaeiBoogaard2024, Chown2025_phangs} in nearby and distant galaxies. In neutral gas, PAHs drive the heating through the photoelectric effect \citep{Wolfire1995} and regulate the ionization balance via nanoparticle-mediated recombination of ions and electrons.

One of the main results from previous IR telescopes is the strong dependence of PAH flux on metallicity \citep[e.g.,][]{Engelbracht2008, Jackson2006, Wu2006, Draine2007, Gordon2008, Hunt2010, Sandstrom2010, Chastenet2019, Aniano2020, Whitcomb2024}. The fraction of PAHs in the overall dust mass drastically decreases around a metallicity of 12 + log(O/H) $\sim$ 8.2 ($\sim 30\% \, \rm Z_{\odot}$) \citep{Draine2007, Aniano2020, Whitcomb2024}. While the origin of this \changes{weak PAH emission} is unknown, it is generally attributed to either enhanced destruction and/or inhibited formation of these small grains. In the enhanced destruction scenario, low dust abundances in metal-poor systems may lead to reduced shielding of PAHs from far-ultraviolet radiation \citep{Remy-Ruyer2014}. PAH compounds are more fragile than larger dust grains \citep{Micelotta2010}, which can result in preferential destruction of PAHs in these metal-poor systems \cite{Engelbracht2008, Jackson2006, Gordon2008, Hunt2010}. Similarly, PAHs may also be destroyed through interactions with electrons in ionized gas \citep{Draine2007, Micelotta2010} or interstellar shocks \citep{Micelotta2010a}. The other scenario to explain the \changes{low abundance of PAHs in metal-poor systems} is through the suppression of formation mechanisms. PAH growth in the ISM may be inhibited at low metallicities due to an under-abundance of carbon, and/or less dense gas available to support efficient grain growth \citep{Greenberg2000, Sandstrom2012, Whitcomb2024}. 
% Another theory suggests that PAHs form through the shattering of larger dust grains, and that reduced dust abundances therefore suppress the creation of PAHs from fragmentation \cite{Jones1997, Rau2019}. 

Understanding the \changes{origin of the metallicity dependence of PAH abundance} is therefore directly tied to the formation, destruction, and overall lifecycle of these small grains. Establishing a framework for PAHs at low metallicities is thus vital for interpreting PAH observations, particularly at high redshift where detections are becoming common and average metallicities are lower \cite[e.g.][]{Shipley2016, Spilker2023, ShivaeiBoogaard2024, McKinney2026}. While observations with \textit{Spitzer} made tremendous progress in revealing the \changes{reduced PAH abundance at low metallicity}, the enhanced sensitivity, spectral coverage, and spatial resolution of JWST is transforming our view of nearby galaxies. Several JWST spectroscopic and photometric programs have targeted metal-poor systems, but PAHs have remained undetected below $Z < 10\% \, \rm Z_{\odot}$ \citep{Hirschauer2024, Chown2025_dwarf, Mingozzi2025, Hunt2025b, Telford2025, DelValle-Espinosa2026}. 

In this work, we leverage the exquisite sensitivity and resolution of JWST to target the nearby, star-forming, metal-poor, dwarf irregular galaxy Sextans~A to detect and characterize PAH emission at very low metallicity. Sextans~A has an oxygen abundance of $12 + \rm log(O/H) = 7.54$ \citep{Kniazev2005}, which translates to 7\% Solar metallicity when assuming $12 + \rm log(O/H) = 8.69$ for the Sun \citep{Asplund2009}. Sextans~A resides at the outer edge of the Local Group, at a distance of 1.4~Mpc \citep{Bellazzini2014, McQuinn2017}. The mass of Sextans~A is dominated by atomic gas, with a $M_{\rm HI} = 6.2 \times 10^{7} \, \rm M_{\odot}$ \citep{Ott2012} and a stellar mass of $M_{\rm *} = 4.4 \times 10^{7} \, \rm M_{\odot}$ \citep{McConnachie2012}. The UV-derived SFR of is $\rm SFR = 1.2 \times 10^{-2} \, \mathrm{ M_{\odot} \ yr^{-1}}$ \citep{Lee2009}, indicating there is ongoing star formation within the galaxy. Sextans~A has a previous detection of cold dust from \textit{Herschel} \citep{Shi2014}, making it an ideal target to detect PAH emission with JWST. 
  
\section{Discovery of compact PAH clumps at 7\% Solar metallicity} \label{sec:main}

\begin{figure}[h!]
    \centering
    \includegraphics[width=\textwidth]{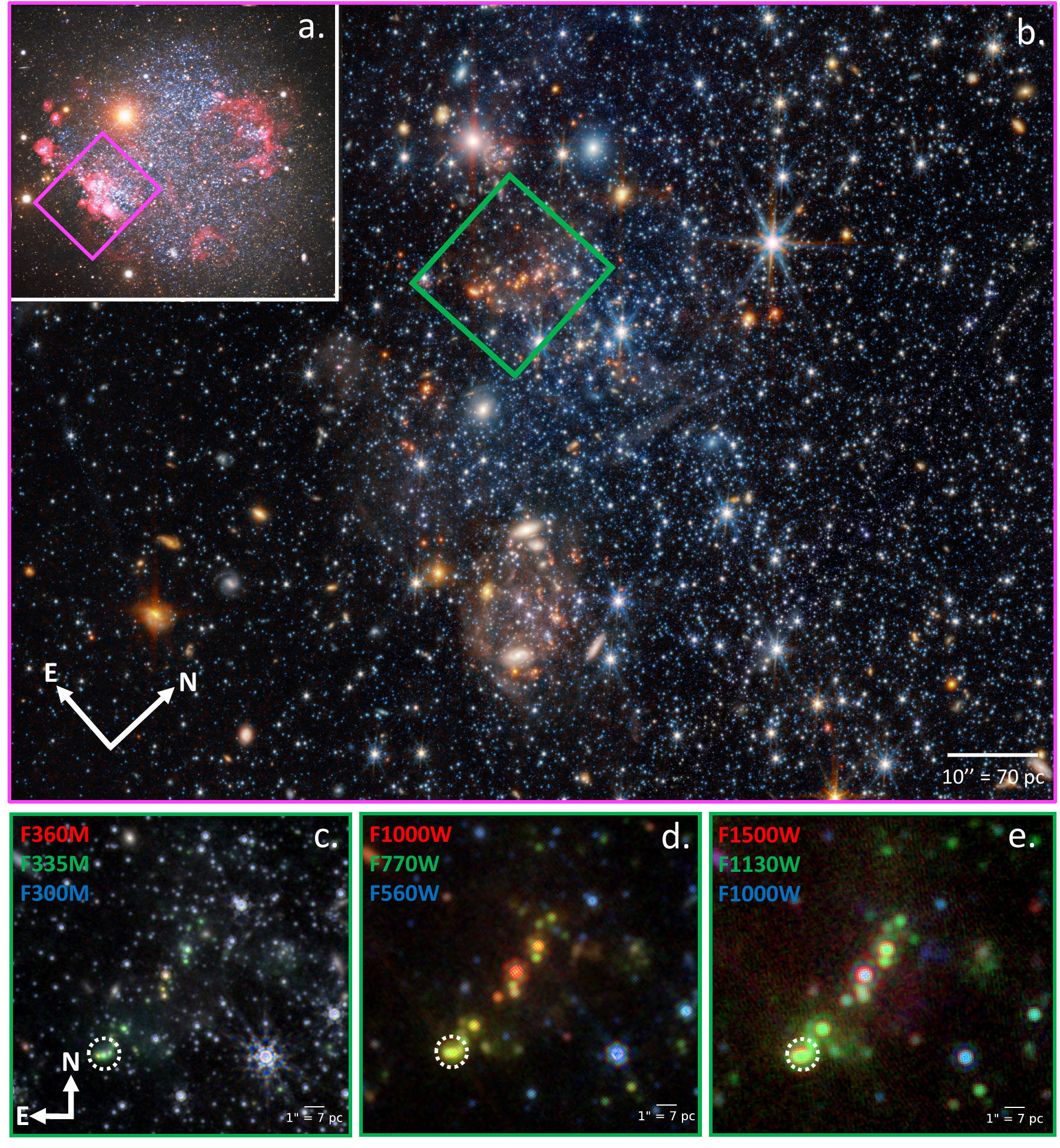}
    \caption{\textbf{JWST NIRCam and MIRI photometry of Sextans~A reveals emission from PAHs} a.) Optical image of Sextans~A (Credit: KPNO/NOIRLab/NSF/AURA) with the JWST field outlined in magenta. b.) JWST color composite image of Sextans~A (Blue: F115W, Cyan: F150W + F200W, Green: F335M, Yellow: F560W, Orange: F770W, Red: F1000W; Credit: A. Pagan, STScI). The region of brightest PAH emission is outlined in green. c.--e.) 3-color zoom-ins on this region, with the PAH filter in green and adjacent continuum filters in red and blue (c. Red: F360M, Green: F335M, Blue: F300M; d. Red: F1000W, Green: F770W, Blue: F560W; e. Red: F1500W, Green: F1130W, Blue: F1000W). We apply the Lupton RGB algorithm \cite{Lupton2004} to visualize the high dynamic range of the JWST data, using the same \textit{asinh} stretch and scaling for all panels. The compact green clumps trace enhanced emission from the PAH 3.3, 7.7, and 11.3 \micron\ features, with the brightest clump highlighted by a circle.} 
    \label{fig:Optical_JWST}
\end{figure}

We obtained photometry of Sextans~A with JWST NIRCam and MIRI imaging (PID 2391; PI: J. Roman-Duval), focusing on filters centered on PAH emission$-$ F335M, F770W, and F1130W$-$ as well as flanking filters to estimate the continuum level (see Section \ref{sec:obs} for details). The observations consist of a single NIRCam and MIRI deep field towards the brightest star-forming region in Sextans~A with a previous dust detection \citep{Shi2014, Jackson2006}, marked by the magenta rectangle in the optical image of Sextans~A shown in Figure~\ref{fig:Optical_JWST}a. The combined NIRCam and MIRI photometric imaging is presented as a multi-color image in \ref{fig:Optical_JWST}b, where the region of brightest PAH emission is encompassed by the green rectangle. The bottom panels (c. -- e.) zoom in on this region, with each PAH filter shown in green and the flanking continuum filters in red and blue. The green color indicates enhanced emission at the 3.3, 7.7, and 11.3 \micron\ PAH features, demonstrating the presence of PAHs in this extremely low-metallicity galaxy. The emission from PAHs is compact, arising from \changes{$0.5-1.5 \arcsec$ ($\rm 3-10~pc$)} sized clumps.

\begin{figure}[h!]
    \centering
    \includegraphics[width=\textwidth]{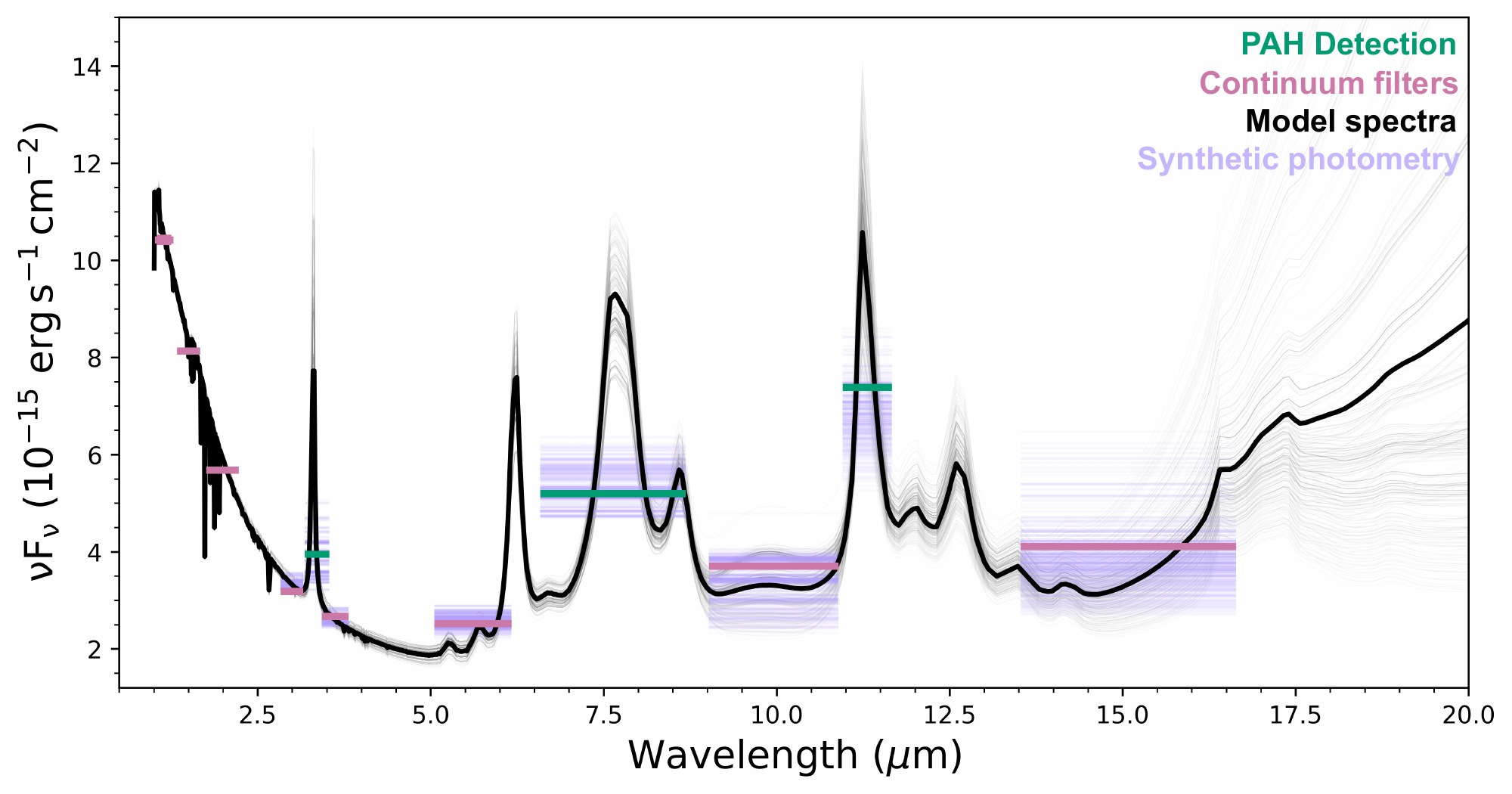}
    \caption{\textbf{SED of the brightest clump of PAH emission in Sextans~A}. The flux from clump 1 (encircled in Figure \ref{fig:Optical_JWST}c.--e) is shown by magenta and green horizontal lines, with green denoting the PAH-centered filters (F335M, F770W, F1130W). \changes{The horizontal line widths} indicate the wavelength range where the filter transmission falls to 50\% of its maximum. Errorbars are smaller than the \changes{vertical line thickness}. The excess flux in the PAH filters (green) relative to the continuum filters (magenta) demonstrates the detection of PAHs. PHOENIX stellar models \cite{Husser2013} and \citetalias{Draine2021} PAH/dust emission spectral models that fit the data within uncertainties are presented as the gray spectra, with the transparency scaled by the $\chi^2$ and the average spectrum presented in black. The spread in models illustrates how different ISM conditions (e.g., radiation field, neutral gas column density, PAH size and charge distribution) can produce equally valid fits to the photometric data (see Section~\ref{sec:D21_SED}). The purple horizontal lines show the synthetic photometry of these models for direct comparison with the observations.}
    % The average of all the well-fitting models is the black spectrum. The blue horizontal lines show the synthetic photometry of the gray Phoenix and \citetalias{Draine2021} model spectra. The excess in emission at the PAH feature filters (green) relative to the bracketing continuum filters (pink) indicates the detection of PAHs. \red{put labels on figure itself}}
    \label{fig:SED}
\end{figure}

To confirm the detection of PAHs in Sextans~A, we analyze the spectral energy distribution (SED) of the brightest clump of PAH emission in Figure \ref{fig:SED} (encircled in Panels \ref{fig:Optical_JWST}c.--e.). We extract the flux density of all filters from the JWST observations and compare the observed SED to MIR model of dust emission spectra from \citet{Draine2021} (hereafter \citetalias{Draine2021}) and near-infrared stellar spectra from PHOENIX models \cite{Husser2013} (see \ref{sec:D21_SED} for more details). This SED shows an uptick in the flux for each PAH filter (green) relative to the bracketing continuum filers (magenta), indicating PAH emission. The detection is seen in the three independent PAH features: 3.3, 7.7, and 11.3 \micron , ensuring that it is robust and not from a spurious high redshift galaxy. At 7\% Solar metallicity, these observations are the lowest metallicity detection of PAH emission to date.

\begin{figure}[h!]
    \centering
    \includegraphics[width=0.99\textwidth]{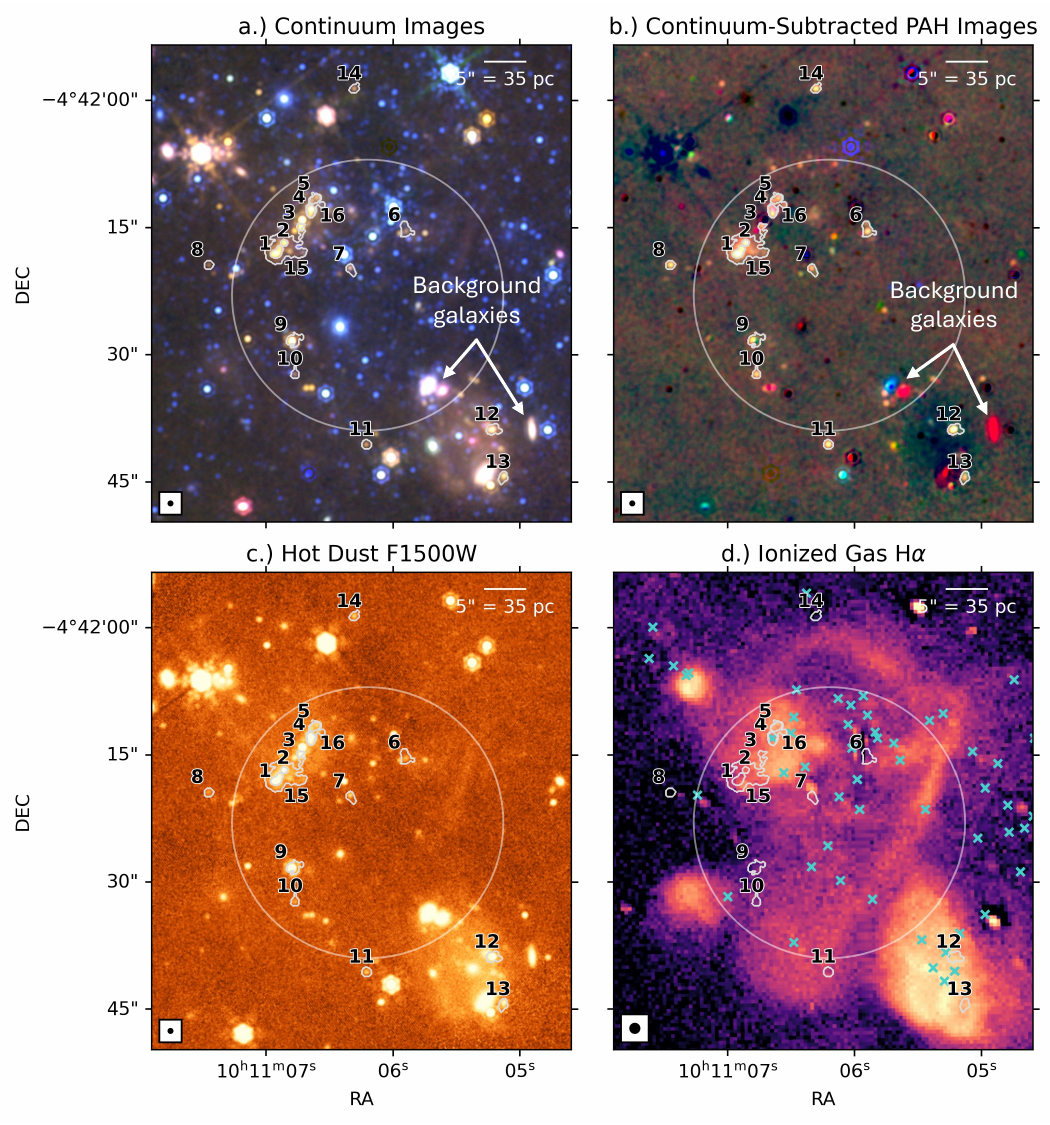}
    \caption{\textbf{PAH emission clumps and their spatial distribution in Sextans~A} Clump structures and labels defined with the dendrogram structure analysis (see Section \ref{sec:dendro}) overlaid on Sextans~A JWST and H$\alpha$ data. Clump properties are described in Table \ref{tab:clump_prop}. We require a 3$\sigma$ detection in all three PAH bands$-$ 3.3, 7.7, and 11.3 \micron$-$ to define a PAH emission clump. We present the clumps on: a.) RGB 3-color images of the stellar and hot dust continuum and b.) the continuum-subtracted PAH flux in each PAH band, where the red color is 11.3 \micron, green is 7.7 \micron, and blue is 3.3 \micron\ all scaled at the same \textit{asinh} stretch to emphasize extended emission. Emission from continuum dominates in Sextans~A, as there are many more features in the continuum image compared to the PAH data. The continuum-subtracted PAH data show the importance of requiring a detection in all three PAH bands, as spurious background galaxies can be bright in a single band, resulting in the red and blue artifacts in the PAH RGB image. PAH clumps are visible as the white clumps of emission outlined from the clump boundary. The white circle marks the region defined in \citet{Shi2014} used to calculate the total infrared luminosity and compare to the PAH flux. c.) Observations of the F1500W filter, a high resolution probe of the hot dust as the reddest available filter in our JWST dataset. PAH clumps frequently appear near regions of hot dust, though the hot dust exhibits a more extended spatial distribution, especially in the southern region near clumps 12 and 13. d.) H$\alpha$ observations of Sextans~A from \citet{Kennicutt2008} that trace the ionized gas, which shows similar spatial distribution with the F1500W hot dust map. We overlay the massive star catalog from \citet{Lorenzo2022}, including O and early-type B stars, as the cyan ``x'' symbols. Some PAH clumps are spatially coincident with regions of H$\alpha$ emission, though they generally do not coincide with the brightest H$\alpha$ peaks, suggesting a complex geometry between the photodissociation region (PDR) and \hii\ region.}
    \label{fig:PAH_clumps}
\end{figure}

We develop a continuum subtraction method that removes the stellar and hot dust continuum contribution to each PAH filter in order to isolate the PAH flux. This method expands on previous work with the F335M NIRCam filter from \citet{Sandstrom2023PAH3um} to also include an approach for the MIRI filters and is described in detail in Section \ref{sec:con_sub}. Figure \ref{fig:PAH_clumps}a shows an RGB 3-color image of the the stellar and hot dust continuum while Figure \ref{fig:PAH_clumps}b shows the continuum-subtracted PAH flux at 11.3 (red), 7.7 (green) and 3.3 (blue) \micron. \changes{Our method measures the total PAH flux within each filter and may include minor contributions from adjacent features (e.g., the 8.3 \micron\ feature in F770W). Because the primary feature dominates the filter flux, we refer to the continuum subtracted measurements as the 3.3, 7.7, and 11.3 \micron\ PAH fluxes.} 
% We also zoom in on the green rectangle highlighted in Figure \ref{fig:Optical_JWST} and show the continuum-subtracted PAH flux for each feature in Figure \ref{fig:consub_PAH_with_F1500W}.

As Figure \ref{fig:PAH_clumps} and Extended Data Figure \ref{fig:consub_PAH_with_F1500W} show, the clumps of PAH emission exhibit hierarchical structure, where smaller clumps are nested within larger ones. We therefore use dendrograms to identify and measure the properties of the PAH clumps (see Section \ref{sec:dendro} for more details). A PAH clump is defined by a $3\sigma$ detection in all three continuum-subtracted PAH images, ensuring the detection is robust and not due to an interloping galaxy. The resulting clump envelopes and labels are outlined in white in Figure \ref{fig:PAH_clumps}. Background galaxies appear bright in only a single band (e.g., red at 11.3~\micron), demonstrating the need for $3\sigma$ detections across all PAH bands. The properties of each clump, including the position, integrated continuum subtracted PAH flux, and size are given in Table \ref{tab:clump_prop}. The \changes{average radii} of these clumps are 0.3$\arcsec$ (2~pc) and range from 0.18 -- 0.72$\arcsec$ (1.22 -- 4.89 pc). Since the spatial resolution of the data (after convolving to 15~\micron\ resolution) is $0.488\arcsec$ (3.3~pc), many of the clumps are unresolved even at the exquisite resolution of JWST.

\begin{table}[h] 
\caption{Clump Properties}\label{tab:clump_prop} 
\begin{tabular}{ccccccccc} 
\toprule 
Clump & RA & Dec. & Radius & Major Axis & Minor Axis & 3.3\,$\mu$m Flux & 7.7\,$\mu$m Flux & 11.3\,$\mu$m Flux \\ 
& (10h 11m s) & (-04$^{\circ} \ 42' \ ''$) & (pc) & (pc) & (pc) & ($\mu$Jy) & ($\mu$Jy) & ($\mu$Jy) \\ 
\midrule 
1 & 06.915 & 17.993 & 1.97 & 2.58 & 1.50 & 1.70 $\pm$ 0.04 & 5.69 $\pm$ 0.06 & 14.15 $\pm$ 0.20 \\
2 & 06.857 & 16.798 & 1.20 & 1.22 & 1.18 & 0.53 $\pm$ 0.02 & 1.24 $\pm$ 0.02 & 4.38 $\pm$ 0.10 \\
3 & 06.727 & 15.057 & 1.99 & 2.96 & 1.33 & 0.13 $\pm$ 0.01 & 0.39 $\pm$ 0.02 & 1.70 $\pm$ 0.05 \\
4 & 06.648 & 13.163 & 1.81 & 2.33 & 1.41 & 0.83 $\pm$ 0.02 & 1.22 $\pm$ 0.02 & 6.84 $\pm$ 0.11 \\
5 & 06.612 & 11.647 & 1.72 & 2.26 & 1.31 & 0.31 $\pm$ 0.01 & 0.76 $\pm$ 0.02 & 2.33 $\pm$ 0.05 \\
6 & 05.898 & 15.334 & 2.17 & 2.57 & 1.84 & 0.38 $\pm$ 0.02 & 1.08 $\pm$ 0.03 & 2.51 $\pm$ 0.06 \\
7 & 06.332 & 19.912 & 1.71 & 2.24 & 1.30 & 0.10 $\pm$ 0.01 & 0.50 $\pm$ 0.02 & 1.28 $\pm$ 0.04 \\
8 & 07.451 & 19.418 & 1.40 & 1.44 & 1.36 & 0.17 $\pm$ 0.01 & 0.76 $\pm$ 0.02 & 1.82 $\pm$ 0.05 \\
9 & 06.790 & 28.257 & 2.09 & 2.53 & 1.74 & 0.27 $\pm$ 0.02 & 1.28 $\pm$ 0.03 & 3.45 $\pm$ 0.08 \\
10 & 06.771 & 32.283 & 1.44 & 1.59 & 1.31 & 0.13 $\pm$ 0.01 & 0.63 $\pm$ 0.02 & 1.78 $\pm$ 0.05 \\
11 & 06.206 & 40.610 & 1.43 & 1.44 & 1.42 & 0.30 $\pm$ 0.01 & 1.03 $\pm$ 0.02 & 2.66 $\pm$ 0.06 \\
12 & 05.217 & 38.846 & 1.73 & 1.97 & 1.52 & 0.76 $\pm$ 0.02 & 2.32 $\pm$ 0.04 & 4.57 $\pm$ 0.10 \\
13 & 05.131 & 44.443 & 1.69 & 2.06 & 1.38 & 0.19 $\pm$ 0.01 & 0.72 $\pm$ 0.02 & 1.06 $\pm$ 0.04 \\
14 & 06.303 & 58.643 & 1.45 & 1.61 & 1.30 & 0.22 $\pm$ 0.01 & 0.89 $\pm$ 0.02 & 2.15 $\pm$ 0.05 \\
15\tnote{\dag} & 06.888 & 17.685 & 4.87 & 6.02 & 3.94 & 3.83 $\pm$ 0.06 & 11.43 $\pm$ 0.09 & 32.89 $\pm$ 0.26 \\
16\tnote{\ddag} & 06.629 & 12.532 & 3.64 & 5.61 & 2.37 & 1.28 $\pm$ 0.03 & 2.34 $\pm$ 0.04 & 10.56 $\pm$ 0.13 \\
\botrule 
\end{tabular} 
\item[\dag]Clump 15 is a larger structure that encompasses clumps 1 and 2
\item[\ddag]Clump 16 is a larger structure that encompasses clumps 4 and 5
\footnotetext{Table Notes: The properties of the PAH clumps defined through dendrogram structure analysis, using the continuum-subtracted 7.7\,$\mu$m data to calculate the center and size of the clumps. The centers of the clumps are reported in right ascension and declination in columns 2 and 3. The radius, major axis, and minor axis of each clump are reported in columns 4 through 6. The continuum-subtracted \changes{PAH flux density in the F335M, F770W, and F1130W filters (see Section \ref{sec:con_sub})}, are reported in columns 7 through 9.}
\end{table} 

The dendrogram analysis as well as Figure \ref{fig:PAH_clumps} all show that the PAH clumps in Sextans~A are small and compact, about $0.5-1.5 \arcsec$, or $\rm 3-10~pc$ in diameter. Higher metallicity galaxies, in contrast, have PAH emitting regions as large as 500~pc or more \cite[e.g.,][]{Smith2007, Sandstrom2023PAHISM, Sutter2024}. Even in the 20\% Solar metallicity Small Magellanic Cloud (SMC), typical sizes of PAH emitting regions are comparable to molecular cloud sizes, around 25~pc or greater \citep{Sandstrom2010, Sandstrom2012, Chastenet2019, Clark2025}. The small size of the PAH clumps in Sextans~A is instead reminiscent of the tiny cores of molecular gas found in $\lesssim 13\%$ Solar metallicity galaxies WLM and Sextans~B \cite{Rubio2015, Shi2020}. Given that PAHs and CO are well correlated in high metallicity galaxies \cite{Chown2025_phangs}, we predict the small clumps of PAH emission trace the molecular cores in Sextans~A and share similar conditions for formation and survival. Because the dust-to-gas ratio decreases with metallicity \citep{Remy-Ruyer2014}, a larger gas column is required to reach the same $\rm{A}_V$ (a measure of the dust shielding) as in a more metal-rich ISM \citep{Bolatto1999, Wolfire2010}, leading to an overall smaller physical scale where dust shielding is effective. The small size of the PAH clumps therefore suggests that dust shielding is playing an important role in survival and growth of PAHs.

\begin{figure}[h!]
    \centering
    \includegraphics[width=\textwidth]{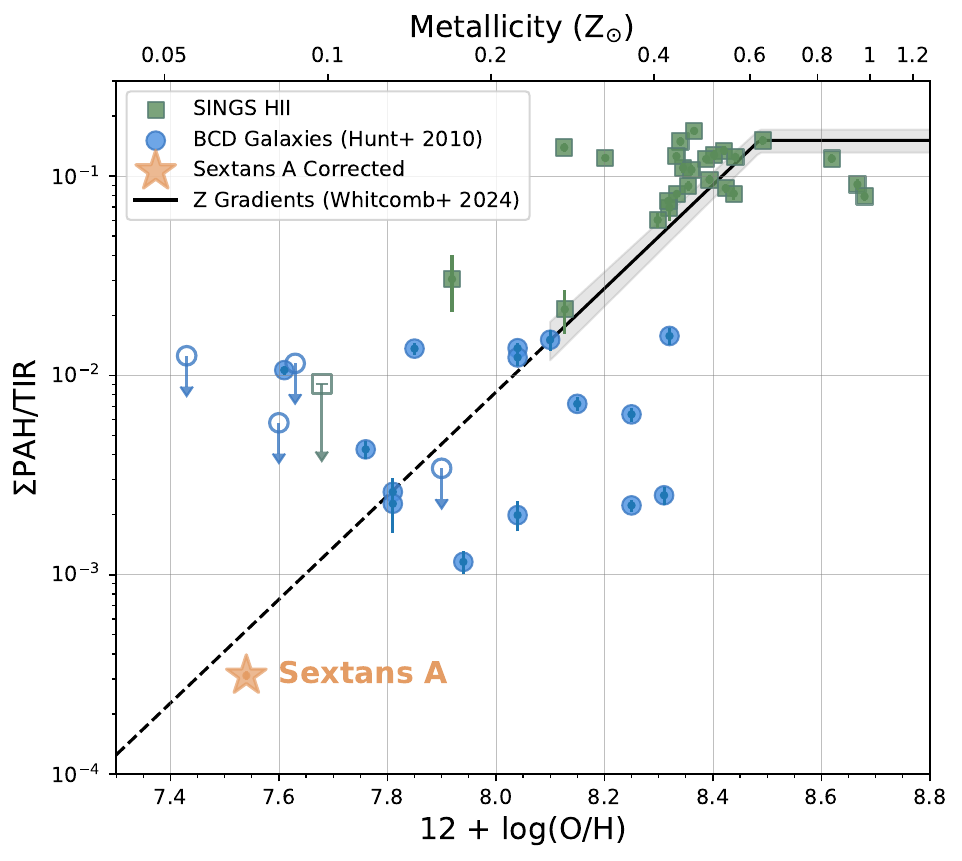}
    \caption{\textbf{The fraction of infrared power emitted by PAH features as a function of metallicity} Comparison of $\Sigma$PAH/TIR, the ratio between the integrated intensity of detected PAH features and the total infrared luminosity, across a range of galaxy samples. Sextans~A is shown as the orange star, the sample of metal-poor blue compact dwarf (BCD) galaxies from \citet{Hunt2010} are displayed as blue circles, and the higher metallicity non-AGN SINGS galaxies from \citet{Smith2007} are the green squares. The errorbars represent the uncertainty on the PAH feature fits and the calculation of TIR. We also plot the empirical fit of $\Sigma$PAH/TIR as a function of the metallicity gradients for the galaxies M101, NGC~628, and NGC~2403 from \citet{Whitcomb2024} in the black line, with the uncertainty presented as gray bars and the extrapolation shown as the dashed line. Sextans~A is the lowest metallicity galaxy with a robust detection of PAH emission and generally $\Sigma$PAH/TIR follows the trends with metallicity seen in the other samples. \changes{We note that $\Sigma$PAH calculated for Sextans~A includes the overall brightest PAH features, which are captured within the F335M, F770W, and F1130W filter bandpasses, but may be slightly underestimated compared to the SINGS and BCD sample that have spectra coverage through 5.5 -- 20 \micron.} Figure adapted from \citet{Hunt2010}.}
    \label{fig:sigmaPAH_TIR}
\end{figure}

To quantify the PAH abundance in Sextans A, we calculate the ratio of the summed PAH feature luminosities to the total infrared luminosity ($\Sigma$PAH/TIR), which captures the fraction of infrared power emitted by PAH features \citep[e.g.,][]{Smith2007, Hunt2010, Whitcomb2024} \changes{in Section \ref{sec:sigma_pah}}. This ratio serves as a proxy for $q_{\rm PAH}$, the fraction of the overall dust budget in PAHs by mass, with small dependence on the radiation field strength at very high radiation fields. In Figure \ref{fig:sigmaPAH_TIR}, we plot $\Sigma$PAH/TIR for Sextans~A (this work) and compare to a sample of low metallicity galaxies presented in \citet{Hunt2010} and the higher metallicity SINGS (The SIRTF Nearby Galaxies Survey) sample \citep{Smith2007}. We follow the procedure in \citet{Draine2007} to calculate the total infrared luminosity in Sextans~A, assuming a distance of 1.4 Mpc \citep{McQuinn2017}. We use the FIR fluxes from \textit{Spitzer} and \textit{Herschel} reported in the $32\arcsec$ region in \citet{Shi2014}, which is displayed as the circle in Figure \ref{fig:PAH_clumps}. After applying a 3$\sigma$ signal-to-noise ratio (SNR) cut to the 3.3, 7.7, and 11.3~\micron\ continuum subtracted PAH data, we sum these three PAH features in this region to calculate $\Sigma$PAH and \changes{plot $\Sigma \rm PAH/TIR = 0.00031$ in Figure \ref{fig:sigmaPAH_TIR}}. \changes{We note that this $\Sigma$PAH value includes the overall brightest PAH features, as captured within the F335M, F770W, and F1130W filter bandpasses, but may miss additional fainter features (e.g., 6.2 \micron, 12.6 \micron) included in the comparison samples shown in Figure \ref{fig:sigmaPAH_TIR}.} We also overplot and extrapolate the trend from \citet{Whitcomb2024}, which fits $\Sigma$PAH/TIR across the metallicity gradients of three nearby galaxies M101, NGC~628, and NGC~2403. The Sextans~A $\Sigma$PAH/TIR measurement is in general agreement with the trend of $\Sigma$PAH/TIR versus metallicity measured in other samples. In higher metallicity star-forming galaxies, $\Sigma$PAH/TIR\,$\sim$\,$10-20\%$ \citep{Smith2007} but declines to a range of $0.1-1.6\%$ in lower metallicity galaxies \cite{Hunt2010}. \changes{Our Sextans~A measurements extends this correlation to $\rm \Sigma PAH/TIR = 0.031\%$ at 7\% Solar metallicities, the lowest metallicity detection to date.}

The difference in spatial resolution between the FIR data ($\sim30\arcsec$) and the PAH clumps ($\sim1\arcsec$) makes it difficult to determine $\Sigma{\mathrm{PAH}}$/TIR at the scale of the PAH clumps. If the overall dust distribution is less concentrated on smaller scales, the resulting $\Sigma{\mathrm{PAH}}$/TIR would likely be higher in the clumps. To test whether the PAHs are more compact than the overall dust distribution, we use the F1500W filter as a proxy for the stochastically heated small dust grain population since F2100W is not available to calculate $\rm R_{PAH}$ \citep[e.g.,][]{Chastenet2023_RPAH, Sutter2024}. While F1500W does contain some PAH emission from the wings of the fainter 12.8, 13.6, 14.8, and 16.4 features \citep{Tielens2008}, as our reddest filter, it is the best probe of the stochastically heated hot dust distribution that has matched resolution to the PAH emission. We use the same TIR area defined in \citet{Shi2014}, which is similar to the FIR beam size ($32\arcsec$), and require a 3$\sigma$ pixel cutoff to sum the emission in the F1500W filter. \changes{We measure $\Sigma$PAH/F1500W = 0.64 for the region.} We then compare $\Sigma$PAH/F1500W for the individual PAH clumps and find that every PAH clump \changes{except clumps 3 and 9} has a $\Sigma$PAH/F1500W greater than the value across the TIR region. We calculate an \changes{average of $\Sigma$PAH/F1500W = 2.3, with a range of $\Sigma$PAH/F1500W = 0.35 - 4.8}, much larger than the value for the $32\arcsec$ circle. We interpret the difference in $\Sigma$PAH/F1500W measured over the FIR beam versus the PAH clumps as evidence that PAHs are confined to the small clumps, while non-PAH small grains reside in a more extended distribution. This matches the visual appearance of the JWST maps (see Figures \ref{fig:PAH_clumps} and Extended Data Figure \ref{fig:consub_PAH_with_F1500W}) and the analysis of a clump radial profile (see Section \ref{sec:radprof}). Given that the spatial distribution of cold dust emitting in the FIR is expected to be more spatially extended than the dust traced by 15 $\mu$m emission, the $\Sigma$PAH/TIR measured for Sextans~A within the FIR beam is likely significantly lower than it would be if measured on the small scale of PAH clumps. This suggests that the \changes{weak PAH emission found} at low metallicities may be partially driven by the small filling factor of PAHs relative to the more extended dust distribution, though it is not the only effect driving the low PAH abundance. The previous \textit{Spitzer}-based measurements in Figure \ref{fig:sigmaPAH_TIR} probe much larger physical scales ($\sim$500 pc or more), meaning that any such PAH beam dilution effect would have been unresolved in earlier studies.

\section{Discussion}

Theoretical models and laboratory experiments indicate that the relative PAH feature strengths depend on the PAH grain size, ionization state, and the incident radiation field \citep{Allamandola1999, Tielens2008, Maragkoudakis2020}. We can estimate the properties of the PAH grains at this very low metallicity with the photometry of the 3.3 \micron , 7.7 \micron , and 11.3~\micron\ features studied in this work. As the shortest wavelength feature, the 3.3 \micron\ band is most sensitive to the smallest PAHs, which can be heated by the absorption of a single photon to temperatures hot enough to emit at the shorter infrared wavelengths \citep{Schutte1993}. Laboratory data also show that the 3.3 and 11.3~\micron\ features originate from predominantly neutral PAHs \citep{Tielens2008}. Therefore, the $\rm 3.3 \, \mu m / 11.3\,  \mu m$ ratio serves as an effective tracer of PAH grain size distribution. In contrast, the 7.7~\micron\ feature is associated with charged PAHs and the $\rm 7.7 \, \mu m / 3.3 \,  \mu m$ ratio probes the degree of ionization in the PAH population \citep{Allamandola1999}. 

\begin{figure}[h!]
    \centering
    \includegraphics[width=\textwidth]{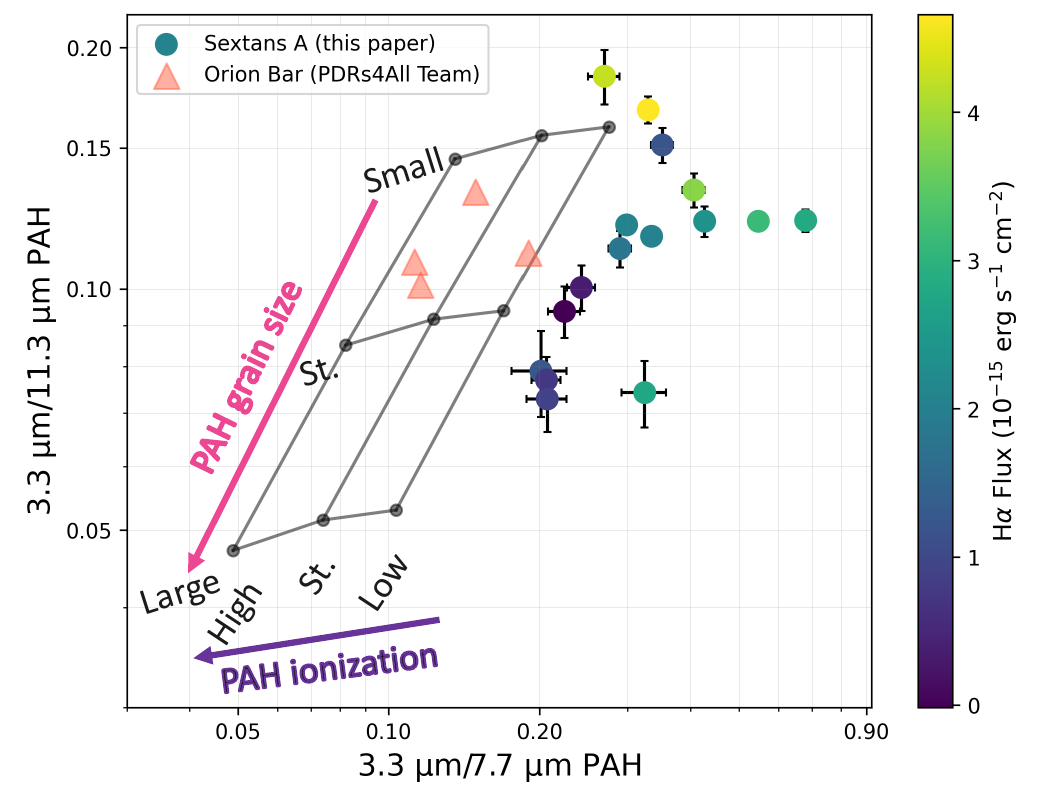}
    \caption{\textbf{Ratios between PAH features in Sextans~A} Band ratios for the PAH emission clumps defined in Table \ref{tab:clump_prop}, using the continuum-subtracted fluxes for the 3.3, 7.7, and 11.3 \micron\ features, \changes{which are measured within the filter bandpasses and therefore may include slight contributions from adjacent PAH features.} The errorbars represent the uncertainty on the 3.3, 7.7, and 11.3 \micron\ fluxes. The points are colored by the H$\alpha$ flux from the LVL survey \citep{Dale2009} and show a correlation ($r_{\rm sp} \sim 0.6$) between the H$\alpha$ flux and the $\rm 3.3 \, \mu m / 11.3\,  \mu m$ and $\rm 3.3 \, \mu m / 7.7\,  \mu m$ ratios. We overlay model predictions from \citetalias{Draine2021} with the low metallicity $Z=0.0004 \approx 0.02 Z_{\odot}$, 10 Myr starburst from \citet{Bruzual2003} as the input spectrum and average the results for radiation field intensities of $\log U=0-3.5$ since there is little scatter. The model grids show band ratio predictions for the size of PAHs, ranging from small ($a_{\rm 01} = \rm 3 \AA$) to large ($a_{\rm 01} = 5 \rm \AA$) size distributions, and the ionization of the PAHs, with standard (St.) values corresponding to the ionization fraction given in \citet{DraineLi2007}. Red triangles are spectroscopy from the PDRs4All team of the Orion Bar \citep{Chown2024, Peeters2024} where we applied synthetic photometry and the same continuum subtraction as the Sextans~A PAH clumps to provide a direct comparison. Overall, the $\rm 3.3 \, \mu m / 11.3\,  \mu m$ band ratio suggests that the PAHs in Sextans~A are similar to the small and standard size distributions assumed in \citetalias{Draine2021}. The ionization grids, however, are inconsistent with the Sextans~A band ratios, suggesting that the PAHs in Sextans~A are more neutral than models assume.}
    \label{fig:band_ratios}
\end{figure}

The comparison between the $\rm 3.3 \, \mu m / 7.7 \,  \mu m$ and $\rm 3.3 \, \mu m / 11.3\,  \mu m$ ratios from the \citetalias{Draine2021} model grids and the PAH clumps observed in Sextans~A is presented in Figure \ref{fig:band_ratios} (see Section \ref{sec:band_ratios} for details). Overall, the $\rm 3.3 \, \mu m / 11.3\,  \mu m$ band ratios from the PAH clumps tend to agree with the ``small'' and ``standard'' \citetalias{Draine2021} size distributions, but there is scatter across the clumps. The $\rm 3.3 \, \mu m / 7.7 \,  \mu m$ band ratios for nearly all the clumps are inconsistent with the \citetalias{Draine2021} ionization grids, suggesting that the PAHs in Sextans~A are more neutral than the ``low ionization'' grid used in the \citetalias{Draine2021} models. \changes{We also test an additional dust model, THEMIS \citep{Jones2017}, which includes a range of grain compositions and sizes, and find that it cannot reproduce the elevated $\rm 3.3 \, \mu m / 7.7 \,  \mu m$ ratios (Section~\ref{sec:themis}).} The preference for smaller, more neutral PAHs is similar to other studies that target low metallicity regions, including the SMC \citep{Sandstrom2012}, M101 \citep{Whitcomb2024, Whitcomb2026}, II~Zw~40 \citep{Lai2025}, and 30 Doradus in the LMC \citep{Zhang2025}. 
We also plot in red triangles results from the PDRs4All spectra from the Orion Bar PDR \citep{Chown2024, Peeters2024} that, for consistency, follow the same continuum subtraction procedure as the PAH clumps in Sextans~A. In contrast to Sextans~A, the PDRs4All spectra are in agreement with both the ionization and size grid from the \citetalias{Draine2021} models, suggesting that the elevated $\rm 3.3 \, \mu m / 7.7 \,  \mu m$ ratio in Sextans~A is significant and not due to inconsistencies with the continuum subtraction.

The Sextans~A PAH clumps in Figure \ref{fig:band_ratios} are colored by H$\alpha$ data from the Local Volume Legacy (LVL) survey \citep[][]{Dale2009} to provide an estimate of the ionizing radiation. There is a moderate correlation between the $\rm 3.3 \, \mu m / 11.3\,  \mu m$ (grain size) and $\rm 3.3 \, \mu m / 7.7 \,  \mu m$ (ionization) ratios and the H$\alpha$ flux, with Spearman rank coefficients of $r_{sp} = 0.61$ and $r_{sp} = 0.68$, respectively. However, there is no correlation between the H$\alpha$ flux and the 7.7/11.3 (ionization) ratio ($r_{sp} = -0.19$). We also test the correlation with \textit{Swift} UVOT W1 data to trace the UV radiation field and find the same trends, with comparable Spearman rank correlation coefficients ($r_{sp} \sim 0.6$), indicating that the results are robust to the choice of radiation field tracer. 

The increasing $\rm 3.3 \, \mu m / 11.3\,  \mu m$ ratio with ionizing flux is opposite to what one might expect, as a common interpretation for \changes{weak PAH emission} at low metallicity is enhanced photodestruction of smaller PAHs in strong UV radiation fields \citep{Engelbracht2008, Gordon2008, Hunt2010}. Because the smallest PAHs are the most easily destroyed \citep{Micelotta2010, Micelotta2010a}, one would anticipate lower $\rm 3.3 \, \mu m / 11.3\,  \mu m$ or $\rm 3.3 \, \mu m / 7.7 \,  \mu m$ ratios in regions with stronger radiation fields. Instead, the positive correlation between these ratios and UV radiation field strength suggests enhancement of the 3.3 \micron\ feature. Rather than attributing this trend to changes in the PAH grain size distribution, we propose that the 3.3 \micron\ feature is particularly sensitive to the hardness of the radiation field (see Figure 16 in \citetalias{Draine2021}). As the radiation field becomes harder, the 3.3~\micron\ flux relative to the TIR increases more rapidly than any other PAH feature.

The lack of evidence of significant processing by the ionizing radiation in Sextans~A contradicts the hypothesis that the \changes{origin of the reduced PAH abundances} at low metallicity is driven solely by enhanced radiation fields destroying PAHs. Instead, our observations favor inhibited PAH grain growth as the origin of the low PAH abundances in Sextans~A. This interpretation aligns with the framework introduced by \citet{Whitcomb2024}, based on \textit{Spitzer} spectroscopy of metallicity gradients in M101, NGC~628, and NGC~2403 ($0.25-1.0 \, \mathrm{Z_{\odot}}$, see Figure \ref{fig:sigmaPAH_TIR}), which revealed that PAH feature power shifts to shorter wavelengths as metallicity decreases (e.g., enhanced 3.3~\micron\ relative to 11.3~\micron), implying smaller PAH grain size distributions. As the lowest metallicity galaxy with detected PAH emission to date, Sextans~A provides a critical test of the inhibited grain growth prediction for the 3.3 \micron\ PAH fraction ($\rm PAH \, 3.3 /\Sigma PAH$) at $\lesssim 10\% \, \mathrm{Z_{\odot}}$. \changes{We derive $\rm PAH , 3.3 / \Sigma PAH = 11.2\%$, slightly above the $6.3 - 9.7\%$ range predicted by \citet{Whitcomb2026}, but consistent with a more quiescent radiation field than assumed in the inhibited grain growth model.}

The ionization state and spatial distribution of PAH emission in Sextans~A also provides a key new observational constraint for understanding the behavior of these important grains at low metallicities. Figure \ref{fig:band_ratios} shows that the 7.7 \micron\ feature is under-luminous compared to the \citetalias{Draine2021} models, suggesting that the PAH population in Sextans~A may be more neutral than those models assume. The PAH ionization state is governed by the balance between photoionization and recombination with electrons and ions in the gas, and can be parameterized as $U \sqrt{T_e} / n_e$ \citep{Draine2021}. Cooler temperatures, higher densities, and weaker radiation fields therefore favor a more neutral PAH population. The faint 7.7~\micron\ feature implies that the PAHs reside in dense, cold, and shielded regions of the ISM, consistent with the compact morphology ($0.5-1.5''$ or $3-10 \rm \, pc$) of the resolved PAH clumps (see Figure \ref{fig:PAH_clumps} and Extended Data Figure \ref{fig:consub_PAH_with_F1500W}), and provide an ideal environment for the formation and survival of molecules. 

We therefore suggest that these dense, shielded clumps of PAH emission in Sextans~A represent active sites of PAH growth in the ISM. This interpretation is supported by increasing evidence that PAHs and their aromatic precursors form within molecular clouds \citep{ McGuire2021, Wenzel2024} and that extinction curve variations in the Milky Way are driven by PAH growth \citep{ZhangHensleyGreen2025}. In Sextans~A, the preference for a small PAH grain size distribution and the agreement with the inhibited PAH grain growth model, combined with the lack of evidence for significant processing by radiation fields, suggests that the PAHs were created in-situ rather than formed elsewhere. The concentration of PAHs in well-shielded clumps supports formation in these dense regions, with PAH nanoparticles surviving only a finite time in more diffuse environments outside of the clumps \citep{Allain1996, Micelotta2010, Micelotta2010a, Sandstrom2010, Sandstrom2012}. Furthermore, since the PAH growth timescale scales roughly with density and inversely with metallicity, growth in Sextans~A is expected to proceed at least 30 times more slowly than at Solar metallicity and requires dense regions. This extended timescale naturally suppresses the formation of large PAHs, consistent with the small-grain size distribution observed in the clumps.

In metal-poor systems with more extreme conditions than Sextans~A, the balance between PAH production and destruction mechanisms can shift. For example, \citet{Hunt2010} report Spitzer spectroscopy of Blue Compact Dwarf (BCD) galaxies, which typically have higher radiation field intensities than Sextans~A, and propose that the \changes{reduced PAH fluxes} at low metallicity is driven by photodestruction rather than inhibited growth. Further, \citet{Lai2025} presents JWST spectroscopy of the BCD galaxy II~Zw~40 and find a bright 3.3~\micron\ feature relative to other PAH bands, consistent with the inhibited PAH grain growth model, but also an anti-correlation between the $\rm 3.3 \, \mu m / 11.3\,  \mu m$ ratio and radiation field hardness, indicating that the PAH size distribution at low metallicity is governed by the interplay between photodestruction and inhibited growth. \changes{JWST MIRI/MRS spectroscopy of an additional BCD galaxy, the 10\% $Z_{\odot}$ CGCG 007-025, robustly detect just the 11.3 \micron\ feature, suggesting that the PAH grains are neutral \citep{DelValle-Espinosa2026}, but} future observations of BCD galaxies targeting the 3.3 \micron\ feature will be essential, as the $\rm 3.3 \, \mu m / 11.3\,  \mu m$ ratio provides the most sensitive diagnostic of PAH size \citep{Maragkoudakis2020}. Along with Sextans~A, these results together suggest that the driver of the \changes{deficit of PAHs} at low metallicity depends on the local physical conditions of a given galaxy. 

The detection of PAHs in Sextans~A has far-reaching implications for future observations of both nearby and high-redshift galaxies. Since PAHs can survive and form in a 7\% Solar metallicity environment, they will play important roles in ISM physics in galaxies out to high redshift where average metallicities are much lower than Solar \cite[e.g.,][]{Shipley2016, Spilker2023, ShivaeiBoogaard2024, Ronayne2024}. We favor the interpretation that the \changes{reduced PAH abundances} at low metallicity is caused by inhibited PAH grain growth and find evidence that PAHs form in-situ in the ISM in the compact clumps in Sextans~A. This implies that the PAH lifecycle is governed by the balance between grain growth and shielding from radiation fields, processes that \changes{are especially important at higher redshifts where harder radiation fields further suppress PAH survival in these systems \citep{Leslie2020}.}

\section{Methods}\label{sec:methods}

\subsection{JWST Data and Processing}\label{sec:obs}

The JWST \citep{Gardner2006, Rigby2023} observations of Sextans~A (GO 02391, PI: J. Roman-Duval) targeted the brightest, most active star-forming region with a far-infrared dust detection \cite{Shi2014}. The observations consist of NIRCam \citep{Rieke2005, Rieke2023} and MIRI \citep{Rieke2015, Wright2023} broadband imaging focused on the filters that isolate the 3.3 $\rm \mu m$ (NIRCam F335M), 7.7 $\rm \mu m$ (MIRI F770W), and 11.3 $\rm \mu m$ (MIRI F1130W) PAH features, as well as filters that bracket each PAH filter to estimate and subtract hot dust and stellar continuum that can overwhelm faint emission. A single MIRI pointing covers the star-forming region that overlaps with Module~B of the NIRCam pointing. Figure \ref{fig:Optical_JWST} presents the data and shows the location of the MIRI and NIRCam coverage on Sextans~A.

\subsubsection{Observations and Data Reduction}\label{sec:obs_red}

The NIRCam observations of Sextans~A were taken on January 2$^{\rm{nd}}$, 2023 and consist of six NIRCam Imaging filters: F115W, F150W, and F200W for the short wavelength detectors and F300M, F335M, and F360M for the long wavelength detectors. The F150W/F300M and F115W/F360M filter combination contained 8 total integrations, with 4 dithers in the INTRAMODULEX pattern, and 5 groups per integration using a MEDIUM8 readout pattern leading to a total integration time of 4166~s. The F200W/F335M filter combination was similar, but used 10 groups per integration to obtain a deeper exposure of 8561~s with the PAH filter, F335M. The majority of the MIRI Imaging data was taken on January 2$^{\rm{nd}}$ and 4$^{\rm{th}}$ 2023, including the filters F560W, F770W, F1000W, and F1500W. JWST observed Sextans~A with the F1130W filter 4 months later, on April 13, 2023, which results in a different orientation with respect to North when compared to the other observations. All of the MIRI observations use the FASTR1 readout pattern, 4 dithers in the 4-Point-Sets optimized for extended emission, but vary on the integrations per exposure as follows: 7 integrations per exposure for F560W (7837~s integration time), 8 integrations per exposure for F770W (8958~s integration time), 25 integrations per exposure for F1000W (28016~s integration time), and 30 integrations per exposure for both F1130W and F1500W (33622~s integration time). The exposure time for each filter was calculated through the Exposure Time Calculator \citep[ETC;][]{Pontoppidan2016} to account for the increase in noise as a function of wavelength due to the JWST backgrounds. 

The NIRCam and MIRI data are reduced with version 1.15.1 (DMS build B11.0rc2) of the JWST Science Calibration Pipeline \citep{Bushouse2025} and CRDS context \texttt{jwst\_1263.pmap} \citep{Greenfield2016} through all three stages of the pipeline. For NIRCam, we follow the default parameters, except in stage one where we allow the frame0 correction to recover stars that saturate in the first group (\texttt{suppress\_one\_group = False}) for the ramp step and we apply the correction for snowballs (\texttt{expand\_large\_events = True}) in the jump step.  In between stage two and three, we corrected the $1/f$ noise that manifests as striping in the NIRCam filters with the algorithm presented in \citet{Willott2022}. For alignment in stage three, we use a catalog of point sources from HST PID 16104 (PI: J. Roman-Duval) which are aligned to GAIA DR3 to obtain a high quality absolute astrometric solution. We modify the tweakreg parameters to have a SNR threshold of 5 and a tolerance of 0.1 for the short wave detectors to improve the alignment. For MIRI, we follow mostly default parameters, except in stage one we slightly increase the cosmic ray rejection threshold from 4 to 5 sigma in the jump step. A high quality alignment in MIRI is more difficult to achieve due to the lack of point sources at longer wavelengths. We therefore take advantage of the excellent relative astrometry for a given visit when using the same guide star \cite{Rigby2023}. First, we align the F560W mosaic to the HST PID 16104 source catalog because F560W will contain the most point sources compared to the longer wavelength MIRI filters. Then, we use the astrometric solutions for F560W and apply them to F770W, F1000W, and F1500W which were all taken within the same day and use the same guide star. For F1130W, which was observed four months later, we create a point source catalog from the F1000W filter to align the F1130W mosaic. The final resulting mosaics have high quality alignments that are well matched from the NIRCam to MIRI filters.

\subsubsection{Background Subtraction}\label{sec:bkgr}

The MIRI imaging sensitivity is background limited \cite{Rieke2015} and the imaging of Sextans~A is dominated by the background from zodiacal light and the thermal background of the telescope. While dedicated background observations in conjunction with MIRI imaging are usually taken to remove the background component, the large integration time required for detecting faint PAHs in Sextans~A made dedicated background observations prohibitive. Instead, we create a background from the data itself by stacking the four dithered exposures used to create each mosaic. The four exposures are stacked together in detector coordinates and sources from the image are removed by sigma clipping and taking the median value for each pixel. The 4-point extended source dither pattern maximizes the dither size to about 20\arcsec, which is $18-27\%$ of the size of the MIRI imager. These large dither sizes enable efficient sigma clipping of the sources in Sextans~A, including point sources and diffuse emission on the scale of 20\arcsec\ or less. We iteratively remove pixels that are three times the standard deviation of a given image pixel. All the structure from the JWST images of Sextans~A is therefore removed, leaving only the background from the MIRI imaging detector. We note that it is possible that faint, extended emission larger than 20\arcsec\ may be subtracted out of the Sextans~A mosaics. Given that the largest structure in the NIRCam imaging is $<10$\arcsec\ in size, it is unlikely that the MIRI background subtraction method is missing emission.

The NIRCam detector sensitivity is not background limited, but does contain a contribution from the background that is wavelength dependent and should be removed in order to compare to the background subtracted MIRI data and for continuum subtracting the PAH emission. We calculate the NIRCam background in a similar fashion to the MIRI background through stacking the four dithered exposures in detector coordinates and sigma clipping as many sources as possible. The dither pattern in NIRCam, however, is much more compact and leads to a few arcsecond differences between the images, which is not large enough to exclude astronomical sources. Instead, we sigma clip out as many sources as possible and fit the resulting image with a smoothly varying two-dimensional first order polynomial to mimic the zodiacal light background that dominates at these wavelengths. We compare both the MIRI and NIRCam background estimates to the JWST background tool \citep{Rigby2023b} and find excellent agreement.

\subsubsection{PSF Matching JWST data to a common resolution}
\label{sec:psf-matching}

The size of the JWST point spread function (PSF) increases as a function of wavelength and changes substantially between data taken with NIRCam and MIRI. The PSF of images taken in each filter therefore needs to be carefully matched across filters in order to subtract the continuum estimated from bracketing filters from the PAH filters. We also compare the 3.3 $\rm \mu m$, 7.7 $\rm \mu m$, and 11.3 $\rm \mu m$ continuum subtracted PAH emission, which necessitates having the same PSF for all filters. We use the procedure outlined in \citet{Aniano2011} to create convolution kernels that transform an image with a narrower PSF to a broader PSF. The PSFs used to create the convolution kernels are taken from STPSF/WebbPSF version~1.3.0 \citep{Perrin2014}, which generates PSF models informed by in-flight wavefront sensing measurements. Since JWST Cycle~1, these models have been updated to incorporate detector effects and match the observed empirical PSFs very well \citep[e.g.,][]{Libralato2024}. The short wavelength NIRCam filters (F115W, F150W, and F200W) are not PSF matched because they are not used in the PAH continuum subtraction procedure. The narrowest PSF we use is for the F300M filter, which has a full width half max (FWHM) of 0.\arcsec097, and the largest PSF is for the F1500W filter, which has a FWHM of 0.\arcsec488. We create custom convolution kernels that match the long wavelength NIRCam filter PSFs (F300M, F335M, F360M) and the MIRI filter PSFs (F560W, F770W, F1000W, F1130W) to the F1500W PSF, as it is the widest. We do not circularize the PSFs as suggested in \citet{Aniano2011}, because the JWST PSF is highly structured and asymmetrical, and we therefore produce much higher quality kernels without circularization. The PSFs are rotated to the orientation angle of each mosaic. This is especially important for filter F1130W, which was observed a few months later than the rest of the data, resulting in the PSF spikes in the native image being rotated compared to the other filters. 

We convolve the images to the PSF of the F1500W filter with the custom kernel created for each filter. We then regrid the mosaics to the F1500W coordinate system. These images are used for all analysis and have a PSF of 0.\arcsec488. This resolution element corresponds to 3.3~pc when assuming a distance of 1.4 Mpc for Sextans~A \cite{McQuinn2017, Bellazzini2014}.

\subsubsection{Continuum Subtraction of the PAH filters}
\label{sec:con_sub}

The near- and mid-infrared spectra of nearby galaxies are complex, comprised of broad PAH emission features, continuum from both starlight and stochastically heated small dust grains, as well as bright ionized gas emission lines. Although spectroscopic observations are most effective for isolating PAH features and removing the underlying continuum, the $\sim 5 \arcsec$ field of view and limited sensitivity of the JWST spectroscopic instruments make observations of star-forming regions in nearby dwarf galaxies prohibitive. In this work, we instead use photometric observations that flank the PAH bands on both the red and blue sides to constrain the continuum slope, allowing an estimate of the underlying continuum emission in the PAH-centered filters. However, the near- and mid-infrared spectrum hosts a rich set of PAH features spanning 3--20 \micron\ \citep[e.g.,][]{Tielens2008, Draine2021}, so even filters chosen to trace the continuum include some PAH emission (see Supplementary Figure 1). Several approaches have focused on continuum subtraction of the F335M filter by accounting for PAH emission in the flanking F360M filter \citep{Lai2020, Sandstrom2023PAH3um, Bolatto2024, Chown2025_pdrs, Whitcomb2026} and empirical calibrations have also been developed for the MIRI F770W and F1130W filters \citep{Chown2025_pdrs, Donnelly2025}. Building upon this work, we generalize the continuum subtraction procedure to all filters centered on PAH emission-- F335M, F770W, and F1130W. Our method is compatible with the approach outlined in \citet{Sandstrom2023PAH3um} and \citet{Whitcomb2026} for the F335M filter, but flexible enough to account for the broader and brighter PAH features probed by MIRI (see Supplementary Figure 1). We outline our assumptions and describe the method below.

Within each filter trio, we assume the following: one filter is centered on and dominated by PAH emission, a second primarily traces the continuum but includes some contamination from nearby PAH complexes, and a third provides a clean measure of the continuum with no PAH contribution. For the F300M/F335M/F360M trio, we take F300M as the clean continuum filter and F360M as the PAH-contaminated flanking filter, affected by the 3.4 and 3.47 \micron\ plateaus of aromatic and aliphatic features \citep{Joblin1996, Hammonds2015, Lai2020}. In the mid-infrared, we assume F1000W is not heavily contaminated by PAH emission. By contrast, F560W is affected by the 5.2, 5.6, and 6.2 \micron\ features and F1500W is contaminated by the wings of the 12.8, 13.6, 14.8, and 16.4 \micron\ features. To illustrate the filter coverage, Supplementary Figure 1 shows the transmission curves overlaid on both a model PAH spectrum \citep{Draine2021} and an empirical spectrum of the PDRs4All Orion Bar JWST/NIRSpec and MIRI/MRS observations \citep{Chown2024, VanDePutte2024, Peeters2024, VanDePutte2025}. We apply synthetic photometry to both spectra, highlighting which PAH features are traced by each filter, and present the PAH-centered filters in green, the flanking filter with PAH contamination in gold, and the clean continuum filters in magenta.

Next, we assume that the level of PAH contamination in the flanking filter scales linearly with the brightness in the main PAH filter. While the brightness of each PAH feature does depend on the size and ionization of the grains, we account for this by varying the size and charge distribution from \citetalias{Draine2021} models and propagate this uncertainty throughout. Lastly, we assume the slope of the continuum can be approximated as linear between the two filters that flank the PAH emission.

With those assumptions, we then define each filter as:

\begin{align}
\begin{split}
\label{eq:sys}
    &f_1 = f_{c1} \\ 
    &f_2 = f_{c2} + f_{p2} \\ 
    &f_3 = f_{c3} + f_{p3} .
\end{split}
\end{align}
Here, $f_1$ is the flux of the filter with only continuum emission. $f_2$ is the flux of the filter centered on the PAH emission and contains contributions from both continuum ($f_{c2}$) and PAH flux ($f_{p2}$). Lastly, $f_3$ is the flux from the filter that is contaminated by adjacent PAH features and therefore contains contributions from both continuum ($f_{c3}$) and PAH emission ($f_{p3}$). We then assume that the PAH emission in the central PAH filter scales linearly (through a proportionality constant $k$) with the contaminating PAH emission in the adjacent continuum filter:

\begin{equation}
\label{eq:k}
    f_{p2} = k f_{p3}.
\end{equation}

With small wavelength differentials ($<5$\micron ), we assume that the slope of the continuum between $f_1$ and $f_3$ is linear and calculate the continuum in $f_2$:

\begin{equation}
\label{eq:con}
    f_{c2} = \frac{f_{c3} - f_{c1}}{\lambda_3 - \lambda_1} \times (\lambda_2 - \lambda_1) + f_{c1}
\end{equation}
where $\lambda$ is the pivot wavelength of a given filter. We define:

\begin{equation}
\label{eq:beta}
    \beta = \frac{\lambda_{2} - \lambda_{1}}{\lambda_{3} - \lambda_{1}}
\end{equation}
to simplify the resulting equations. By combining equations \ref{eq:sys}, \ref{eq:k}, \ref{eq:con}, and \ref{eq:beta}, we calculate $f_{p2}$, the continuum subtracted PAH flux in the main PAH filter for a given $k$:

\begin{equation}
\label{eq:consub_up}
f_{p2} = \frac{k}{k -\beta} \times [ f_{2} - (1 - \beta)f_{1} - \beta f_{3} ],
\end{equation}
which is the same functional form as Equation 1 in \citet{Whitcomb2026} for direct comparison.

Estimating the PAH flux in filter 2, $f_{p2}$, relies on a value for $k$, the degree of PAH emission contaminating the flanking filter. We determine $k$ through a combination of the \citetalias{Draine2021} PAH models and empirical JWST spectra from the PDRs4All program \citep{Chown2024, VanDePutte2024, Peeters2024, VanDePutte2025}. We summarize our procedure to calculate $k$ as follows. First, we separate the PAH features from the underlying continuum emission in the spectrum. Then, we compute synthetic photometry for only the PAH portion of the spectrum, calculating values for the central PAH filter ($f_{p2}$) and the PAH-contaminated flanking filter ($f_{p3}$), yielding: $k = f_{p2} / f_{p3}$. Decomposing the spectrum into PAH and continuum components can be complex and requires careful treatment, however. Below, we outline the strengths and limitations of our various approaches, including the differences between the \citetalias{Draine2021} PAH model spectra and the empirical spectra from PDRs4All.

The \citetalias{Draine2021} models report a PAH emission spectrum as a function of the radiation field, PAH grain size distribution, column density of hydrogen, and PAH charge distribution. Using these models allows us to study how $k$ may change based on the properties of the PAHs, as band ratios of particular PAH features vary depending on the physical conditions of the PAH population. The top panel of Supplementary Figure 1 shows an example of a \citetalias{Draine2021} model, with a radiation field log(U) = 1 and standard PAH grain size and ionization distributions. To fit the continuum, we mask regions of the spectrum containing PAH emission and fit either a first (for F335M and F770W) or sixth order (for F1130W) polynomial to identify and remove the continuum. We repeat this process across the grids of \citetalias{Draine2021} models, varying the size, ionization, and radiation intensity up to $\log U = 4$. Supplementary Figure 2 presents $k$ for each filter trio and shows how it varies depending on the properties of the PAH population. The final $k$ value is the mean of the distribution and the standard deviation is used as the uncertainty to represent the systematic uncertainty in $k$.

The \citetalias{Draine2021} models provide an estimate of how $k$ can change depending on the PAH properties, but not all of the PAH features are accurately captured in the models. Crucially, the models do not include the 3.4 \micron\ aliphatic feature or the broader 3.3 \micron\ PAH complex. Therefore, we additionally estimate $k$ with empirical JWST spectra of the Orion Bar from the PDRs4All ERS project \citep{Peeters2024, Chown2024, VanDePutte2024, VanDePutte2025}, which provide five template spectra from the strip taken along the PDR in Orion from NIRSpec and MIRI-MRS on JWST (see \citet{Chown2024} for more details). The empirical spectra from the Orion Bar also provide an independent estimate of $k$ for the two filter trios that focus on the 7.7 $\mu$m and 11.3 $\mu$m PAH features. We work with template spectra that have emission lines removed through the line identification presented in \citet{VanDePutte2024, VanDePutte2025} and provide two methods of fitting the spectra. The first follows the same approach to fitting the \citetalias{Draine2021} models, where we mask out the regions of the spectrum that contain PAH emission and fit either a first (for F335M and F770W) or sixth order (for F1130W) polynomial to identify and remove the continuum. The second approach uses the code \texttt{PAHFIT} \citep{Smith2007} to simultaneously fit the complex infrared spectrum, including the hot dust continuum, individual PAH features, the stellar continuum, and emission lines. The details of the PDRs4All PAHFITs are presented in \citet{VanDePutte2025}. We then apply synthetic photometry on the PAH portion of each spectrum to calculate $k$ for each filter trio for the five template spectra and report the results in Supplementary Figure 2, where the mean and standard deviation is shown in the spread along the y-axis.

Since the infrared spectrum is quite complex and challenging to fit, we evaluate two approaches for calculating $k$ values for each filter trio, depending on how well each method fits the particular area of the spectrum. The rationale for each choice is described below, and the resulting values are listed in Supplementary Table 1. We also provide the conversion factor needed to translate PAH fluxes between the two adopted $k$ values. We note that Equation~\ref{eq:consub_up} shows that the relationship between $k$ and the PAH flux is non-linear: smaller $k$ values indicate greater contamination in the flanking filter, making the derived PAH flux more sensitive to changes in $k$ when the value is small. By providing two estimates of $k$, we aim to capture one of the primary sources of systematic uncertainty in the continuum subtraction. Future spectroscopic observations will be required to determine the true PAH flux in each filter.

\paragraph{F300M/F335M/F360M} 

The \citetalias{Draine2021} $k$ estimates are not useful for these filters because the models do not contain a 3.4 \micron\ feature or broader 3.3 \micron\ PAH complex (see Supplementary Figures 1 and 2). Instead, we use the polynomial fit and PAHFIT of the PDRs4All data to estimate $k$. With a value of $k = \rm F335M_{PAH} / F360M_{PAH} = 2.07 \pm 0.30$, the PAHFIT method estimates more than a factor of two higher level of contamination than the polynomial fit, which report a value of $k = \rm F335M_{PAH} / F360M_{PAH} = 4.45 \pm 0.39$ (lower k values implies larger amounts of contamination). This is because the region between 3.3 - 5 \micron\ contains a variety of low amplitude, broad, PAH features that create a ``PAH continuum" or plateau of PAH emission \citep{Boersma2023, Allamandola1989, Allamandola2021, Esposito2024}. For the polynomial fit, these features are considered continuum and are removed, but they are treated as PAH features for the PAHFIT method. The PAHFIT approach is also close to the slope of 1.6 reported in \citet{Sandstrom2023PAH3um}. In this work, the PAH contamination in the F360M filter is found by identifying the F335M/F360M ratio in PAH dominated regions through comparing the expected color of PAH emission vs the color of hot dust and stellar continuum. The slope of 1.6 is comparable to the $k$ value calculated through PAHFIT, which is consistent with the explanation of low amplitude PAH features contributing to the F360M PAH filter flux. 

We select $k = \rm F335M_{PAH} / F360M_{PAH} = 2.07 \pm 0.30$ as the contamination constant to use throughout this work in order to be consistent with the method in \citet{Sandstrom2023PAH3um} and the implementation in \citet{Dale2025}. The 3.3 \micron\ feature is 21\% brighter with this choice of $k$ when compared to PAHFIT $k = \rm F335M_{PAH} / F360M_{PAH} = 4.45 \pm 0.39$, leading to a systematic uncertainty on the PAH flux assumed through this work. 

\paragraph{F560W/F770W/F1000W} 

For this filter trio, we use the $k$ based on the \citetalias{Draine2021} models and the PAHFIT of the PDRs4All template spectra, which exhibit a difference of 35\% depending on the properties of the PAHs. The variation stems from the PDRs4All spectra having systematically brighter 5.3, 5.6, and 6.3 \micron\ features, while the 7.7~\micron\ complex is fainter compared to the \citetalias{Draine2021} models. It is likely that the PAH population in the Orion Bar differs from the population assumed in \citetalias{Draine2021}, which was tuned to observations from the extragalactic sample SINGS \citep{Smith2007}. However, the contamination level in F560W is not as high as it is in F360M, so the precise choice of $k$ does not have as large of an impact on the final PAH flux. The PDRs4All PAHFIT data yields $k = \rm F770W_{PAH} / F560W_{PAH} = 4.33 \pm 0.35$ while the \citetalias{Draine2021} yields $k = \rm F770W_{PAH} / F560W_{PAH} = 5.84 \pm 0.73$. Propagating these values through Equation \ref{eq:consub_up} changes the resulting 7.7~\micron\ flux by only $\sim$4\%, indicating that this systematic uncertainty does not significantly affect our results. We adopt the PDRs4All PAHFIT value, as the 7.7~\micron\ feature in Sextans~A is under-luminous (see Figure \ref{fig:band_ratios} and more consistent with the spectra in the PDRs4All sample.

We compare our approach to \citet{Donnelly2025}, an independent, empirically-derived continuum subtraction method focused on MIRI filters benchmarked by MIRI/MRS spectra of luminous infrared galaxies. The two methods agree well, with differences of only 5.5\% when using the F560W/F770W/F1000W filter trio to derive the total PAH flux in the F770W filter.

\paragraph{F1000W/F1130W/F1500W} 

Here, we report $k$ values from \citetalias{Draine2021} and the polynomial fit of the PDRs4All data. The PAHFIT for this trio is not used because the modified blackbody in PAHFIT does not adequately fit the Orion Bar in the 15 -- 20 \micron\ range \citep{VanDePutte2025}, leading to over-subtraction of PAH features. The sixth order polynomial fits this region nicely, however, so the simple polynomial fit is adopted instead. The difference between $k = \rm F1130W_{PAH} / F1500W_{PAH} = 10.17 \pm 1.24$ for the \citetalias{Draine2021} models and $k = \rm F1130W_{PAH} / F1500W_{PAH} = 7.21 \pm 0.92$ is due to the \citetalias{Draine2021} models predicting slightly brighter and broader 11.3 \micron\ features than is seen in the PDRs4All data. This difference in $k$ is negligible in the flux calculation of 11.3 \micron\ PAH flux, however, as it introduces a $<1\%$ change. We adopt the PDRs4All polynomial fit for this work, but note that the F1000W/F1130W/F1500W trio is the most robust to this choice of $k$. 

Our continuum subtraction method for F1130W also agrees very well with the approach in \citet{Donnelly2025}, deriving a difference of only 5.1\% between F1130W fluxes from the two methods when using the F1000W/F1130W/F1500W trio. 

\paragraph{Limitations of the continuum subtraction method}

While our continuum subtraction method provides a robust estimate of PAH fluxes that is consistent with other approaches \citep[e.g.][]{Sandstrom2023PAH3um, Bolatto2024, Chown2024, Donnelly2025, Whitcomb2026}, several limitations should be noted. First, the template spectra used to estimate the continuum may not accurately represent the conditions in Sextans~A; for example, Figure \ref{fig:band_ratios} shows that the 7.7 \micron\ feature is under-luminous compared to both the \citetalias{Draine2021} model grids and the PDRs4All data. Second, some filters assumed to measure just the continuum are not entirely free of PAH emission or other features: the F1000W filter, for instance, captures low-level PAH emission in its wings and may also include a weak silicate feature \citep[e.g.,][]{Smith2007}. Additionally, the continuum may not be well approximated as linear, particularly in the F1000W/F1130W/F1500W filter trio where stochastically heated hot dust can dominate and manifests as a powerlaw rather than linear continuum. Furthermore, the method presented here reports \textit{all} the PAH flux within a given filter and does not explicitly separate contributions from adjacent PAH features that are partially captured in the filter bandpass. For example, the F770W filter also includes flux from the 8.3~\micron\ feature, and the 3.3~\micron\ filter may contain a contribution from the 3.4~\micron\ aliphatic feature, if present. Lastly, ionized gas emission lines will contaminate some filters, although the contribution is generally low $<10\%$ except in extreme cases near very bright \hii\ regions \citep[e.g.][]{Whitcomb2023a, Misselt2025, Lai2025}. While our photometric continuum subtraction provides robust estimates of PAH fluxes, it cannot fully account for all systematic uncertainties arising from the effects discussed above. Upcoming Cycle 4 JWST NIRSpec and MIRI MRS spectroscopic observations (GO 7396, PI: E. Tarantino) will help address these systematics, enabling a more precise determination of the PAH band strengths.

\subsubsection{Continuum Subtracted PAH images}
\changes{We present the continuum subtracted Sextans~A PAH data in Extended Data Figure \ref{fig:consub_PAH_with_F1500W} that zooms in on the area of the brightest, most concentrated PAH emission (green rectangle in Figure \ref{fig:Optical_JWST}). The full continuum and PAH continuum subtracted images of the JWST coverage in Sextans~A is displayed in Figure \ref{fig:PAH_clumps}. Overall, the continuum in the PAH filters dominates the bulk of emission throughout the JWST coverage of the galaxy. In the bright PAH clumps, the PAH emission contributes a median of 20\% of the total filter flux for F335M, 35\% in F770W, and 50\% in F1130W, demonstrating the necessity of continuum subtraction when analyzing the PAH emission in metal-poor environments where the features are faint relative to the continuum.}

\subsubsection{Estimate of Uncertainties} \label{sec:unc}

While the JWST pipeline reports a drizzled variance array, the noise properties of the JWST images depend on the spatial resolution and become covariant after PSF matching and resampling to a common grid (see Section \ref{sec:psf-matching}) since the noise is correlated at a pixel level. Therefore, we calculate the noise and resulting uncertainties in the PSF matched images themselves, similar to the approach in \citet{Chown2025_dwarf} and explored in \citet{Williams2024}. We identify a region in the image that is relatively signal free, besides a few stars (centered at RA=10h11m08.2s; DEC= -4d42m41.6s). We sigma-clip with 5 iterations to remove these stars and calculate the root mean squared (RMS) noise within the region, which serves as the $1\sigma$ level uncertainty for all PSF matched images.

We propagate the uncertainties of each PSF-matched filter when calculating the continuum subtracted PAH images, described in Section \ref{sec:con_sub}, \changes{leading to $1\sigma$ uncertainties of $ 0.0036 \ \rm MJy \, sr^{-1}$ for the 3.3 \micron\ PAH flux density, $0.0066 \ \rm MJy \, sr^{-1}$ for the 7.7 \micron\ PAH flux density and $ 0.0138 \ \rm MJy \, sr^{-1}$ for the 11.3 \micron\ PAH flux density}. We also consider the uncertainty in the estimated value of $k$ by including the standard deviation of the variation in $k$ values as an additional uncertainty that is propagated through the continuum subtraction equation.

\subsection{Ancillary Data} \label{sec:anc_data}

We present H$\alpha$ observations of Sextans~A from the LVL survey \citep{Kennicutt2008, Dale2009} as well as \textit{Swift} UVOT observations \citep{Hagen2017} to provide a high resolution ($\sim$1.1\arcsec\ for H$\alpha$; $\sim$2.5\arcsec\ for UV) comparison to the ionized gas and the UV radiation field in Sextans~A. We converted the LVL H$\alpha$ map from counts into $\mathrm{erg \, s^{-1} \, cm^{-2}}$ using the documentation outlined for the DR5 LVL data release. We then removed the contribution of the [NII] doublet from the narrowband imaging by assuming a ratio of $\mathrm{[NII]/H\alpha} = 0.04$ from \citet{Kennicutt2008}. The H$\alpha$ image is displayed in Figure \ref{fig:PAH_clumps}. The \textit{Swift} UVOT data were reduced with the method outlined in \citet{Hagen2017} and we use this UV data as a secondary measure of the radiation field near the PAH clumps.

\subsection{Clump identification} \label{sec:dendro}

The spatial distribution of PAH emission in Sextans~A closely resembles the hierarchical structure captured by dendrograms, where a larger ``trunk" encompasses several ``leaf" structures that lack further substructure (see Figures \ref{fig:PAH_clumps} and Extended Data Figure \ref{fig:consub_PAH_with_F1500W}). Dendrograms are commonly employed to characterize the fragmentation of molecular clouds, dust, and other ISM tracers \cite[e.g.,][]{Rosolowsky2008, Goodman2009, Kirk2013}. Given this, it is reasonable to represent the PAH morphology in Sextans~A using a dendrogram-based framework. The main purpose of using dendrograms is to have a uniform procedure in defining the clump boundaries and properties. We utilize the \texttt{astrodendro} package to define the boundaries of these clumps and calculate their properties, such as the shape, integrated flux, and size of the clumps. 

First, we identify an initial catalog by requiring each clump of emission to have a 3$\sigma$ detection or higher in each of the 3.3 \micron , 7.7 \micron , and 11.3 \micron\ continuum subtracted PAH images. This ensures that the PAH clump detection is robust as it is seen in all three features and not a spurious high redshift galaxy. We adopt the 7.7 \micron\ continuum subtracted PAH data to define the clump structure, as clump sizes differ slightly between bands, and 7.7 \micron\ provides a consistent standard being central between 3.3 and 11.3 \micron . The minimum value of the dendrogram is set to be the 1$\sigma$ noise level of the 7.7 \micron\ continuum subtracted PAH data, ensuring the noise is pruned from the final structure. The minimum significance for structures (\texttt{min delta}) is set to the 3$\sigma$ level, requiring each structure and sub-structure to have at least three times the noise level to be defined as a clump. Each clump is required to be at least 4 pixels in size, about the size of the FWHM of all the PSF matched images. We cross match the dendrogram structure list with the initial clump catalog and include only the clumps that pass the initial cut of a $> 3\sigma$ detection in all three continuum subtracted PAH data. 

The resulting catalog of clumps is presented in Table \ref{tab:clump_prop} and the clumps are plotted on the JWST images in Figure \ref{fig:PAH_clumps}. We use the 7.7 \micron\ dendrogram clump boundaries and calculate the integrated PAH flux of the other two features from the 3.3 \micron\ and 11.3 \micron\ continuum subtracted PAH data, ensuring equivalent area and boundaries for each clump definition. The dendrogram analysis mostly identifies ``leaf" structures, which do not have additional substructure, with the exception of clumps 15 and 16 that are the larger ``trunks" encompassing the two of the smaller leaves. Interestingly, most of the PAH structures identified in the dendrogram are individual leaves, which contrasts with observations of higher metallicity galaxies, where PAH emission is typically more extended, highly structured, and closely traces the bulk of the ISM \cite[e.g.,][]{Chastenet2019, Sandstrom2023PAHISM, Lind-Thomsen2025, Chown2025_phangs}. In Sextans~A, however, the PAH distribution is compact and point-like, with only two resolved PAH structures located within the JWST observations of the galaxy.

\subsection{Modeling the infrared spectrum} \label{sec:D21_SED}
The infrared spectrum from $1-15$ \micron\ is complex, containing contributions from stellar light, PAH emission, and hot dust. The multiband JWST NIRCam and MIRI photometry of Sextans~A comprises eleven filters that sample the spectrum across the $1-15$ \micron\ baseline. We model the flux in clump 1 (see Table \ref{tab:clump_prop}) using a combination of stellar and dust models. We note that our goal is to provide an approximate fit to the photometric data for Figure \ref{fig:SED}, a more in depth evaluation of the plausible parameter space for the models will be explored in future work. First, we select the short wavelength NIRCam filters (F115W, F150W, and F200W) and identify the best fitting stellar models using grids from high-resolution synthetic spectra based on the stellar atmosphere code PHOENIX \cite{Husser2013}. We apply synthetic photometry through the \texttt{synphot} package, which implements Equation 5 in \cite{Gordon2022} to calculate the photon weighted average flux density through a given photometric filter. For clump 1, the best fitting model with a $\chi^2 = 3$ has a metallicity of $[M/H]=-1.15$, a temperature of $T = 6600 \, \rm K$, surface gravity of $\log(g) = 0.0 \rm \, cm \, s^{-1}$, and a normalization constant of $\log(C) -35.1$. We set this PHOENIX stellar model as the stellar portion of the SED, which dominates the $1-5$ \micron\ portion of the spectrum.

For the PAH emission and hot dust, we use the physically motivated \citetalias{Draine2021} models, which contain PAH particles of varying sizes and a population of larger ``astrodust'' grains from \cite{Hensley2023} for the hot dust continuum. These models predict a MIR spectrum from a variety of input illuminating radiation fields and model the influence of size and ionization distribution of the PAHs on their emission features. We use the low metallicity $Z=0.0004 \approx 0.02 Z_{\odot}$, 10 Myr burst from \citet{Bruzual2003} as the input radiation spectral shape to match the low metallicity conditions in Sextans~A. We then create a grid of \citetalias{Draine2021} models by varying the radiation field intensity ($U$), PAH-to-dust mass fraction ($\rm q_{PAH}$), column density of hydrogen ($\rm N_{HI}$), the ionization state of the PAHs, and the size of the PAHs. The PHOENIX stellar model derived above is added to each \citetalias{Draine2021} spectra and we apply synthetic photometry for the rest of the JWST filters (F300M, F335M, F360M, F560W, F770W, F1000W, F1130W, and F1500W). We calculate the best fit through $\chi^2$, which tend to be large, as the uncertainties on the filter fluxes are quite low. We identify models that have $\chi^2 < 10000$ to be the models that provide the best fit and plot the MIR spectra in Figure \ref{fig:SED} as the gray lines, where the transparency corresponds to the $\chi^2$. The average of all the spectra is presented as the dark black line.  

There isn't a single \citetalias{Draine2021} model that well produces the observed SED in Sextans~A, as one that matches one filter may not match another, i.e., fitting F770W well leads to an over prediction in F1000W and F1500W. The mismatch between the JWST photometry and \citetalias{Draine2021} models could be due to variety of effects, including: emission from bright emission lines contaminating a given filter (e.g., [SIV] in F1000W), absorption from silicates at 10 $\mu$m, and that PAHs fluxes or profiles at these very low 7\% Solar metallicities vary significantly from their high metallicity counterparts. The \citetalias{Draine2021} models were tested with spectra from the Spitzer Infrared Nearby Galaxy Survey (SINGS) \citep{Kennicutt2003, Smith2007} and AKARI \citep{Lai2020}, which are biased towards the centers of typical, high metallicity star-forming galaxies. Future Cycle 4 JWST spectroscopic observations of Sextans~A (GO 7396, PI: E. Tarantino) will help disentangle the origin of these differences. 

The \citetalias{Draine2021} fits to the SED favor low values of $\rm q_{PAH}$, ranging from $\rm q_{PAH} = 0.2 - 0.9$, consistent with observations of metal-poor dwarf galaxies \citep{Hunt2010, Sandstrom2010, Chastenet2019, Wu2006}. The intensity of the radiation field varies from $\log U=0 - 3.5$, without much dependence for a particular value, but models with $\rm log\, U>3.5$ do not fit the SED well. These reported properties should be interpreted with caution, as the \citetalias{Draine2021} models do not fully reproduce the JWST observations. Future work that incorporates spectroscopic data and applies a statistically rigorous treatment of the free parameters in the \citetalias{Draine2021} models will enable a more accurate assessment of the PAH properties in Sextans~A. 

\subsection{Calculating $\Sigma$PAH for Sextans A} \label{sec:sigma_pah}

\changes{We calculate the sum of the PAH features captured by our Sextans~A JWST photometry, $\Sigma$PAH, in order to compute $\Sigma$PAH/TIR, a measure of the fraction of infrared power emitted by PAH features. Comparing to the sample of spectroscopically-derived $\Sigma$PAH/TIR from \textit{Spitzer} in Figure \ref{fig:sigmaPAH_TIR} requires careful treatment of the conversion between the average flux density from JWST photometry to integrated flux similar to the \textit{Spitzer} IRS spectra. JWST photometry reports the photon-weighted average flux density within a filter bandpass at the effective wavelength (equation 5 in \citet{Gordon2022}):}

\begin{equation}
    \langle F(\lambda) \rangle = \frac{\int F(\lambda) R(\lambda) \lambda d \lambda}{\int R(\lambda) \lambda d \lambda }
\end{equation}

\noindent \changes{where $F(\lambda)$ is the flux density, $R(\lambda)$ is the bandpass function, and the integration is performed over $\lambda d \lambda$ in photon units. Further, while the JWST calibration is defined in $F_{\lambda}$ units, the units given for observations are in $F_{\nu}$, in particular $\mathrm{MJy \, sr^{-1}}$. We therefore use $\lambda_{\mathrm{pivot}}$, which is defined to translate between $F_{\lambda}$ and $F_{\nu}$ units:}

\begin{equation}
    F_{\lambda} = F_{\nu} \times c/\lambda_{\mathrm{pivot}}^2
\end{equation}

\changes{The spectroscopic samples from \citet{Smith2007, Hunt2010, Whitcomb2023} shown in Figure \ref{fig:sigmaPAH_TIR} all report integrated PAH fluxes through the code PAHFIT, which integrates over each individual PAH feature, $F_{\mathrm{int}} = \int F(\lambda) d \lambda $, and is typically in units of $\mathrm{erg \ s^{-1} \ cm^{-2}}$. Converting the JWST reported average flux densities to integrated fluxes is not possible without knowledge of the shape of the spectrum within the filter bandpass. We therefore define a constant called the ``modified bandwidth'', $BW_{\mathrm{mod}}$, to translate between the JWST average flux density to an integrated flux comparable to the spectroscopic samples:}

\begin{equation}
    F_{\mathrm{int}} = \langle F(\nu) \rangle \times c/\lambda_{\mathrm{pivot}}^2 \times BW_{\mathrm{mod}}
\end{equation}

\changes{In the limiting case where the underlying spectrum is a delta function centered on the effective wavelength of the filter, the modified bandwidth is $BW_{\mathrm{mod}} = BW = \int R(\lambda) d \lambda / R_{\mathrm{max}}$ and is defined in the NIRCam and MIRI documentation. Therefore, multiplying by the bandwidth of the filter to convert average flux densities to integrated fluxes can be effective for emission lines that are centered in the filter. However, PAH features are broad and complex and this simplification can lead to incorrect flux conversions. We therefore forward model the \citetalias{Draine2021} template spectra to calculate $BW_{\mathrm{mod}}$, following a similar procedure to Section \ref{sec:D21_SED}. We report $BW_{\mathrm{mod}}$ for the three PAH-centered filters in Supplementary Table 2, which are substantially larger than the bandwidth. For the PAH template spectra considered here, disregarding the spectral shape within the filter bandpass can lead to systematic offsets of factors of 2-3 in the inferred integrated fluxes. }

\subsection{Radial profile of a PAH clump} \label{sec:radprof}

To examine how the spatial distribution of the PAH emission compares to that of the hot dust, we calculate radial profiles of the PAH clumps defined by the dendrogram structures in Section \ref{sec:dendro}. We use the F1500W filter as a measure of the hot dust continuum. In Supplementary Figure 3, we present the radial profile of clump 15, defined as the azimuthally-averaged flux in circular annuli apertures as a function of radius from the clump center (see Table \ref{tab:clump_prop}). Clump 15 (see Figure \ref{fig:PAH_clumps}) is the largest and brightest clump that encompasses two smaller PAH structures and is the prominent structure highlighted in Extended Data Figure \ref{fig:consub_PAH_with_F1500W}. We plot the radial profile of the 11.3 \micron\ PAH emission, the continuum at 11.3 \micron, and the F1500W hot dust continuum emission and normalize each profile so the sum is equal to one. The resulting radial profiles show that emission from PAHs peaks at the center, decreases slightly, then increases again due to the second PAH structure internal in the larger clump of emission. The PAH emission then falls off to the 1$\sigma$ noise level at larger radii. In contrast, the F1500W emission increases in between the peaks of the two PAH clumps encompassed in this structure (likely due to the extended spatial distribution of the dust), then the emission overall decreases but to a level 2.5$\sigma$ above the noise floor in F1500W, signifying that there is extended emission in the hot dust continuum not present in the PAH emission. A similar profile is seen in the continuum at 11.3 \micron, but does not peak as high in between the two smaller clump structures. 

The extended nature of both the F1500W filter and continuum at 11.3 \micron\ signifies that PAH emission is more compact than the dust and supports the argument in Section \ref{sec:main} that the resolved $\Sigma$PAH/TIR ratio is likely affected by beam smearing from the large $\sim 30\arcsec$ spatial resolution of the FIR data. However, it is difficult to quantify the degree to which the PAH emission is more compact than the overall dust distribution. Future work involving high resolution dust extinction mapping \citep[e.g.][]{Dalcanton2015, Gordon2016, Lindberg2025} and observations of the F2100W filter from GO 7396 (PI: E. Tarantino) will help illuminate the spatial scale of the PAH emission relative to other ISM tracers.

\subsection{The $\rm 3.3 \, \mu m / 11.3\,  \mu m$ and $\rm 3.3 \, \mu m / 7.7 \,  \mu m$ Band Ratios} \label{sec:band_ratios}

\subsubsection{\citetalias{Draine2021} Models} \label{sec:d21_band_ratio}

We investigate the size and ionization state of the PAHs in Sextans~A by calculating the continuum subtracted $\rm 3.3 \, \mu m / 11.3\,  \mu m$ and $\rm 3.3 \, \mu m / 7.7 \,  \mu m$ band ratios for the clump catalog in Table \ref{tab:clump_prop}. We compare measured band ratios with the models from \citetalias{Draine2021} that use the low metallicity $Z=0.0004 \approx 0.02 Z_{\odot}$, 10 Myr burst from \citet{Bruzual2003} as the input spectrum. The PAH size distribution in \citetalias{Draine2021} is modeled as a sum of two log normal size distributions, where the ``small'', ``standard'', and ``large'' models correspond to a peak size in the distribution at $3 \rm \AA$, $4 \rm \AA$, and $5 \rm \AA$, respectively (see Figure 9 and Equation 15 in \citetalias{Draine2021}). The charge of the PAHs are separated into the ``standard'' model, which uses the ionization fraction given in \citet{DraineLi2007}, as well as ``low'' and ``high'' models which have a factor of two shift lower and higher in the $f_{\rm ion}$ respectively. We average the \citetalias{Draine2021} model grids from $\rm log \, U = 0 - 3.5$, since the radiation field does not impact the PAH band ratios until high radiation fields at $\rm log \, U \gtrsim 4$ \citep{Draine2021}.

We apply synthetic photometry to the \citetalias{Draine2021} models for all the filters centered on PAH emission (F335M, F770W, and F1130W), as well as the bracketing continuum filters, and calculate the continuum subtracted PAH emission for the 3.3~$\mu$m, 7.7~$\mu$m, and 11.3~$\mu$m features. The same continuum subtraction method is used for both the \citetalias{Draine2021} models and the JWST data; however, for the models, we adopt the contamination constant, $k$, derived specifically from the \citetalias{Draine2021} spectra. This choice has a negligible effect on the continuum subtraction at 7.7~$\mu$m and 11.3~$\mu$m. However, it significantly affects the 3.3~$\mu$m measurement: because the \citetalias{Draine2021} models do not include the nearby 3.4~$\mu$m feature, applying a $k$ value derived from PDRs4All data (which does include 3.4~$\mu$m emission) would result in a negative continuum estimate at 3.3~$\mu$m in the \citetalias{Draine2021} models. \changes{We instead use $k = 23$ for the continuum subtraction of the 3.3 feature} and present the results in Figure \ref{fig:band_ratios}.

\changes{Throughout this work, we use the $k_1$ values for the contamination constant, as described in Section \ref{sec:con_sub} and in Supplementary Table 1. In Supplementary Figure 4, we show how the Sextans~A $\rm 3.3 \, \mu m / 11.3\,  \mu m$ and $\rm 3.3 \, \mu m / 7.7 \,  \mu m$ band ratios would shift when assuming $k_2$ rather than $k_1$. Overall, our main conclusions are unaffected, but the ratios shift slightly closer to the \citetalias{Draine2021} model grids. The preference for the small or standard grain size distribution remains and the $\rm 3.3 \, \mu m / 7.7 \,  \mu m$ continues to be elevated relative to the \citetalias{Draine2021} models and PDRs4All Orion Bar data. The majority of the shift is driven from the systematic uncertainty in the 3.3 \micron\ flux, which varies by $\sim 20\%$, depending on the contamination constant assumed.}

\subsubsection{Correlations with radiation field}

\changes{The H$\alpha$ \citep{Kennicutt2008, Dale2009} and \textit{Swift} UVOT \citep{Hagen2017} observations of Sextans~A are compared to the PAH feature ratios in Figure \ref{fig:band_ratios}. There is a moderate correlation between the $\rm 3.3 \, \mu m / 11.3\,  \mu m$ (grain size) and $\rm 3.3 \, \mu m / 7.7 \,  \mu m$ (ionization) ratios and the H$\alpha$ flux, with Spearman rank coefficients of $r_{sp} = 0.61$ and $r_{sp} = 0.68$, respectively. The correlation persists when using the ionized gas fraction, defined as H$\alpha$ flux normalized by the 21\,cm atomic gas emission \citep{Hunter2011}. Since the 21\,cm data is much more coarse, we report the H$\alpha$ based coefficients. This correlation is robust to the choice of radiation field tracer, as the \textit{Swift} UVOT shares a similar value of $r_{sp} \sim 0.6$.}

\changes{We interpret this correlation is due to the increasing hardness of the radiation field, which raises the 3.3 \micron\ flux substantially when compared to the other PAH features. There is also a similar effect at very high intensity radiation fields ($U > 10^4$) where the 3.3 \micron\ PAH flux increases relative to the 11.3 and 7.7 \micron\ features. Since the H$\alpha$ and \textit{Swift} UVOT W1 data will trace a combination of both the hardness and the intensity of the incident radiation field, it is difficult to determine which effect is driving this correlation, but we note that the higher radiation field intensities ($\rm U > 10^4$) are not preferred after fitting the observed SEDs (see Section \ref{sec:D21_SED}), suggesting that the hardness of the radiation field may contribute to the correlation more. }

\subsubsection{THEMIS Models} \label{sec:themis}

\changes{Since the \citetalias{Draine2021} models do not predict the parameter space covered by the Sextans~A clumps (specifically the 3.3/7.7 ratio), we also explore whether THEMIS \citep[The Heterogeneous dust Evolution Model for Interstellar Solids,][]{Jones2017} dust models better match our data. We create THEMIS grids using grain properties for the diffuse ISM \citep{Jones2017} and the Orion Bar PDR \citep{Elyajouri2024}. The incident radiation field shape includes both the Milky Way interstellar radiation field (MW ISRF) from \citet{Mathis1983} and the \citet{Bruzual2003} stellar population model (Z=0.0004, 10 Myr, Chabrier IMF) adopted in the \citet{Draine2021} grids in this paper. Within the grids, we varied the band gap ($E_g$), the minimum grain size ($a_{\rm min}$), grain size distribution slope ($\alpha$), and the radiation field intensity ($G_0$) in these models. We produce dust and PAH emission spectra using DustEM 4.3 \citep{Compiegne2010}. The spectra are then processed through the JWST synthetic photometry pipeline where we apply the same continuum subtraction procedure described in Section \ref{sec:con_sub}. Unlike for the \citetalias{Draine2021} models, we use the same contamination constant values as we do for the data, since the THEMIS models predict the aliphatic features in addition to the aromatic (i.e., there is a strong 3.4 \micron\ feature). }

\changes{The THEMIS models compared to the \citetalias{Draine2021} grids and the Sextans~A PAH clumps are presented in Supplementary Figure 5. The THEMIS models span a parameter space similar to that of the \citetalias{Draine2021} models and likewise fail to reproduce the elevated 3.3/7.7 ratios observed in Sextans A. While both models are physically grounded in laboratory and astrophysical data, they adopt different treatments of the small carbonaceous grain population. The \citetalias{Draine2021} models PAHs as a distinct population of carbonaceous nanoparticles, whereas in THEMIS the aromatic emission carriers arise from the smallest ($< 20$~nm) amorphous hydrocarbon (a-C(:H)) grains rather than a chemically distinct species. Further, the THEMIS models span a broader parameter space and provide a prediction for the 3.4~\micron\ aliphatic feature. Despite these differences, the agreement between the predicted band ratios in Supplementary Figure 5 suggests that the under-luminous 7.7 feature is robust and not an artifact of a particular modeling framework.}

\subsubsection{Origin of the under-luminous 7.7 \micron\ feature}

\changes{Both the \citetalias{Draine2021} and THEMIS models are unable to reproduce the fainter 7.7 \micron\ feature observed in Sextans A. A weak or absent 7.7 \micron\ feature is not uncommon in metal-poor system, however. In the SMC, the low ratios of 7.7/11.2 and 7.7/6.2 compared to the SINGS sample are attributed to the weakness of the 7.7 feature \citep{Sandstrom2012}. JWST observations of II~Zw~40 show detections of the 3.3, 6.2, and 11.3 features but no emission at 7.7 \micron\ \citep{Lai2025}. This pattern suggests that suppressed 7.7 emission may be a common characteristic of low metallicity environments, although larger samples are needed to establish this trend. }

\changes{A natural consequence of metal-poor systems favoring smaller grain size distributions is an increased fraction of neutral grains \citep{Li2001}. The PAH ionization is set by the balance between the photoionization and electron capture rate, which scales as $U a \sqrt{T_e} / n_e$, where $U$ is the radiation field, $a$ is the grain radius, $T_e$ is the temperature, and $n_e$ is the electron density \citep{Draine2021}. Therefore, the smaller grain size distributions lead to more neutral PAHs and therefore an under-luminous 7.7 \micron\ feature in these systems. Further, as Section \ref{sec:d21_band_ratio} states, the \citetalias{Draine2021} models assume PAH ionization fractions that are only a factor of two above or below the standard value, defined as the weighted average between the three phases of the ISM-- the cold neutral medium (CNM), warm neutral medium (WNM), and warm ionized medium \citep[see Figure 7 in][]{Li2001}. Given the compact nature of the PAHs in Sextans~A, it is likely that they reside in the CNM, which would drive the ionization fraction of the PAHs even lower. The treatment of charge state of PAHs in the \citetalias{Draine2021} models also may not capture the complexity of the PAHs in Sextans~A, as the models do not distinguish among the PAH ionization states (e.g., anions, cations, dications) and do not allow for variations in aliphatic/aromatic fraction or other structural characteristics. Another possibility is that the 7.7 \micron\ feature arises primarily from C-C stretching modes, its relative weakness may reflect reduced carbon availability and/or differences in grain structure at low metallicity. Future JWST spectroscopy of Sextans A will help us understand the nature of the 7.7 \micron\ feature and the origin of the model discrepancy. }

\section{Extended Data}

\setcounter{figure}{0}
\renewcommand{\thefigure}{ED\arabic{figure}}
\makeatletter
\renewcommand{\theHfigure}{ED\arabic{figure}}
\makeatother

\begin{figure}[h!]
    \centering
    \includegraphics[width=\textwidth]{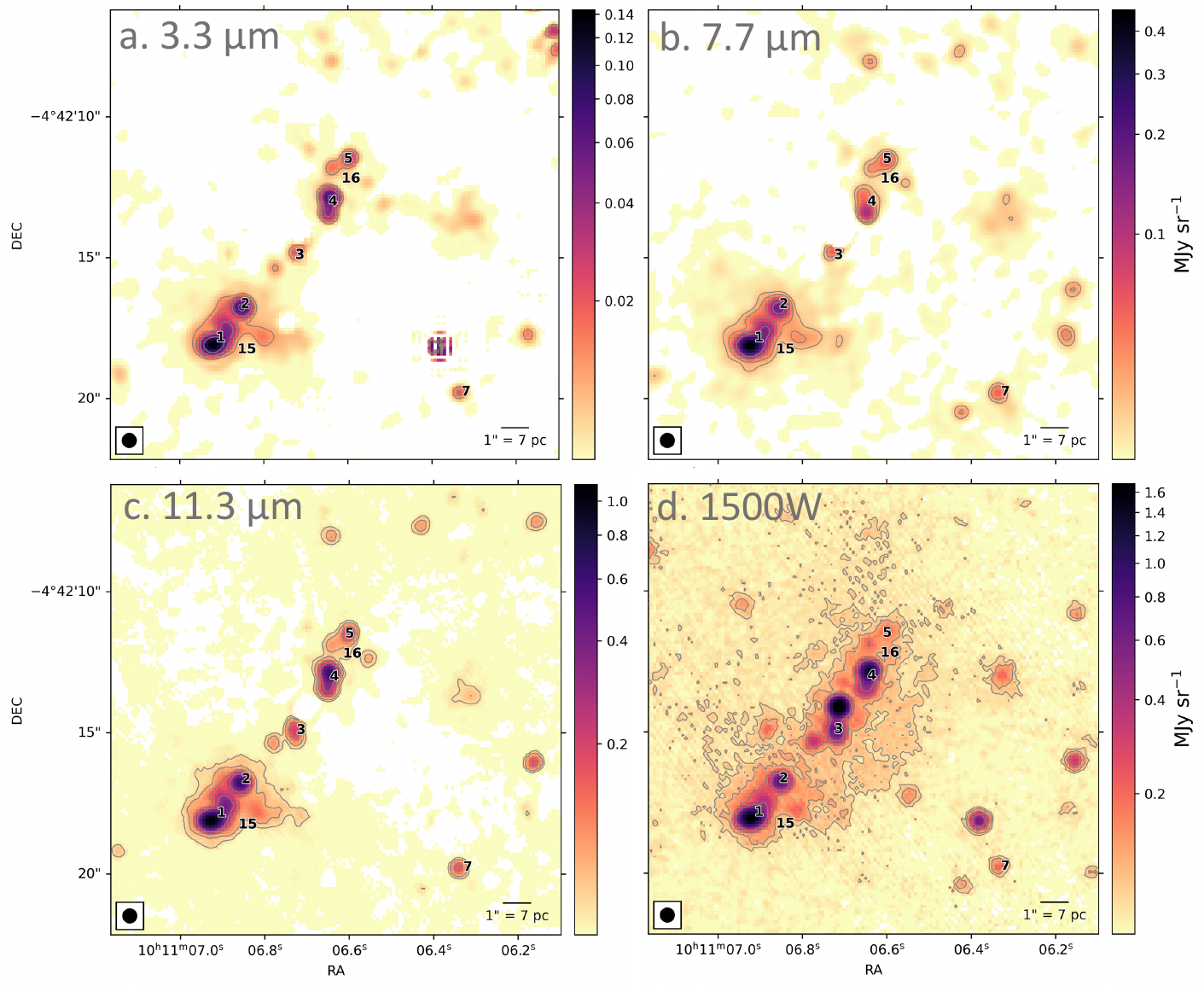}
    \caption{\textbf{Continuum subtracted images of the 3.3, 7.7, and 11.3 \micron\ features reveal compact, clumpy structure} The PAH emission flux in the three PAH filters, as well as the hot dust from the F1500W filter, targeted through the photometric imaging in JWST GO 2391. \changes{The 3.3, 7.7, and 11.3 \micron\ PAH fluxes are measured within the full filter bandpasses and may include small contributions from adjacent features.} The region is defined by the green square in Figure \ref{fig:Optical_JWST} and is the area with the brightest and most concentrated PAH emission. \changes{Clumps are numbered based off of their definitions in Table \ref{tab:clump_prop}.} All data is PSF matched to the F1500W filter prior to continuum subtraction, ensuring each image has a resolution of $\sim$0.48\arcsec (PSF in bottom left corners). PAH emission clumps are compact, comparable to the PSF, with sizes of $0.5-1.5''$ or $3-10 \rm \, pc$. Contours are at the 3$\sigma$, 5$\sigma$, 10$\sigma$, and 20$\sigma$ level.}
    \label{fig:consub_PAH_with_F1500W}
\end{figure}

\section*{Declarations}

\bmhead{Data Availability}
The JWST NIRCam and MIRI imaging data used in this work can be obtained in the MAST archive through this DOI link: \url{https://dx.doi.org/10.17909/432w-5204} with proposal ID 2391 (PI: J. Roman-Duval). The H$\alpha$ data for Sextans~A was obtained through the Local Volume Legacy (LVL) survey, which can be accessed through this link: \url{https://irsa.ipac.caltech.edu/data/SPITZER/LVL/images/}. The properties of the PAH clumps are provided in a machine readable table.

\bmhead{Code Availability}
The code used to reduce the raw JWST data, calibrate the PAH continuum subtraction method, analyze the resulting data, and produce all figures and tables included in this manuscript can be found through this Github link: \url{https://github.com/liztino/sextansA_PAH_paper/}. 

\bmhead{Acknowledgments}
This work is based on observations made with the NASA/ESA/CSA James Webb Space Telescope. The data were obtained from the Mikulski Archive for Space Telescopes at the Space Telescope Science Institute, which is operated by the Association of Universities for Research in Astronomy, Inc., under NASA contract NAS5-03127 for JWST. We acknowledge Interstellar Institute programmes and Paris-Saclay University’s Institut Pascal for hosting discussions that nourished the development of the ideas behind this work. We also acknowledge the NRAO and Research Corporation for their support of the PAH Workshop in Charlottesville, Virginia. 

\bmhead{Funding Statement}
These observations are associated with program GO-2391. Support for program 2391 was provided by NASA through a grant from the Space Telescope Science Institute, which is operated by the Association of Universities for Research in Astronomy, Inc., under NASA contract NAS 5-03127. This research was carried out in part at the Jet Propulsion Laboratory, California Institute of Technology, under a contract with the National Aeronautics and Space Administration (80NM0018D0004). O.G.T. acknowledges support from a Carnegie-Princeton Fellowship through Princeton University and the Carnegie Observatories.

\bmhead{Peer Review}
We greatly appreciate the anonymous reviewers for their thoughtful comments that greatly improved the work.

\bmhead{Author Contributions}

E.J.T. led the analysis and scientific interpretation of the work, including reduction of the JWST data, development of the PAH continuum subtraction method, discovery of the small Sextans~A PAH clumps, interpretation of the PAH features ratios, implications of the results, and is the primary author of the manuscript. J.R.-D. conceptualized the study, wrote the proposal to acquire the JWST data, and provided scientific expertise throughout the analysis. K.M.S. aided in the discovery of the PAH clumps using the 3.3 \micron\ continuum subtraction method from \citep{Sandstrom2023PAH3um} and helped develop the continuum subtraction method used in this work. C.M.W. and J.-D.T.S. produced the prediction for the inhibited PAH grain growth model for the metallicity of Sextans~A \citep{Whitcomb2026}. J.-D.T.S also provided the data for the SINGS sample in Figure \ref{fig:sigmaPAH_TIR}. B.T.D. and B.S.H. provided the models from \citetalias{Draine2021} and expertise in implementing them for this paper. M.L.B. and K.D.G. helped with data reduction of the NIRCam and MIRI imaging data. D.V.D.P. provided the PAHFIT results for the PDRs4All data, aiding in the continuum subtraction method. M.E. provided the THEMIS models to compare to the Sextans~A PAH clump band ratios. J.C., R.C., C.J.R.C., T.S.-Y.L., C.W.L., K.B.W.M., M.J.B.N., O.G.T., and B.F.W. all provided feedback during the analysis, scientific interpretation, and writing of this article. All authors gave feedback on initial drafts to improve the manuscript.

\bmhead{Corresponding Author}
Correspondence to Elizabeth J. Tarantino (etarantino@stsci.edu)

\bmhead{Competing Interests}
The authors declare no competing interests.

\section{Supplementary Information}

% Redefine the figure/table caption prefixes
\renewcommand{\figurename}{Supplementary Figure}
\renewcommand{\tablename}{Supplementary Table}

% Reset counters so numbering starts fresh at 1 for the supplement
\setcounter{figure}{0}
\setcounter{table}{0}

\setcounter{figure}{0}
\renewcommand{\thefigure}{\arabic{figure}}
\makeatletter
\renewcommand{\theHfigure}{\arabic{figure}}
\makeatother

\begin{figure}[h!]
    \centering
    \includegraphics[width=\textwidth]{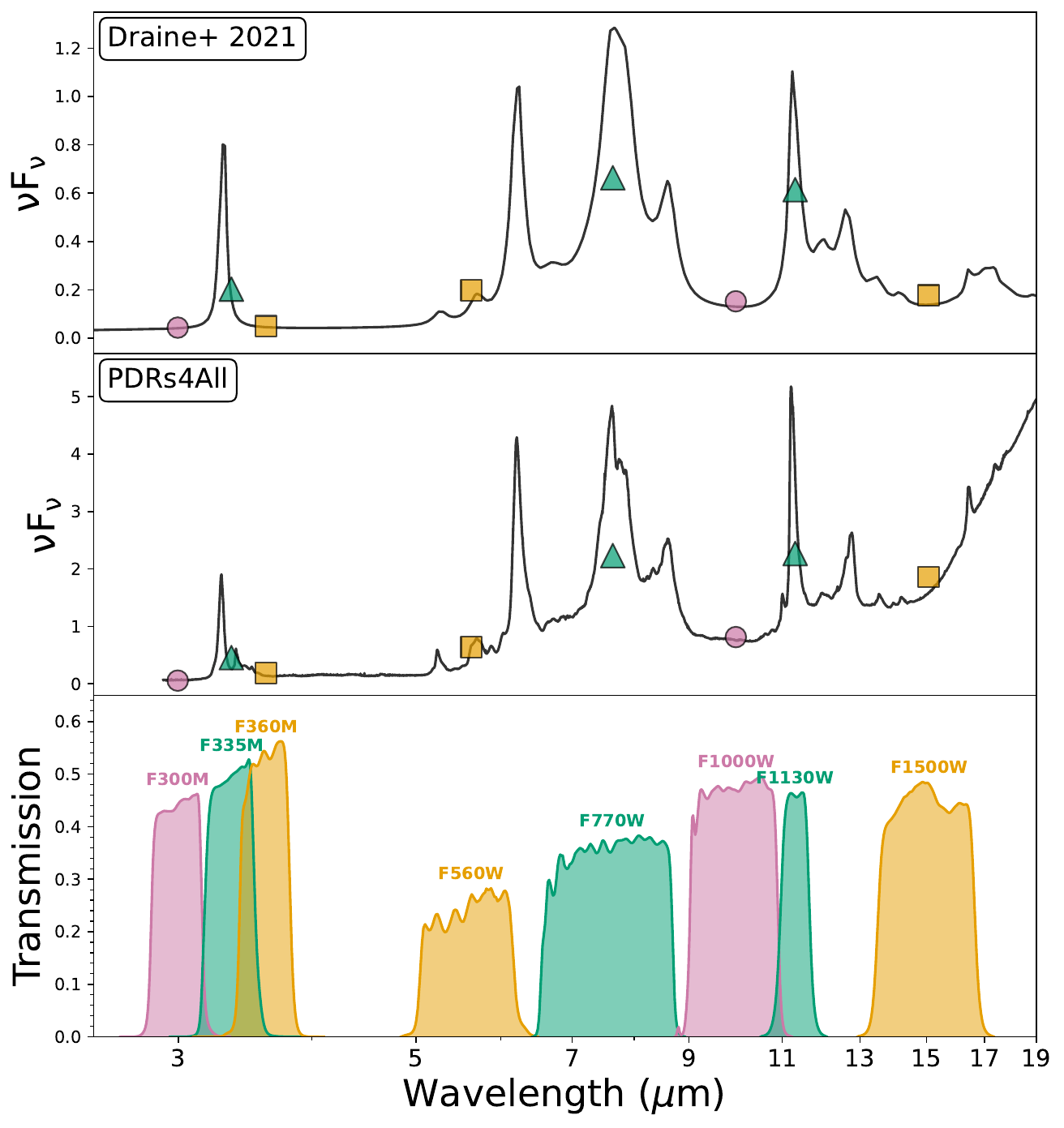}
    \caption{\textbf{Filters and spectral templates used for PAH continuum subtraction} JWST filter transmission curves used for the PAH continuum subtraction in this work (bottom panel) and two example mid-infrared spectra that show the PAH features in arbitrary flux units. The top panel is a model from \citetalias{Draine2021} ($\log \rm U = 1$; standard size and ionization distributions) and the middle panel is empirical data from PDRs4All with lines removed (dissociation front 1 template spectrum) \citep{Chown2024, VanDePutte2024, Peeters2024, VanDePutte2025}. The wavelength coverage of each filter shows which PAH features are contained in a given filter and are color coded by whether they are defined as the main PAH filter (green and triangle; F335M, F770W, and F1130W), the continuum filter without substantial PAH contamination (magenta and circle; F300M and F1000W), or the continuum filter that is contaminated by PAH complexes (gold and square; F360M, F560W, and F1500W). These filters are used in conjunction to calculate the continuum subtracted PAH flux in the main PAH filter (see Section 4.1.4 and Equation 1). We also apply synthetic photometry for each filter on the two example spectra.}
    \label{fig:filters}
\end{figure}

\begin{figure}[h!]
    \centering
    \includegraphics[width=\textwidth]{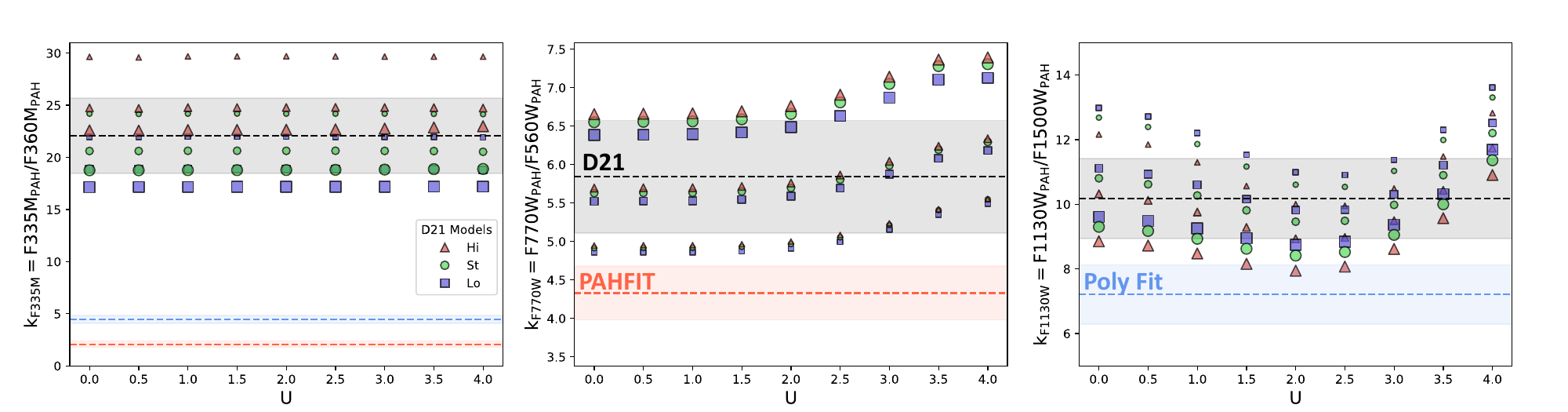}
    \caption{\textbf{Estimating the contamination in the flanking PAH filters for continuum subtraction} The MIR spectrum is rich in PAH features, making it difficult to identify a true ``continuum'' filter required for an accurate continuum subtraction of PAH emission. We estimate the degree of PAH contamination using template spectra from \citetalias{Draine2021} models and the Orion Bar PDRs4All team \citep{Peeters2024, Chown2024, VanDePutte2024, VanDePutte2025}. The contamination constant, $k$, is equal to the amount of PAH emission in the main PAH filter-- F335M (left), F770W (middle), or F1130W (right)-- divided by the amount of PAH emission in the contaminated continuum filter-- F360M (left), F560W (middle), or F1500W (right). We report the values of $k$ from the \citetalias{Draine2021} grids as a function of ionizing intensity, $\log U$. The properties of the PAHs are represented by the points, where the red triangle shows high ionization PAHs, the green circle is standard, and the blue square is low. The size of the points corresponds to the PAH grain size distribution grid. The dashed black line is the average of the \citetalias{Draine2021} models and the standard deviation is given in the gray region. We also show results from the PDRs4All template spectra fits, where the red dashed line uses the PAHFITs from \citet{VanDePutte2025} and the blue dashed line uses a simple polynomial to fit and remove the continuum.}
    \label{fig:k_values}
\end{figure}

\begin{table}[h] 
\caption{PAH Contamination Constants for Continuum Subtraction}\label{tab:k_vals} 
\begin{tabular}{cccc} 
\toprule 
Filter Trio & $k_{1}$ & $k_{2}$ & Conversion Factor \\ 
\midrule 
F300M/F335M/F360M & PDRs4All PAHFIT & PDRs4All Polyfit & $0.826$ \\
&  $k = 2.07 \pm 0.30$ &  $k = 4.45 \pm 0.39$ &  \\
F560W/F770W/F1000W & PDRs4All PAHFIT & Draine2021 & $0.965$ \\
&  $k = 4.33 \pm 0.35$ &  $k = 5.84 \pm 0.73$ &  \\
F1000W/F1130W/F1500W & PDRs4All Polyfit & Draine2021 & $0.991$ \\
&  $k = 7.21 \pm 0.92$ &  $k = 10.17 \pm 1.24$ &  \\
\end{tabular} 
\footnotetext{Table Notes: The contamination constants used for each filter trio in this work and the method used to derive them; $k_{1}$ is the adopted value used throughout the paper. PAH fluxes derived from $k_1$ can be converted to $k_2$ using the provided factor.}
\end{table}

\begin{table}[h] 
\caption{Modified Bandwidth Values Calculated from \citetalias{Draine2021} models}\label{tab:bandwidth} 
\begin{tabular}{ccc} 
\toprule 
Filter & BW ($\mathrm{\AA}$) & $\mathrm{BW_{mod}}$ ($\mathrm{\AA}$) \\ 
\midrule 
F360M & 3474 & 4124 $\pm$ 42 \\
F770W & 19736 & 28690 $\pm$ 1325 \\
F1130W & 7337 & 18295 $\pm$ 1554 \\
\end{tabular} 
\footnotetext{Table Notes: The conversion between average flux density within a JWST filter and integrated flux requires knowledge of the astrophysical signal convolved with the filter bandpass. BW is the bandwidth defined in the NIRCam and MIRI JWST documentation. $\mathrm{BW_{mod}}$ is calculated through the \citetalias{Draine2021} models, where the uncertainty is the $1\sigma$ standard deviation between models.}
\end{table}

\begin{figure}[h!]
    \centering
    \includegraphics[width=\textwidth]{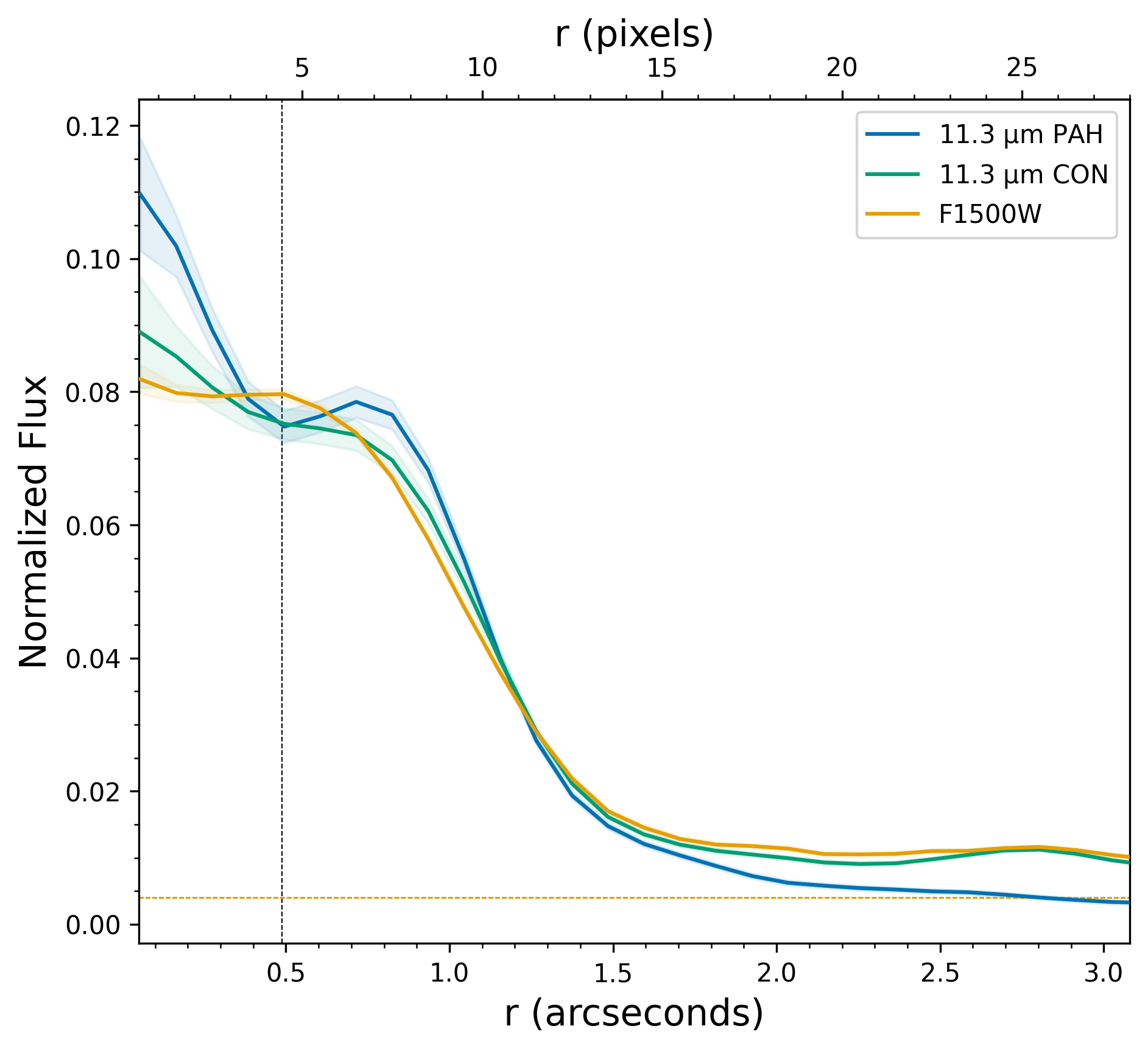}
    \caption{\textbf{Radial profile of the most extended and brightest PAH emission structure, clump 15}  We present the 11.3 \micron\ PAH flux in blue, the continuum at 11.3 \micron\ in green, and the F1500W filter as a tracer of hot dust in orange. Each profile is the azimuthally averaged flux in a circular annular aperture, where the center is the clump defined by the dendrogram structure analysis (see Section 4.3). The profiles are normalized so the sum is equal to one. The FWHM of the 0.\arcsec488 PSF is given by the vertical dashed black line and the 1$\sigma$ level uncertainty for each profile are the horizontal dashed lines (they all overlap in this scaling). The PAH emission profile peaks towards the center of the PAH clump, falls off before reaching another peak when the azimuthal averages passes through a clump local maximum, and then declines to the 1$\sigma$ noise floor. The continuum and F1500W profiles instead continue to a level 2.5$\sigma$ above, signifying that they trace more extended emission than the PAH profile. }
    \label{fig:radprof}
\end{figure}

\begin{figure}[h!]
    \centering
    \includegraphics[width=0.5\textwidth]{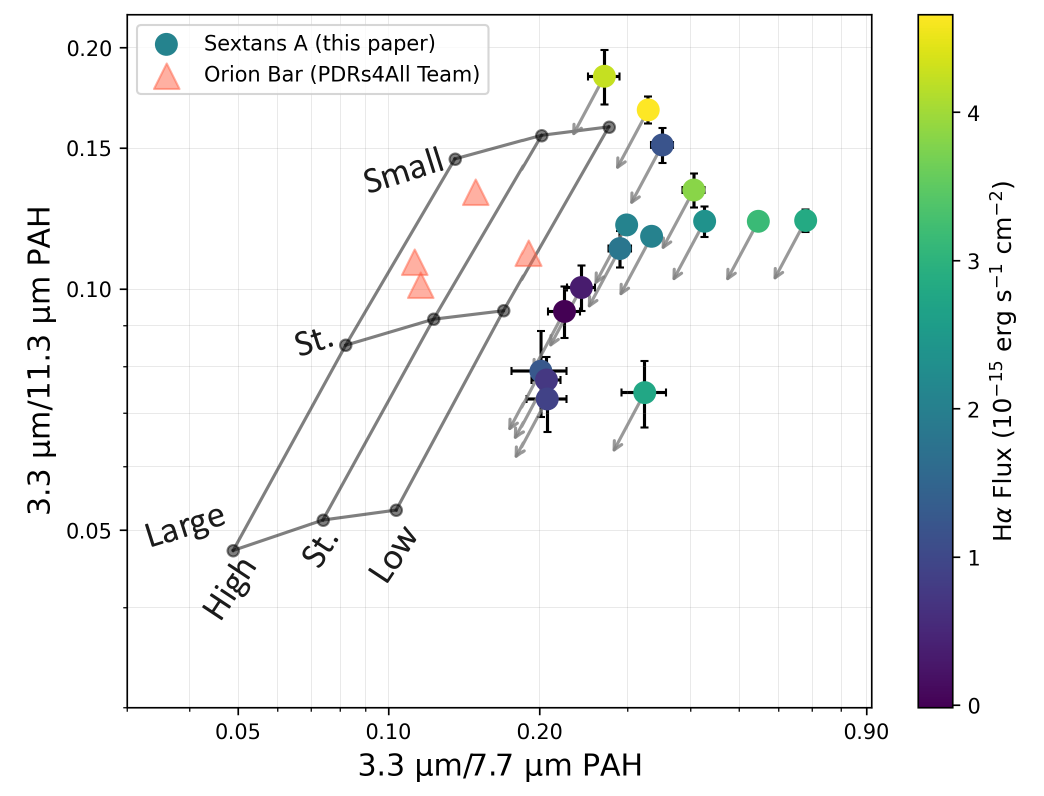}
    \caption{\textbf{Shift in the Sextans~A PAH feature ratios} Ratios between PAH features in Sextans~A where the arrows show how the ratios would shift when assuming the $k_2$ contamination constant rather than $k_1$ (see Table \ref{tab:k_vals}). The magnitude of the shift is mostly driven by the difference in the 3.3 \micron\ flux, which varies by 20\% between $k_1$ and $k_2$. The band ratios move slightly closer to the \citetalias{Draine2021} grids, but overall do not impact our conclusions. The errorbars represent the uncertainty on the 3.3, 7.7, and 11.3 \micron\ fluxes and the points are colored by the H$\alpha$ flux from the LVL survey \citep{Dale2009}. We overlay model predictions from \citetalias{Draine2021} with the low metallicity $Z=0.0004 \approx 0.02 Z_{\odot}$, 10 Myr starburst from \citet{Bruzual2003} as the input spectrum and average the results for radiation field intensities of $\log U=0-3.5$ since there is little scatter. The model grids show band ratio predictions for the size of PAHs, ranging from small ($a_{\rm 01} = \rm 3 \AA$) to large ($a_{\rm 01} = 5 \rm \AA$) size distributions, and the ionization of the PAHs, with standard (St.) values corresponding to the ionization fraction given in \citet{DraineLi2007}. Red triangles are spectroscopy from the PDRs4All team of the Orion Bar \citep{Chown2024, Peeters2024} where we applied synthetic photometry and the same continuum subtraction as the Sextans~A PAH clumps to provide a direct comparison.}
    \label{fig:band_ratios_k2}
\end{figure}

\begin{figure}[h!]
    \centering
    \includegraphics[width=1\textwidth]{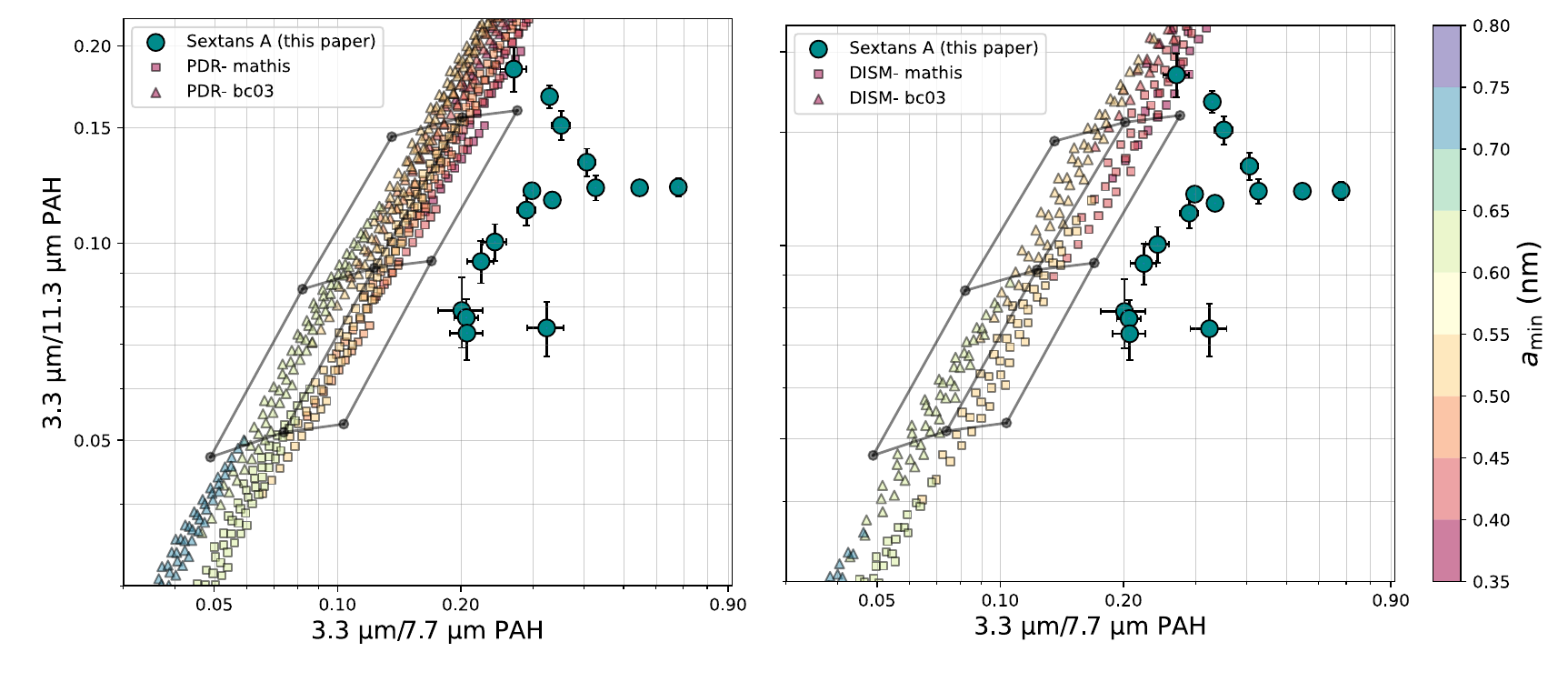}
    \caption{\textbf{Sextans~A PAH feature ratios compared to THEMIS models} The predicted 3.3/7.7 and 3.3/11.3 band ratios from THEMIS models, with the results from Sextans~A and the \citet{Draine2021} grids from Supplementary Figure \ref{fig:band_ratios_k2} overlaid. The left plot uses the PDR model from the Orion Bar \citep{Elyajouri2024} while the right uses the diffuse interstellar medium model \citep{Jones2017}. The square points are for the \citet{Mathis1983} input spectrum while the triangles represent the \citet{Bruzual2003} low metallicity burst model. We vary the band gap ($E_g$), the minimum grain size ($a_{\rm min}$), grain size distribution slope ($\alpha$), and the radiation field intensity ($G_0$) in the THEMIS models. The color of the points represents $a_{\rm min}$ and the Sextans~A PAH clumps agree with lower minimum grain size values. The errorbars represent the uncertainty on the 3.3, 7.7, and 11.3 \micron\ fluxes. Although the THEMIS models cover a broad and diverse parameter space, they occupy a similar range as \citet{Draine2021} and fail to predict the elevated 3.3/7.7 ratio seen in Sextans~A.}
    \label{fig:THEMIS}
\end{figure}

\clearpage



\begin{thebibliography}{120}
\ifx \bisbn   \undefined \def \bisbn  #1{ISBN #1}\fi
\ifx \binits  \undefined \def \binits#1{#1}\fi
\ifx \bauthor  \undefined \def \bauthor#1{#1}\fi
\ifx \batitle  \undefined \def \batitle#1{#1}\fi
\ifx \bjtitle  \undefined \def \bjtitle#1{#1}\fi
\ifx \bvolume  \undefined \def \bvolume#1{\textbf{#1}}\fi
\ifx \byear  \undefined \def \byear#1{#1}\fi
\ifx \bissue  \undefined \def \bissue#1{#1}\fi
\ifx \bfpage  \undefined \def \bfpage#1{#1}\fi
\ifx \blpage  \undefined \def \blpage #1{#1}\fi
\ifx \burl  \undefined \def \burl#1{\textsf{#1}}\fi
\ifx \doiurl  \undefined \def \doiurl#1{\url{https://doi.org/#1}}\fi
\ifx \betal  \undefined \def \betal{\textit{et al.}}\fi
\ifx \binstitute  \undefined \def \binstitute#1{#1}\fi
\ifx \binstitutionaled  \undefined \def \binstitutionaled#1{#1}\fi
\ifx \bctitle  \undefined \def \bctitle#1{#1}\fi
\ifx \beditor  \undefined \def \beditor#1{#1}\fi
\ifx \bpublisher  \undefined \def \bpublisher#1{#1}\fi
\ifx \bbtitle  \undefined \def \bbtitle#1{#1}\fi
\ifx \bedition  \undefined \def \bedition#1{#1}\fi
\ifx \bseriesno  \undefined \def \bseriesno#1{#1}\fi
\ifx \blocation  \undefined \def \blocation#1{#1}\fi
\ifx \bsertitle  \undefined \def \bsertitle#1{#1}\fi
\ifx \bsnm \undefined \def \bsnm#1{#1}\fi
\ifx \bsuffix \undefined \def \bsuffix#1{#1}\fi
\ifx \bparticle \undefined \def \bparticle#1{#1}\fi
\ifx \barticle \undefined \def \barticle#1{#1}\fi
\bibcommenthead
\ifx \bconfdate \undefined \def \bconfdate #1{#1}\fi
\ifx \botherref \undefined \def \botherref #1{#1}\fi
\ifx \url \undefined \def \url#1{\textsf{#1}}\fi
\ifx \bchapter \undefined \def \bchapter#1{#1}\fi
\ifx \bbook \undefined \def \bbook#1{#1}\fi
\ifx \bcomment \undefined \def \bcomment#1{#1}\fi
\ifx \oauthor \undefined \def \oauthor#1{#1}\fi
\ifx \citeauthoryear \undefined \def \citeauthoryear#1{#1}\fi
\ifx \endbibitem  \undefined \def \endbibitem {}\fi
\ifx \bconflocation  \undefined \def \bconflocation#1{#1}\fi
\ifx \arxivurl  \undefined \def \arxivurl#1{\textsf{#1}}\fi
\csname PreBibitemsHook\endcsname

%%% 1
\bibitem[\protect\citeauthoryear{Knacke}{1977}]{Knacke1977}
\begin{barticle}
\bauthor{\bsnm{Knacke}, \binits{R.F.}}:
\batitle{Carbonaceous compounds in interstellar dust}.
\bjtitle{Nature}
\bvolume{269}(\bissue{5624}),
\bfpage{132}--\blpage{134}
(\byear{1977})
\doiurl{10.1038/269132a0}
\end{barticle}
\endbibitem

%%% 2
\bibitem[\protect\citeauthoryear{Allamandola et~al.}{1985}]{Allamandola1985}
\begin{barticle}
\bauthor{\bsnm{Allamandola}, \binits{L.J.}},
\bauthor{\bsnm{Tielens}, \binits{A.G.G.M.}},
\bauthor{\bsnm{Barker}, \binits{J.R.}}:
\batitle{Polycyclic aromatic hydrocarbons and the unidentified infrared
  emission bands: Auto exhaust along the milky way.}
\bjtitle{Astrophys. J. Lett.}
\bvolume{290},
\bfpage{25}--\blpage{28}
(\byear{1985})
\doiurl{10.1086/184435}
\end{barticle}
\endbibitem

%%% 3
\bibitem[\protect\citeauthoryear{Tielens}{2008}]{Tielens2008}
\begin{barticle}
\bauthor{\bsnm{Tielens}, \binits{A.G.G.M.}}:
\batitle{Interstellar polycyclic aromatic hydrocarbon molecules.}
\bjtitle{Annu. Rev. Astron. Astrophys.}
\bvolume{46},
\bfpage{289}--\blpage{337}
(\byear{2008})
\doiurl{10.1146/annurev.astro.46.060407.145211}
\end{barticle}
\endbibitem

%%% 4
\bibitem[\protect\citeauthoryear{Li}{2020}]{Li2020}
\begin{barticle}
\bauthor{\bsnm{Li}, \binits{A.}}:
\batitle{Spitzer's perspective of polycyclic aromatic hydrocarbons in
  galaxies}.
\bjtitle{Nat. Astron.}
\bvolume{4},
\bfpage{339}--\blpage{351}
(\byear{2020})
\doiurl{10.1038/s41550-020-1051-1}
\end{barticle}
\endbibitem

%%% 5
\bibitem[\protect\citeauthoryear{Smith et~al.}{2007}]{Smith2007}
\begin{barticle}
\bauthor{\bsnm{Smith}, \binits{J.D.T.}},
\bauthor{\bsnm{Draine}, \binits{B.T.}},
\bauthor{\bsnm{Dale}, \binits{D.A.}},
\bauthor{\bsnm{Moustakas}, \binits{J.}},
\bauthor{\bsnm{Kennicutt}, \binits{{\relax Jr}.} \bsuffix{R.~C.}},
\bauthor{\bsnm{Helou}, \binits{G.}},
\bauthor{\bsnm{Armus}, \binits{L.}},
\bauthor{\bsnm{Roussel}, \binits{H.}},
\bauthor{\bsnm{Sheth}, \binits{K.}},
\bauthor{\bsnm{Bendo}, \binits{G.J.}},
\bauthor{\bsnm{Buckalew}, \binits{B.A.}},
\bauthor{\bsnm{Calzetti}, \binits{D.}},
\bauthor{\bsnm{Engelbracht}, \binits{C.W.}},
\bauthor{\bsnm{Gordon}, \binits{K.D.}},
\bauthor{\bsnm{Hollenbach}, \binits{D.J.}},
\bauthor{\bsnm{Li}, \binits{A.}},
\bauthor{\bsnm{Malhotra}, \binits{S.}},
\bauthor{\bsnm{Murphy}, \binits{E.J.}},
\bauthor{\bsnm{Walter}, \binits{F.}}:
\batitle{The {{Mid-Infrared Spectrum}} of {{Star-forming Galaxies}}: {{Global
  Properties}} of {{Polycyclic Aromatic Hydrocarbon Emission}}}.
\bjtitle{Astrophys. J.}
\bvolume{656}(\bissue{2}),
\bfpage{770}--\blpage{791}
(\byear{2007})
\doiurl{10.1086/510549}
\end{barticle}
\endbibitem

%%% 6
\bibitem[\protect\citeauthoryear{Shipley et~al.}{2016}]{Shipley2016}
\begin{barticle}
\bauthor{\bsnm{Shipley}, \binits{H.V.}},
\bauthor{\bsnm{Papovich}, \binits{C.}},
\bauthor{\bsnm{Rieke}, \binits{G.H.}},
\bauthor{\bsnm{Brown}, \binits{M.J.I.}},
\bauthor{\bsnm{Moustakas}, \binits{J.}}:
\batitle{A {{New Star Formation Rate Calibration}} from {{Polycyclic Aromatic
  Hydrocarbon Emission Features}} and {{Application}} to {{High-redshift
  Galaxies}}}.
\bjtitle{Astrophys. J.}
\bvolume{818}(\bissue{1}),
\bfpage{60}
(\byear{2016})
\doiurl{10.3847/0004-637X/818/1/60}
\end{barticle}
\endbibitem

%%% 7
\bibitem[\protect\citeauthoryear{Ronayne et~al.}{2024}]{Ronayne2024}
\begin{barticle}
\bauthor{\bsnm{Ronayne}, \binits{K.}},
\bauthor{\bsnm{Papovich}, \binits{C.}},
\bauthor{\bsnm{Yang}, \binits{G.}},
\bauthor{\bsnm{Shen}, \binits{L.}},
\bauthor{\bsnm{Dickinson}, \binits{M.}},
\bauthor{\bsnm{Kennicutt}, \binits{R.}},
\bauthor{\bsnm{Alavi}, \binits{A.}},
\bauthor{\bsnm{Arrabal~Haro}, \binits{P.}},
\bauthor{\bsnm{Bagley}, \binits{M.B.}},
\bauthor{\bsnm{Burgarella}, \binits{D.}},
\bauthor{\bsnm{Le~Bail}, \binits{A.}},
\bauthor{\bsnm{Bell}, \binits{E.F.}},
\bauthor{\bsnm{Cleri}, \binits{N.J.}},
\bauthor{\bsnm{Cole}, \binits{J.}},
\bauthor{\bsnm{Costantin}, \binits{L.}},
\bauthor{\bsnm{{de la Vega}}, \binits{A.}},
\bauthor{\bsnm{Daddi}, \binits{E.}},
\bauthor{\bsnm{Elbaz}, \binits{D.}},
\bauthor{\bsnm{Finkelstein}, \binits{S.L.}},
\bauthor{\bsnm{Grogin}, \binits{N.A.}},
\bauthor{\bsnm{Holwerda}, \binits{B.W.}},
\bauthor{\bsnm{Kartaltepe}, \binits{J.S.}},
\bauthor{\bsnm{Kirkpatrick}, \binits{A.}},
\bauthor{\bsnm{Koekemoer}, \binits{A.M.}},
\bauthor{\bsnm{Lucas}, \binits{R.A.}},
\bauthor{\bsnm{Magnelli}, \binits{B.}},
\bauthor{\bsnm{Mobasher}, \binits{B.}},
\bauthor{\bsnm{{P{\'e}rez-Gonz{\'a}lez}}, \binits{P.G.}},
\bauthor{\bsnm{Prichard}, \binits{L.}},
\bauthor{\bsnm{Rafelski}, \binits{M.}},
\bauthor{\bsnm{Rodighiero}, \binits{G.}},
\bauthor{\bsnm{Sunnquist}, \binits{B.}},
\bauthor{\bsnm{Teplitz}, \binits{H.I.}},
\bauthor{\bsnm{Wang}, \binits{X.}},
\bauthor{\bsnm{Windhorst}, \binits{R.A.}},
\bauthor{\bsnm{Yung}, \binits{L.Y.A.}}:
\batitle{{{CEERS}}: 7.7 {$M$}m {{PAH Star Formation Rate Calibration}} with
  {{JWST MIRI}}}.
\bjtitle{Astrophys. J.}
\bvolume{970},
\bfpage{61}
(\byear{2024})
\doiurl{10.3847/1538-4357/ad5006}
\end{barticle}
\endbibitem

%%% 8
\bibitem[\protect\citeauthoryear{Cortzen et~al.}{2019}]{Cortzen2019}
\begin{barticle}
\bauthor{\bsnm{Cortzen}, \binits{I.}},
\bauthor{\bsnm{Garrett}, \binits{J.}},
\bauthor{\bsnm{Magdis}, \binits{G.}},
\bauthor{\bsnm{Rigopoulou}, \binits{D.}},
\bauthor{\bsnm{Valentino}, \binits{F.}},
\bauthor{\bsnm{{Pereira-Santaella}}, \binits{M.}},
\bauthor{\bsnm{Combes}, \binits{F.}},
\bauthor{\bsnm{{Alonso-Herrero}}, \binits{A.}},
\bauthor{\bsnm{Toft}, \binits{S.}},
\bauthor{\bsnm{Daddi}, \binits{E.}},
\bauthor{\bsnm{Elbaz}, \binits{D.}},
\bauthor{\bsnm{{G{\'o}mez-Guijarro}}, \binits{C.}},
\bauthor{\bsnm{Stockmann}, \binits{M.}},
\bauthor{\bsnm{Huang}, \binits{J.}},
\bauthor{\bsnm{Kramer}, \binits{C.}}:
\batitle{{{PAHs}} as tracers of the molecular gas in star-forming galaxies}.
\bjtitle{Mon. Not. R. Astron. Soc.}
\bvolume{482}(\bissue{2}),
\bfpage{1618}--\blpage{1633}
(\byear{2019})
\doiurl{10.1093/mnras/sty2777}
\end{barticle}
\endbibitem

%%% 9
\bibitem[\protect\citeauthoryear{Shivaei and
  Boogaard}{2024}]{ShivaeiBoogaard2024}
\begin{barticle}
\bauthor{\bsnm{Shivaei}, \binits{I.}},
\bauthor{\bsnm{Boogaard}, \binits{L.A.}}:
\batitle{The tight correlation between {{PAH}} and {{CO}} emission from z
  {$\sim$} 0 to 4}.
\bjtitle{Astron. Astrophys.}
\bvolume{691},
\bfpage{2}
(\byear{2024})
\doiurl{10.1051/0004-6361/202451826}
\end{barticle}
\endbibitem

%%% 10
\bibitem[\protect\citeauthoryear{Chown et~al.}{2025}]{Chown2025_phangs}
\begin{barticle}
\bauthor{\bsnm{Chown}, \binits{R.}},
\bauthor{\bsnm{Leroy}, \binits{A.K.}},
\bauthor{\bsnm{Sandstrom}, \binits{K.}},
\bauthor{\bsnm{Chastenet}, \binits{J.}},
\bauthor{\bsnm{Sutter}, \binits{J.}},
\bauthor{\bsnm{Koch}, \binits{E.W.}},
\bauthor{\bsnm{Koziol}, \binits{H.B.}},
\bauthor{\bsnm{Neumann}, \binits{L.}},
\bauthor{\bsnm{Sun}, \binits{J.}},
\bauthor{\bsnm{Williams}, \binits{T.G.}},
\bauthor{\bsnm{Baron}, \binits{D.}},
\bauthor{\bsnm{Anand}, \binits{G.S.}},
\bauthor{\bsnm{Barnes}, \binits{{\relax Ashley}.T.}},
\bauthor{\bsnm{Bazzi}, \binits{Z.}},
\bauthor{\bsnm{Belfiore}, \binits{F.}},
\bauthor{\bsnm{Bigiel}, \binits{F.}},
\bauthor{\bsnm{Bolatto}, \binits{A.}},
\bauthor{\bsnm{Boquien}, \binits{M.}},
\bauthor{\bsnm{Cao}, \binits{Y.}},
\bauthor{\bsnm{Chevance}, \binits{M.}},
\bauthor{\bsnm{Colombo}, \binits{D.}},
\bauthor{\bsnm{Dale}, \binits{D.A.}},
\bauthor{\bsnm{{den Brok}}, \binits{J.}},
\bauthor{\bsnm{Egorov}, \binits{O.V.}},
\bauthor{\bsnm{Eibensteiner}, \binits{C.}},
\bauthor{\bsnm{Emsellem}, \binits{E.}},
\bauthor{\bsnm{Hassani}, \binits{H.}},
\bauthor{\bsnm{Henshaw}, \binits{J.D.}},
\bauthor{\bsnm{He}, \binits{H.}},
\bauthor{\bsnm{Kim}, \binits{J.}},
\bauthor{\bsnm{Klessen}, \binits{R.S.}},
\bauthor{\bsnm{Kreckel}, \binits{K.}},
\bauthor{\bsnm{Larson}, \binits{K.L.}},
\bauthor{\bsnm{Lee}, \binits{J.C.}},
\bauthor{\bsnm{Meidt}, \binits{S.E.}},
\bauthor{\bsnm{Murphy}, \binits{E.J.}},
\bauthor{\bsnm{Oakes}, \binits{E.K.}},
\bauthor{\bsnm{Ostriker}, \binits{E.C.}},
\bauthor{\bsnm{Pan}, \binits{H.-A.}},
\bauthor{\bsnm{Pathak}, \binits{D.}},
\bauthor{\bsnm{Rosolowsky}, \binits{E.}},
\bauthor{\bsnm{Sarbadhicary}, \binits{S.K.}},
\bauthor{\bsnm{Schinnerer}, \binits{E.}},
\bauthor{\bsnm{Teng}, \binits{Y.-H.}},
\bauthor{\bsnm{Thilker}, \binits{D.A.}},
\bauthor{\bsnm{Weinbeck}, \binits{T.D.}},
\bauthor{\bsnm{Watkins}, \binits{E.J.}}:
\batitle{Polycyclic {{Aromatic Hydrocarbon}} and {{CO}}(2--1) {{Emission}} at
  50--150 pc {{Scales}} in 70 {{Nearby Galaxies}}}.
\bjtitle{Astrophys. J.}
\bvolume{983},
\bfpage{64}
(\byear{2025})
\doiurl{10.3847/1538-4357/adbd40}
\end{barticle}
\endbibitem

%%% 11
\bibitem[\protect\citeauthoryear{Wolfire et~al.}{1995}]{Wolfire1995}
\begin{barticle}
\bauthor{\bsnm{Wolfire}, \binits{M.G.}},
\bauthor{\bsnm{Hollenbach}, \binits{D.}},
\bauthor{\bsnm{McKee}, \binits{C.F.}},
\bauthor{\bsnm{Tielens}, \binits{A.G.G.M.}},
\bauthor{\bsnm{Bakes}, \binits{E.L.O.}}:
\batitle{The {{Neutral Atomic Phases}} of the {{Interstellar Medium}}}.
\bjtitle{Astrophys. J.}
\bvolume{443},
\bfpage{152}
(\byear{1995})
\doiurl{10.1086/175510}
\end{barticle}
\endbibitem

%%% 12
\bibitem[\protect\citeauthoryear{Engelbracht et~al.}{2008}]{Engelbracht2008}
\begin{barticle}
\bauthor{\bsnm{Engelbracht}, \binits{C.W.}},
\bauthor{\bsnm{Rieke}, \binits{G.H.}},
\bauthor{\bsnm{Gordon}, \binits{K.D.}},
\bauthor{\bsnm{Smith}, \binits{J.-D.T.}},
\bauthor{\bsnm{Werner}, \binits{M.W.}},
\bauthor{\bsnm{Moustakas}, \binits{J.}},
\bauthor{\bsnm{Willmer}, \binits{C.N.A.}},
\bauthor{\bsnm{Vanzi}, \binits{L.}}:
\batitle{Metallicity {{Effects}} on {{Dust Properties}} in {{Starbursting
  Galaxies}}}.
\bjtitle{Astrophys. J.}
\bvolume{678}(\bissue{2}),
\bfpage{804}--\blpage{827}
(\byear{2008})
\doiurl{10.1086/529513}
\end{barticle}
\endbibitem

%%% 13
\bibitem[\protect\citeauthoryear{Jackson et~al.}{2006}]{Jackson2006}
\begin{barticle}
\bauthor{\bsnm{Jackson}, \binits{D.C.}},
\bauthor{\bsnm{Cannon}, \binits{J.M.}},
\bauthor{\bsnm{Skillman}, \binits{E.D.}},
\bauthor{\bsnm{Lee}, \binits{H.}},
\bauthor{\bsnm{Gehrz}, \binits{R.D.}},
\bauthor{\bsnm{Woodward}, \binits{C.E.}},
\bauthor{\bsnm{Polomski}, \binits{E.}}:
\batitle{Hot {{Dust}} and {{Polycyclic Aromatic Hydrocarbon Emission}} at {{Low
  Metallicity}}: {{A Spitzer Survey}} of {{Local Group}} and {{Other Nearby
  Dwarf Galaxies}}}.
\bjtitle{Astrophys. J.}
\bvolume{646}(\bissue{1}),
\bfpage{192}--\blpage{204}
(\byear{2006})
\doiurl{10.1086/504707}
\end{barticle}
\endbibitem

%%% 14
\bibitem[\protect\citeauthoryear{Wu et~al.}{2006}]{Wu2006}
\begin{barticle}
\bauthor{\bsnm{Wu}, \binits{Y.}},
\bauthor{\bsnm{Charmandaris}, \binits{V.}},
\bauthor{\bsnm{Hao}, \binits{L.}},
\bauthor{\bsnm{Brandl}, \binits{B.R.}},
\bauthor{\bsnm{{Bernard-Salas}}, \binits{J.}},
\bauthor{\bsnm{Spoon}, \binits{H.W.W.}},
\bauthor{\bsnm{Houck}, \binits{J.R.}}:
\batitle{Mid-{{Infrared Properties}} of {{Low-Metallicity Blue Compact Dwarf
  Galaxies}} from the {{Spitzer Infrared Spectrograph}}}.
\bjtitle{Astrophys. J.}
\bvolume{639}(\bissue{1}),
\bfpage{157}--\blpage{172}
(\byear{2006})
\doiurl{10.1086/499226}
\end{barticle}
\endbibitem

%%% 15
\bibitem[\protect\citeauthoryear{Draine et~al.}{2007}]{Draine2007}
\begin{barticle}
\bauthor{\bsnm{Draine}, \binits{B.T.}},
\bauthor{\bsnm{Dale}, \binits{D.A.}},
\bauthor{\bsnm{Bendo}, \binits{G.}},
\bauthor{\bsnm{Gordon}, \binits{K.D.}},
\bauthor{\bsnm{Smith}, \binits{J.D.T.}},
\bauthor{\bsnm{Armus}, \binits{L.}},
\bauthor{\bsnm{Engelbracht}, \binits{C.W.}},
\bauthor{\bsnm{Helou}, \binits{G.}},
\bauthor{\bsnm{Kennicutt}, \binits{{\relax Jr}.} \bsuffix{R.~C.}},
\bauthor{\bsnm{Li}, \binits{A.}},
\bauthor{\bsnm{Roussel}, \binits{H.}},
\bauthor{\bsnm{Walter}, \binits{F.}},
\bauthor{\bsnm{Calzetti}, \binits{D.}},
\bauthor{\bsnm{Moustakas}, \binits{J.}},
\bauthor{\bsnm{Murphy}, \binits{E.J.}},
\bauthor{\bsnm{Rieke}, \binits{G.H.}},
\bauthor{\bsnm{Bot}, \binits{C.}},
\bauthor{\bsnm{Hollenbach}, \binits{D.J.}},
\bauthor{\bsnm{Sheth}, \binits{K.}},
\bauthor{\bsnm{Teplitz}, \binits{H.I.}}:
\batitle{Dust {{Masses}}, {{PAH Abundances}}, and {{Starlight Intensities}} in
  the {{SINGS Galaxy Sample}}}.
\bjtitle{Astrophys. J.}
\bvolume{663}(\bissue{2}),
\bfpage{866}--\blpage{894}
(\byear{2007})
\doiurl{10.1086/518306}
\end{barticle}
\endbibitem

%%% 16
\bibitem[\protect\citeauthoryear{Gordon et~al.}{2008}]{Gordon2008}
\begin{barticle}
\bauthor{\bsnm{Gordon}, \binits{K.D.}},
\bauthor{\bsnm{Engelbracht}, \binits{C.W.}},
\bauthor{\bsnm{Rieke}, \binits{G.H.}},
\bauthor{\bsnm{Misselt}, \binits{K.A.}},
\bauthor{\bsnm{Smith}, \binits{J.-D.T.}},
\bauthor{\bsnm{Kennicutt}, \binits{R.C.} \bsuffix{Jr.}}:
\batitle{The {{Behavior}} of the {{Aromatic Features}} in {{M101
  H}}{\textsc{ii}}{{Regions}}: {{Evidence}} for {{Dust Processing}}}.
\bjtitle{Astrophys. J.}
\bvolume{682}(\bissue{1}),
\bfpage{336}--\blpage{354}
(\byear{2008})
\doiurl{10.1086/589567}
\end{barticle}
\endbibitem

%%% 17
\bibitem[\protect\citeauthoryear{Hunt et~al.}{2010}]{Hunt2010}
\begin{barticle}
\bauthor{\bsnm{Hunt}, \binits{L.K.}},
\bauthor{\bsnm{Thuan}, \binits{T.X.}},
\bauthor{\bsnm{Izotov}, \binits{Y.I.}},
\bauthor{\bsnm{Sauvage}, \binits{M.}}:
\batitle{The {{Spitzer View}} of {{Low-Metallicity Star Formation}}. {{III}}.
  {{Fine-Structure Lines}}, {{Aromatic Features}}, and {{Molecules}}}.
\bjtitle{Astrophys. J.}
\bvolume{712}(\bissue{1}),
\bfpage{164}--\blpage{187}
(\byear{2010})
\doiurl{10.1088/0004-637X/712/1/164}
\end{barticle}
\endbibitem

%%% 18
\bibitem[\protect\citeauthoryear{Sandstrom et~al.}{2010}]{Sandstrom2010}
\begin{barticle}
\bauthor{\bsnm{Sandstrom}, \binits{K.M.}},
\bauthor{\bsnm{Bolatto}, \binits{A.D.}},
\bauthor{\bsnm{Draine}, \binits{B.T.}},
\bauthor{\bsnm{Bot}, \binits{C.}},
\bauthor{\bsnm{Stanimirovi{\'c}}, \binits{S.}}:
\batitle{The {{Spitzer Survey}} of the {{Small Magellanic Cloud}}
  ({{S}}{\textsuperscript{3}}{{MC}}): {{Insights}} into the {{Life Cycle}} of
  {{Polycyclic Aromatic Hydrocarbons}}}.
\bjtitle{Astrophys. J.}
\bvolume{715}(\bissue{2}),
\bfpage{701}--\blpage{723}
(\byear{2010})
\doiurl{10.1088/0004-637X/715/2/701}
\end{barticle}
\endbibitem

%%% 19
\bibitem[\protect\citeauthoryear{Chastenet et~al.}{2019}]{Chastenet2019}
\begin{barticle}
\bauthor{\bsnm{Chastenet}, \binits{J.}},
\bauthor{\bsnm{Sandstrom}, \binits{K.}},
\bauthor{\bsnm{Chiang}, \binits{I.-D.}},
\bauthor{\bsnm{Leroy}, \binits{A.K.}},
\bauthor{\bsnm{Utomo}, \binits{D.}},
\bauthor{\bsnm{Bot}, \binits{C.}},
\bauthor{\bsnm{Gordon}, \binits{K.D.}},
\bauthor{\bsnm{Draine}, \binits{B.T.}},
\bauthor{\bsnm{Fukui}, \binits{Y.}},
\bauthor{\bsnm{Onishi}, \binits{T.}},
\bauthor{\bsnm{Tsuge}, \binits{K.}}:
\batitle{The {{Polycyclic Aromatic Hydrocarbon Mass Fraction}} on a 10 pc
  {{Scale}} in the {{Magellanic Clouds}}}.
\bjtitle{Astrophys. J.}
\bvolume{876}(\bissue{1}),
\bfpage{62}
(\byear{2019})
\doiurl{10.3847/1538-4357/ab16cf}
\end{barticle}
\endbibitem

%%% 20
\bibitem[\protect\citeauthoryear{Aniano et~al.}{2020}]{Aniano2020}
\begin{barticle}
\bauthor{\bsnm{Aniano}, \binits{G.}},
\bauthor{\bsnm{Draine}, \binits{B.T.}},
\bauthor{\bsnm{Hunt}, \binits{L.K.}},
\bauthor{\bsnm{Sandstrom}, \binits{K.}},
\bauthor{\bsnm{Calzetti}, \binits{D.}},
\bauthor{\bsnm{Kennicutt}, \binits{R.C.}},
\bauthor{\bsnm{Dale}, \binits{D.A.}},
\bauthor{\bsnm{Galametz}, \binits{M.}},
\bauthor{\bsnm{Gordon}, \binits{K.D.}},
\bauthor{\bsnm{Leroy}, \binits{A.K.}},
\bauthor{\bsnm{Smith}, \binits{J.-D.T.}},
\bauthor{\bsnm{Roussel}, \binits{H.}},
\bauthor{\bsnm{Sauvage}, \binits{M.}},
\bauthor{\bsnm{Walter}, \binits{F.}},
\bauthor{\bsnm{Armus}, \binits{L.}},
\bauthor{\bsnm{Bolatto}, \binits{A.D.}},
\bauthor{\bsnm{Boquien}, \binits{M.}},
\bauthor{\bsnm{Crocker}, \binits{A.}},
\bauthor{\bsnm{De~Looze}, \binits{I.}},
\bauthor{\bsnm{Donovan~Meyer}, \binits{J.}},
\bauthor{\bsnm{Helou}, \binits{G.}},
\bauthor{\bsnm{Hinz}, \binits{J.}},
\bauthor{\bsnm{Johnson}, \binits{B.D.}},
\bauthor{\bsnm{Koda}, \binits{J.}},
\bauthor{\bsnm{Miller}, \binits{A.}},
\bauthor{\bsnm{Montiel}, \binits{E.}},
\bauthor{\bsnm{Murphy}, \binits{E.J.}},
\bauthor{\bsnm{Rela{\~n}o}, \binits{M.}},
\bauthor{\bsnm{Rix}, \binits{H.-W.}},
\bauthor{\bsnm{Schinnerer}, \binits{E.}},
\bauthor{\bsnm{Skibba}, \binits{R.}},
\bauthor{\bsnm{Wolfire}, \binits{M.G.}},
\bauthor{\bsnm{Engelbracht}, \binits{C.W.}}:
\batitle{Modeling {{Dust}} and {{Starlight}} in {{Galaxies Observed}} by
  {{Spitzer}} and {{Herschel}}: {{The KINGFISH Sample}}}.
\bjtitle{Astrophys. J.}
\bvolume{889}(\bissue{2}),
\bfpage{150}
(\byear{2020})
\doiurl{10.3847/1538-4357/ab5fdb}
\end{barticle}
\endbibitem

%%% 21
\bibitem[\protect\citeauthoryear{Whitcomb et~al.}{2024}]{Whitcomb2024}
\begin{barticle}
\bauthor{\bsnm{Whitcomb}, \binits{C.M.}},
\bauthor{\bsnm{Smith}, \binits{J.-D.T.}},
\bauthor{\bsnm{Sandstrom}, \binits{K.}},
\bauthor{\bsnm{Starkey}, \binits{C.A.}},
\bauthor{\bsnm{Donnelly}, \binits{G.P.}},
\bauthor{\bsnm{Draine}, \binits{B.T.}},
\bauthor{\bsnm{Skillman}, \binits{E.D.}},
\bauthor{\bsnm{Dale}, \binits{D.A.}},
\bauthor{\bsnm{Armus}, \binits{L.}},
\bauthor{\bsnm{Hensley}, \binits{B.S.}},
\bauthor{\bsnm{Lai}, \binits{T.S.-Y.}},
\bauthor{\bsnm{Kennicutt}, \binits{R.C.}}:
\batitle{The {{Metallicity Dependence}} of {{PAH Emission}} in {{Galaxies}}.
  {{I}}. {{Insights}} from {{Deep Radial Spitzer Spectroscopy}}}.
\bjtitle{Astrophys. J.}
\bvolume{974},
\bfpage{20}
(\byear{2024})
\doiurl{10.3847/1538-4357/ad66c8}
\end{barticle}
\endbibitem

%%% 22
\bibitem[\protect\citeauthoryear{{R{\'e}my-Ruyer}
  et~al.}{2014}]{Remy-Ruyer2014}
\begin{barticle}
\bauthor{\bsnm{{R{\'e}my-Ruyer}}, \binits{A.}},
\bauthor{\bsnm{Madden}, \binits{S.C.}},
\bauthor{\bsnm{Galliano}, \binits{F.}},
\bauthor{\bsnm{Galametz}, \binits{M.}},
\bauthor{\bsnm{Takeuchi}, \binits{T.T.}},
\bauthor{\bsnm{Asano}, \binits{R.S.}},
\bauthor{\bsnm{Zhukovska}, \binits{S.}},
\bauthor{\bsnm{Lebouteiller}, \binits{V.}},
\bauthor{\bsnm{Cormier}, \binits{D.}},
\bauthor{\bsnm{Jones}, \binits{A.}},
\bauthor{\bsnm{Bocchio}, \binits{M.}},
\bauthor{\bsnm{Baes}, \binits{M.}},
\bauthor{\bsnm{Bendo}, \binits{G.J.}},
\bauthor{\bsnm{Boquien}, \binits{M.}},
\bauthor{\bsnm{Boselli}, \binits{A.}},
\bauthor{\bsnm{DeLooze}, \binits{I.}},
\bauthor{\bsnm{{Doublier-Pritchard}}, \binits{V.}},
\bauthor{\bsnm{Hughes}, \binits{T.}},
\bauthor{\bsnm{Karczewski}, \binits{O.{\L}.}},
\bauthor{\bsnm{Spinoglio}, \binits{L.}}:
\batitle{Gas-to-dust mass ratios in local galaxies over a 2 dex metallicity
  range}.
\bjtitle{Astron. Astrophys.}
\bvolume{563},
\bfpage{31}
(\byear{2014})
\doiurl{10.1051/0004-6361/201322803}
\end{barticle}
\endbibitem

%%% 23
\bibitem[\protect\citeauthoryear{Micelotta et~al.}{2010a}]{Micelotta2010}
\begin{barticle}
\bauthor{\bsnm{Micelotta}, \binits{E.R.}},
\bauthor{\bsnm{Jones}, \binits{A.P.}},
\bauthor{\bsnm{Tielens}, \binits{A.G.G.M.}}:
\batitle{Polycyclic aromatic hydrocarbon processing in a hot gas}.
\bjtitle{Astron. Astrophys.}
\bvolume{510},
\bfpage{37}
(\byear{2010})
\doiurl{10.1051/0004-6361/200911683}
\end{barticle}
\endbibitem

%%% 24
\bibitem[\protect\citeauthoryear{Micelotta et~al.}{2010b}]{Micelotta2010a}
\begin{barticle}
\bauthor{\bsnm{Micelotta}, \binits{E.R.}},
\bauthor{\bsnm{Jones}, \binits{A.P.}},
\bauthor{\bsnm{Tielens}, \binits{A.G.G.M.}}:
\batitle{Polycyclic aromatic hydrocarbon processing in interstellar shocks}.
\bjtitle{Astron. Astrophys.}
\bvolume{510},
\bfpage{36}
(\byear{2010})
\doiurl{10.1051/0004-6361/200911682}
\end{barticle}
\endbibitem

%%% 25
\bibitem[\protect\citeauthoryear{Greenberg et~al.}{2000}]{Greenberg2000}
\begin{barticle}
\bauthor{\bsnm{Greenberg}, \binits{J.M.}},
\bauthor{\bsnm{Gillette}, \binits{J.S.}},
\bauthor{\bsnm{Mu{\~n}oz~Caro}, \binits{G.M.}},
\bauthor{\bsnm{Mahajan}, \binits{T.B.}},
\bauthor{\bsnm{Zare}, \binits{R.N.}},
\bauthor{\bsnm{Li}, \binits{A.}},
\bauthor{\bsnm{Schutte}, \binits{W.A.}},
\bauthor{\bsnm{{de Groot}}, \binits{M.}},
\bauthor{\bsnm{{Mendoza-G{\'o}mez}}, \binits{C.}}:
\batitle{Ultraviolet {{Photoprocessing}} of {{Interstellar Dust Mantles}} as a
  {{Source}} of {{Polycyclic Aromatic Hydrocarbons}} and {{Other Conjugated
  Molecules}}}.
\bjtitle{Astrophys. J.}
\bvolume{531},
\bfpage{71}--\blpage{73}
(\byear{2000})
\doiurl{10.1086/312526}
\end{barticle}
\endbibitem

%%% 26
\bibitem[\protect\citeauthoryear{Sandstrom et~al.}{2012}]{Sandstrom2012}
\begin{barticle}
\bauthor{\bsnm{Sandstrom}, \binits{K.M.}},
\bauthor{\bsnm{Bolatto}, \binits{A.D.}},
\bauthor{\bsnm{Bot}, \binits{C.}},
\bauthor{\bsnm{Draine}, \binits{B.T.}},
\bauthor{\bsnm{Ingalls}, \binits{J.G.}},
\bauthor{\bsnm{Israel}, \binits{F.P.}},
\bauthor{\bsnm{Jackson}, \binits{J.M.}},
\bauthor{\bsnm{Leroy}, \binits{A.K.}},
\bauthor{\bsnm{Li}, \binits{A.}},
\bauthor{\bsnm{Rubio}, \binits{M.}},
\bauthor{\bsnm{Simon}, \binits{J.D.}},
\bauthor{\bsnm{Smith}, \binits{J.D.T.}},
\bauthor{\bsnm{Stanimirovi{\'c}}, \binits{S.}},
\bauthor{\bsnm{Tielens}, \binits{A.G.G.M.}},
\bauthor{\bsnm{{van Loon}}, \binits{J.T.}}:
\batitle{The {{Spitzer Spectroscopic Survey}} of the {{Small Magellanic Cloud}}
  ({{S4MC}}): {{Probing}} the {{Physical State}} of {{Polycyclic Aromatic
  Hydrocarbons}} in a {{Low-metallicity Environment}}}.
\bjtitle{Astrophys. J.}
\bvolume{744},
\bfpage{20}
(\byear{2012})
\doiurl{10.1088/0004-637X/744/1/20}
\end{barticle}
\endbibitem

%%% 27
\bibitem[\protect\citeauthoryear{Spilker et~al.}{2023}]{Spilker2023}
\begin{barticle}
\bauthor{\bsnm{Spilker}, \binits{J.S.}},
\bauthor{\bsnm{Phadke}, \binits{K.A.}},
\bauthor{\bsnm{Aravena}, \binits{M.}},
\bauthor{\bsnm{Archipley}, \binits{M.}},
\bauthor{\bsnm{Bayliss}, \binits{M.B.}},
\bauthor{\bsnm{Birkin}, \binits{J.E.}},
\bauthor{\bsnm{B{\'e}thermin}, \binits{M.}},
\bauthor{\bsnm{Burgoyne}, \binits{J.}},
\bauthor{\bsnm{Cathey}, \binits{J.}},
\bauthor{\bsnm{Chapman}, \binits{S.C.}},
\bauthor{\bsnm{Dahle}, \binits{H.}},
\bauthor{\bsnm{Gonzalez}, \binits{A.H.}},
\bauthor{\bsnm{Gururajan}, \binits{G.}},
\bauthor{\bsnm{Hayward}, \binits{C.C.}},
\bauthor{\bsnm{Hezaveh}, \binits{Y.D.}},
\bauthor{\bsnm{Hill}, \binits{R.}},
\bauthor{\bsnm{Hutchison}, \binits{T.A.}},
\bauthor{\bsnm{Kim}, \binits{K.J.}},
\bauthor{\bsnm{Kim}, \binits{S.}},
\bauthor{\bsnm{Law}, \binits{D.}},
\bauthor{\bsnm{Legin}, \binits{R.}},
\bauthor{\bsnm{Malkan}, \binits{M.A.}},
\bauthor{\bsnm{Marrone}, \binits{D.P.}},
\bauthor{\bsnm{Murphy}, \binits{E.J.}},
\bauthor{\bsnm{Narayanan}, \binits{D.}},
\bauthor{\bsnm{Navarre}, \binits{A.}},
\bauthor{\bsnm{Olivier}, \binits{G.M.}},
\bauthor{\bsnm{Rich}, \binits{J.A.}},
\bauthor{\bsnm{Rigby}, \binits{J.R.}},
\bauthor{\bsnm{Reuter}, \binits{C.}},
\bauthor{\bsnm{Rhoads}, \binits{J.E.}},
\bauthor{\bsnm{Sharon}, \binits{K.}},
\bauthor{\bsnm{Smith}, \binits{J.D.T.}},
\bauthor{\bsnm{Solimano}, \binits{M.}},
\bauthor{\bsnm{Sulzenauer}, \binits{N.}},
\bauthor{\bsnm{Vieira}, \binits{J.D.}},
\bauthor{\bsnm{Vizgan}, \binits{D.}},
\bauthor{\bsnm{Wei{\ss}}, \binits{A.}},
\bauthor{\bsnm{Whitaker}, \binits{K.E.}}:
\batitle{Spatial variations in aromatic hydrocarbon emission in a dust-rich
  galaxy}.
\bjtitle{Nature}
\bvolume{618}(\bissue{7966}),
\bfpage{708}--\blpage{711}
(\byear{2023})
\doiurl{10.1038/s41586-023-05998-6}
\end{barticle}
\endbibitem

%%% 28
\bibitem[\protect\citeauthoryear{{McKinney} et~al.}{2026}]{McKinney2026}
\begin{barticle}
\bauthor{\bsnm{{McKinney}}, \binits{J.}},
\bauthor{\bsnm{{Eleazer}}, \binits{M.}},
\bauthor{\bsnm{{Pope}}, \binits{A.}},
\bauthor{\bsnm{{Sajina}}, \binits{A.}},
\bauthor{\bsnm{{Alberts}}, \binits{S.}},
\bauthor{\bsnm{{Stone}}, \binits{M.}},
\bauthor{\bsnm{{Sajkov}}, \binits{L.}},
\bauthor{\bsnm{{Vanicek}}, \binits{V.}},
\bauthor{\bsnm{{Kirkpatrick}}, \binits{A.}},
\bauthor{\bsnm{{Lai}}, \binits{T.S.-Y.}},
\bauthor{\bsnm{{Casey}}, \binits{C.M.}},
\bauthor{\bsnm{{Armus}}, \binits{L.}},
\bauthor{\bsnm{{D{\'\i}az-Santos}}, \binits{T.}},
\bauthor{\bsnm{{Korkus}}, \binits{A.}},
\bauthor{\bsnm{{Cooper}}, \binits{O.}},
\bauthor{\bsnm{{House}}, \binits{L.R.}},
\bauthor{\bsnm{{Akins}}, \binits{H.}},
\bauthor{\bsnm{{Lambrides}}, \binits{E.}},
\bauthor{\bsnm{{Long}}, \binits{A.S.}},
\bauthor{\bsnm{{Yan}}, \binits{L.}}:
\batitle{{A JWST MIRI LRS Survey of 37 Massive Star-forming Galaxies and Active
  Galactic Nuclei at Cosmic Noon: Overview and First Results}}.
\bjtitle{Astrophys. J.}
\bvolume{1001}(\bissue{1}),
\bfpage{93}
(\byear{2026})
\doiurl{10.3847/1538-4357/ae4d46}
{\href{https://arxiv.org/abs/2510.07365}{{arXiv:2510.07365}}}
{[astro-ph.GA]}
\end{barticle}
\endbibitem

%%% 29
\bibitem[\protect\citeauthoryear{Hirschauer et~al.}{2024}]{Hirschauer2024}
\begin{barticle}
\bauthor{\bsnm{Hirschauer}, \binits{A.S.}},
\bauthor{\bsnm{Crouzet}, \binits{N.}},
\bauthor{\bsnm{Habel}, \binits{N.}},
\bauthor{\bsnm{Lenki{\'c}}, \binits{L.}},
\bauthor{\bsnm{Nally}, \binits{C.}},
\bauthor{\bsnm{Jones}, \binits{O.C.}},
\bauthor{\bsnm{Bortolini}, \binits{G.}},
\bauthor{\bsnm{Boyer}, \binits{M.L.}},
\bauthor{\bsnm{Justtanont}, \binits{K.}},
\bauthor{\bsnm{Meixner}, \binits{M.}},
\bauthor{\bsnm{{\"O}stlin}, \binits{G.}},
\bauthor{\bsnm{Wright}, \binits{G.S.}},
\bauthor{\bsnm{Azzollini}, \binits{R.}},
\bauthor{\bsnm{Blommaert}, \binits{J.A.D.L.}},
\bauthor{\bsnm{Brandl}, \binits{B.}},
\bauthor{\bsnm{Decin}, \binits{L.}},
\bauthor{\bsnm{Nayak}, \binits{O.}},
\bauthor{\bsnm{Royer}, \binits{P.}},
\bauthor{\bsnm{Sargent}, \binits{B.A.}},
\bauthor{\bsnm{{van der Werf}}, \binits{P.}}:
\batitle{Imaging of {{I Zw}} 18 by {{JWST}}. {{I}}. {{Detecting Dusty Stellar
  Populations}}}.
\bjtitle{Astron. J.}
\bvolume{168},
\bfpage{23}
(\byear{2024})
\doiurl{10.3847/1538-3881/ad4967}
\end{barticle}
\endbibitem

%%% 30
\bibitem[\protect\citeauthoryear{Chown et~al.}{2025}]{Chown2025_dwarf}
\begin{barticle}
\bauthor{\bsnm{Chown}, \binits{R.}},
\bauthor{\bsnm{Leroy}, \binits{A.K.}},
\bauthor{\bsnm{Bolatto}, \binits{A.D.}},
\bauthor{\bsnm{Chastenet}, \binits{J.}},
\bauthor{\bsnm{Glover}, \binits{S.C.O.}},
\bauthor{\bsnm{Indebetouw}, \binits{R.}},
\bauthor{\bsnm{Koch}, \binits{E.W.}},
\bauthor{\bsnm{Donovan~Meyer}, \binits{J.}},
\bauthor{\bsnm{Pingel}, \binits{N.M.}},
\bauthor{\bsnm{Rosolowsky}, \binits{E.}},
\bauthor{\bsnm{Sandstrom}, \binits{K.}},
\bauthor{\bsnm{Sutter}, \binits{J.}},
\bauthor{\bsnm{Tarantino}, \binits{E.}},
\bauthor{\bsnm{Bigiel}, \binits{F.}},
\bauthor{\bsnm{Boquien}, \binits{M.}},
\bauthor{\bsnm{Chiang}, \binits{I.-D.}},
\bauthor{\bsnm{Dale}, \binits{D.A.}},
\bauthor{\bsnm{Dalcanton}, \binits{J.J.}},
\bauthor{\bsnm{Egorov}, \binits{O.V.}},
\bauthor{\bsnm{Eibensteiner}, \binits{C.}},
\bauthor{\bsnm{Grasha}, \binits{K.}},
\bauthor{\bsnm{Hassani}, \binits{H.}},
\bauthor{\bsnm{He}, \binits{H.}},
\bauthor{\bsnm{Kim}, \binits{J.}},
\bauthor{\bsnm{Meidt}, \binits{S.}},
\bauthor{\bsnm{Pathak}, \binits{D.}},
\bauthor{\bsnm{Sarbadhicary}, \binits{S.K.}},
\bauthor{\bsnm{Stanimirovic}, \binits{S.}},
\bauthor{\bsnm{Villanueva}, \binits{V.}},
\bauthor{\bsnm{Williams}, \binits{T.G.}}:
\batitle{Relationships between {{Polycyclic Aromatic Hydrocarbons}}, {{Small
  Dust Grains}}, {{H2}}, and {{H I}} in {{Local Group Dwarf Galaxies NGC}} 6822
  and {{WLM Using JWST}}, {{ALMA}}, and the {{VLA}}}.
\bjtitle{Astrophys. J.}
\bvolume{987},
\bfpage{91}
(\byear{2025})
\doiurl{10.3847/1538-4357/add73a}
\end{barticle}
\endbibitem

%%% 31
\bibitem[\protect\citeauthoryear{Mingozzi et~al.}{2025}]{Mingozzi2025}
\begin{barticle}
\bauthor{\bsnm{Mingozzi}, \binits{M.}},
\bauthor{\bsnm{{Garcia del Valle-Espinosa}}, \binits{M.}},
\bauthor{\bsnm{James}, \binits{B.L.}},
\bauthor{\bsnm{Rickards~Vaught}, \binits{R.J.}},
\bauthor{\bsnm{Hayes}, \binits{M.}},
\bauthor{\bsnm{Amor{\'i}n}, \binits{R.O.}},
\bauthor{\bsnm{Leitherer}, \binits{C.}},
\bauthor{\bsnm{Aloisi}, \binits{A.}},
\bauthor{\bsnm{Hunt}, \binits{L.}},
\bauthor{\bsnm{Law}, \binits{D.}},
\bauthor{\bsnm{Richardson}, \binits{C.T.}},
\bauthor{\bsnm{Pidgeon}, \binits{A.}},
\bauthor{\bsnm{{Arellano-C{\'o}rdova}}, \binits{K.Z.}},
\bauthor{\bsnm{Berg}, \binits{D.A.}},
\bauthor{\bsnm{Chisholm}, \binits{J.}},
\bauthor{\bsnm{Hernandez}, \binits{S.}},
\bauthor{\bsnm{Jones}, \binits{L.}},
\bauthor{\bsnm{Kumari}, \binits{N.}},
\bauthor{\bsnm{Martin}, \binits{C.L.}},
\bauthor{\bsnm{Ravindranath}, \binits{S.}},
\bauthor{\bsnm{Vallini}, \binits{L.}},
\bauthor{\bsnm{Xu}, \binits{X.}}:
\batitle{Exploring the {{Mysterious High-ionization Source Powering}} [{{Ne
  V}}] in {{High-z Analog SBS0335-052 E}} with {{JWST}}/{{MIRI}}}.
\bjtitle{Astrophys. J.}
\bvolume{985},
\bfpage{253}
(\byear{2025})
\doiurl{10.3847/1538-4357/adc996}
\end{barticle}
\endbibitem

%%% 32
\bibitem[\protect\citeauthoryear{Hunt et~al.}{2025}]{Hunt2025b}
\begin{barticle}
\bauthor{\bsnm{Hunt}, \binits{L.K.}},
\bauthor{\bsnm{Draine}, \binits{B.T.}},
\bauthor{\bsnm{Navarro}, \binits{M.G.}},
\bauthor{\bsnm{Aloisi}, \binits{A.}},
\bauthor{\bsnm{Vaught}, \binits{R.J.R.}},
\bauthor{\bsnm{Adamo}, \binits{A.}},
\bauthor{\bsnm{Annibali}, \binits{F.}},
\bauthor{\bsnm{Calzetti}, \binits{D.}},
\bauthor{\bsnm{Hernandez}, \binits{S.}},
\bauthor{\bsnm{James}, \binits{B.L.}},
\bauthor{\bsnm{Mingozzi}, \binits{M.}},
\bauthor{\bsnm{Schneider}, \binits{R.}},
\bauthor{\bsnm{Tosi}, \binits{M.}},
\bauthor{\bsnm{Brandl}, \binits{B.}},
\bauthor{\bsnm{{del Valle-Espinosa}}, \binits{M.G.}},
\bauthor{\bsnm{Donnan}, \binits{F.}},
\bauthor{\bsnm{Hirschauer}, \binits{A.S.}},
\bauthor{\bsnm{Meixner}, \binits{M.}},
\bauthor{\bsnm{Rigopoulou}, \binits{D.}}:
\batitle{The {{Interstellar Medium}} in {{I Zw}} 18 {{Seen}} with
  {{JWST}}/{{MIRI}}. {{II}}. {{Warm Molecular Hydrogen}} and {{Warm Dust}}}.
\bjtitle{Astrophys. J.}
\bvolume{993},
\bfpage{84}
(\byear{2025})
\doiurl{10.3847/1538-4357/ae0191}
\end{barticle}
\endbibitem

%%% 33
\bibitem[\protect\citeauthoryear{Telford et~al.}{2025}]{Telford2025}
\begin{barticle}
\bauthor{\bsnm{Telford}, \binits{O.G.}},
\bauthor{\bsnm{Sandstrom}, \binits{K.M.}},
\bauthor{\bsnm{McQuinn}, \binits{K.B.W.}},
\bauthor{\bsnm{Glover}, \binits{S.C.O.}},
\bauthor{\bsnm{Tarantino}, \binits{E.J.}},
\bauthor{\bsnm{Bolatto}, \binits{A.D.}},
\bauthor{\bsnm{Rickards~Vaught}, \binits{R.J.}}:
\batitle{Molecular hydrogen in the extremely metal- and dust-poor galaxy {{Leo
  P}}}.
\bjtitle{Nature}
\bvolume{642}(\bissue{8069}),
\bfpage{900}--\blpage{904}
(\byear{2025})
\doiurl{10.1038/s41586-025-09115-7}
\end{barticle}
\endbibitem

%%% 34
\bibitem[\protect\citeauthoryear{{del Valle-Espinosa}
  et~al.}{2026}]{DelValle-Espinosa2026}
\begin{barticle}
\bauthor{\bsnm{{del Valle-Espinosa}}, \binits{M.G.}},
\bauthor{\bsnm{Mingozzi}, \binits{M.}},
\bauthor{\bsnm{James}, \binits{B.}},
\bauthor{\bsnm{{S{\'a}nchez-Janssen}}, \binits{R.}},
\bauthor{\bsnm{{Fern{\'a}ndez-Ontiveros}}, \binits{J.A.}},
\bauthor{\bsnm{Vaught}, \binits{R.J.}},
\bauthor{\bsnm{Amor{\'i}n}, \binits{R.O.}},
\bauthor{\bsnm{Hunt}, \binits{L.}},
\bauthor{\bsnm{Aloisi}, \binits{A.}},
\bauthor{\bsnm{{Arellano-C{\'o}rdova}}, \binits{K.Z.}},
\bauthor{\bsnm{Berg}, \binits{D.A.}},
\bauthor{\bsnm{Chisholm}, \binits{J.}},
\bauthor{\bsnm{Hayes}, \binits{M.}},
\bauthor{\bsnm{Hernandez}, \binits{S.}},
\bauthor{\bsnm{Hirschauer}, \binits{A.S.}},
\bauthor{\bsnm{Jones}, \binits{L.}},
\bauthor{\bsnm{Martin}, \binits{C.L.}},
\bauthor{\bsnm{Vallini}, \binits{L.}},
\bauthor{\bsnm{Xu}, \binits{X.}}:
\batitle{{{JWST}}/{{MIRI-MRS View}} of the {{Metal-poor Galaxy CGCG}} 007-025:
  {{The Spatial Location}} of {{Polycyclic Aromatic Hydrocarbons}} and {{Very
  Highly Ionized Gas}}}.
\bjtitle{Astrophys. J. Lett.}
\bvolume{998}(\bissue{2}),
\bfpage{45}
(\byear{2026})
\doiurl{10.3847/2041-8213/ae43ed}
\end{barticle}
\endbibitem

%%% 35
\bibitem[\protect\citeauthoryear{Kniazev et~al.}{2005}]{Kniazev2005}
\begin{barticle}
\bauthor{\bsnm{Kniazev}, \binits{A.Y.}},
\bauthor{\bsnm{Grebel}, \binits{E.K.}},
\bauthor{\bsnm{Pustilnik}, \binits{S.A.}},
\bauthor{\bsnm{Pramskij}, \binits{A.G.}},
\bauthor{\bsnm{Zucker}, \binits{D.B.}}:
\batitle{Spectrophotometry of {{Sextans A}} and {{B}}: {{Chemical Abundances}}
  of {{H II Regions}} and {{Planetary Nebulae}}}.
\bjtitle{Astron. J.}
\bvolume{130}(\bissue{4}),
\bfpage{1558}--\blpage{1573}
(\byear{2005})
\doiurl{10.1086/432931}
\end{barticle}
\endbibitem

%%% 36
\bibitem[\protect\citeauthoryear{Asplund et~al.}{2009}]{Asplund2009}
\begin{barticle}
\bauthor{\bsnm{Asplund}, \binits{M.}},
\bauthor{\bsnm{Grevesse}, \binits{N.}},
\bauthor{\bsnm{Sauval}, \binits{A.J.}},
\bauthor{\bsnm{Scott}, \binits{P.}}:
\batitle{The {{Chemical Composition}} of the {{Sun}}}.
\bjtitle{Annu. Rev. Astron. Astrophys.}
\bvolume{47},
\bfpage{481}--\blpage{522}
(\byear{2009})
\doiurl{10.1146/annurev.astro.46.060407.145222}
\end{barticle}
\endbibitem

%%% 37
\bibitem[\protect\citeauthoryear{Bellazzini et~al.}{2014}]{Bellazzini2014}
\begin{barticle}
\bauthor{\bsnm{Bellazzini}, \binits{M.}},
\bauthor{\bsnm{Beccari}, \binits{G.}},
\bauthor{\bsnm{Fraternali}, \binits{F.}},
\bauthor{\bsnm{Oosterloo}, \binits{T.A.}},
\bauthor{\bsnm{Sollima}, \binits{A.}},
\bauthor{\bsnm{Testa}, \binits{V.}},
\bauthor{\bsnm{Galleti}, \binits{S.}},
\bauthor{\bsnm{Perina}, \binits{S.}},
\bauthor{\bsnm{Faccini}, \binits{M.}},
\bauthor{\bsnm{Cusano}, \binits{F.}}:
\batitle{The extended structure of the dwarf irregular galaxies {{Sextans A}}
  and {{Sextans B}}. {{Signatures}} of tidal distortion in the outskirts of the
  {{Local Group}}}.
\bjtitle{Astron. Astrophys.}
\bvolume{566},
\bfpage{44}
(\byear{2014})
\doiurl{10.1051/0004-6361/201423659}
\end{barticle}
\endbibitem

%%% 38
\bibitem[\protect\citeauthoryear{McQuinn et~al.}{2017}]{McQuinn2017}
\begin{barticle}
\bauthor{\bsnm{McQuinn}, \binits{K.B.W.}},
\bauthor{\bsnm{Boyer}, \binits{M.L.}},
\bauthor{\bsnm{Mitchell}, \binits{M.B.}},
\bauthor{\bsnm{Skillman}, \binits{E.D.}},
\bauthor{\bsnm{Gehrz}, \binits{R.D.}},
\bauthor{\bsnm{Groenewegen}, \binits{M.A.T.}},
\bauthor{\bsnm{McDonald}, \binits{I.}},
\bauthor{\bsnm{Sloan}, \binits{G.C.}},
\bauthor{\bsnm{{van Loon}}, \binits{J.T.}},
\bauthor{\bsnm{Whitelock}, \binits{P.A.}},
\bauthor{\bsnm{Zijlstra}, \binits{A.A.}}:
\batitle{{{DUSTiNGS}}. {{III}}. {{Distribution}} of {{Intermediate-age}} and
  {{Old Stellar Populations}} in {{Disks}} and {{Outer Extremities}} of {{Dwarf
  Galaxies}}}.
\bjtitle{Astrophys. J.}
\bvolume{834}(\bissue{1}),
\bfpage{78}
(\byear{2017})
\doiurl{10.3847/1538-4357/834/1/78}
\end{barticle}
\endbibitem

%%% 39
\bibitem[\protect\citeauthoryear{Ott et~al.}{2012}]{Ott2012}
\begin{barticle}
\bauthor{\bsnm{Ott}, \binits{J.}},
\bauthor{\bsnm{Stilp}, \binits{A.M.}},
\bauthor{\bsnm{Warren}, \binits{S.R.}},
\bauthor{\bsnm{Skillman}, \binits{E.D.}},
\bauthor{\bsnm{Dalcanton}, \binits{J.J.}},
\bauthor{\bsnm{Walter}, \binits{F.}},
\bauthor{\bsnm{{de Blok}}, \binits{W.J.G.}},
\bauthor{\bsnm{Koribalski}, \binits{B.}},
\bauthor{\bsnm{West}, \binits{A.A.}}:
\batitle{{{VLA-ANGST}}: {{A High-resolution H I Survey}} of {{Nearby Dwarf
  Galaxies}}}.
\bjtitle{Astron. J.}
\bvolume{144},
\bfpage{123}
(\byear{2012})
\doiurl{10.1088/0004-6256/144/4/123}
\end{barticle}
\endbibitem

%%% 40
\bibitem[\protect\citeauthoryear{McConnachie}{2012}]{McConnachie2012}
\begin{barticle}
\bauthor{\bsnm{McConnachie}, \binits{A.W.}}:
\batitle{The {{Observed Properties}} of {{Dwarf Galaxies}} in and around the
  {{Local Group}}}.
\bjtitle{Astron. J.}
\bvolume{144},
\bfpage{4}
(\byear{2012})
\doiurl{10.1088/0004-6256/144/1/4}
\end{barticle}
\endbibitem

%%% 41
\bibitem[\protect\citeauthoryear{Lee et~al.}{2009}]{Lee2009}
\begin{barticle}
\bauthor{\bsnm{Lee}, \binits{J.C.}},
\bauthor{\bsnm{{Gil de Paz}}, \binits{A.}},
\bauthor{\bsnm{Tremonti}, \binits{C.}},
\bauthor{\bsnm{Kennicutt}, \binits{R.C.}},
\bauthor{\bsnm{Salim}, \binits{S.}},
\bauthor{\bsnm{Bothwell}, \binits{M.}},
\bauthor{\bsnm{Calzetti}, \binits{D.}},
\bauthor{\bsnm{Dalcanton}, \binits{J.}},
\bauthor{\bsnm{Dale}, \binits{D.}},
\bauthor{\bsnm{Engelbracht}, \binits{C.}},
\bauthor{\bsnm{Jos{\'e} G.~Funes}, \binits{S.J.}},
\bauthor{\bsnm{Johnson}, \binits{B.}},
\bauthor{\bsnm{Sakai}, \binits{S.}},
\bauthor{\bsnm{Skillman}, \binits{E.}},
\bauthor{\bsnm{{van Zee}}, \binits{L.}},
\bauthor{\bsnm{Walter}, \binits{F.}},
\bauthor{\bsnm{Weisz}, \binits{D.}}:
\batitle{{{COMPARISON OF H$\alpha$ AND UV STAR FORMATION RATES IN THE LOCAL
  VOLUME}}: {{SYSTEMATIC DISCREPANCIES FOR DWARF GALAXIES}}}.
\bjtitle{Astrophys. J.}
\bvolume{706}(\bissue{1}),
\bfpage{599}
(\byear{2009})
\doiurl{10.1088/0004-637X/706/1/599}
\end{barticle}
\endbibitem

%%% 42
\bibitem[\protect\citeauthoryear{Shi et~al.}{2014}]{Shi2014}
\begin{barticle}
\bauthor{\bsnm{Shi}, \binits{Y.}},
\bauthor{\bsnm{Armus}, \binits{L.}},
\bauthor{\bsnm{Helou}, \binits{G.}},
\bauthor{\bsnm{Stierwalt}, \binits{S.}},
\bauthor{\bsnm{Gao}, \binits{Y.}},
\bauthor{\bsnm{Wang}, \binits{J.}},
\bauthor{\bsnm{Zhang}, \binits{Z.-Y.}},
\bauthor{\bsnm{Gu}, \binits{Q.}}:
\batitle{Inefficient star formation in extremely metal poor galaxies}.
\bjtitle{Nature}
\bvolume{514}(\bissue{7522}),
\bfpage{335}--\blpage{338}
(\byear{2014})
\doiurl{10.1038/nature13820}
\end{barticle}
\endbibitem

%%% 43
\bibitem[\protect\citeauthoryear{Lupton et~al.}{2004}]{Lupton2004}
\begin{barticle}
\bauthor{\bsnm{Lupton}, \binits{R.}},
\bauthor{\bsnm{Blanton}, \binits{M.R.}},
\bauthor{\bsnm{Fekete}, \binits{G.}},
\bauthor{\bsnm{Hogg}, \binits{D.W.}},
\bauthor{\bsnm{O'Mullane}, \binits{W.}},
\bauthor{\bsnm{Szalay}, \binits{A.}},
\bauthor{\bsnm{Wherry}, \binits{N.}}:
\batitle{Preparing {{Red-Green-Blue Images}} from {{CCD Data}}}.
\bjtitle{Publ. Astron. Soc. Pac.}
\bvolume{116},
\bfpage{133}--\blpage{137}
(\byear{2004})
\doiurl{10.1086/382245}
\end{barticle}
\endbibitem

%%% 44
\bibitem[\protect\citeauthoryear{Draine et~al.}{2021}]{Draine2021}
\begin{barticle}
\bauthor{\bsnm{Draine}, \binits{B.T.}},
\bauthor{\bsnm{Li}, \binits{A.}},
\bauthor{\bsnm{Hensley}, \binits{B.S.}},
\bauthor{\bsnm{Hunt}, \binits{L.K.}},
\bauthor{\bsnm{Sandstrom}, \binits{K.}},
\bauthor{\bsnm{Smith}, \binits{J.-D.T.}}:
\batitle{Excitation of {{Polycyclic Aromatic Hydrocarbon Emission}}:
  {{Dependence}} on {{Size Distribution}}, {{Ionization}}, and {{Starlight
  Spectrum}} and {{Intensity}}}.
\bjtitle{Astrophys. J.}
\bvolume{917}(\bissue{1}),
\bfpage{3}
(\byear{2021})
\doiurl{10.3847/1538-4357/abff51}
\end{barticle}
\endbibitem

%%% 45
\bibitem[\protect\citeauthoryear{Husser et~al.}{2013}]{Husser2013}
\begin{barticle}
\bauthor{\bsnm{Husser}, \binits{T.-O.}},
\bauthor{\bsnm{Berg}, \binits{S.W.-v.}},
\bauthor{\bsnm{Dreizler}, \binits{S.}},
\bauthor{\bsnm{Homeier}, \binits{D.}},
\bauthor{\bsnm{Reiners}, \binits{A.}},
\bauthor{\bsnm{Barman}, \binits{T.}},
\bauthor{\bsnm{Hauschildt}, \binits{P.H.}}:
\batitle{A new extensive library of {{PHOENIX}} stellar atmospheres and
  synthetic spectra}.
\bjtitle{Astron. Astrophys.}
\bvolume{553},
\bfpage{6}
(\byear{2013})
\doiurl{10.1051/0004-6361/201219058}
\end{barticle}
\endbibitem

%%% 46
\bibitem[\protect\citeauthoryear{Sandstrom et~al.}{2023}]{Sandstrom2023PAH3um}
\begin{barticle}
\bauthor{\bsnm{Sandstrom}, \binits{K.M.}},
\bauthor{\bsnm{Chastenet}, \binits{J.}},
\bauthor{\bsnm{Sutter}, \binits{J.}},
\bauthor{\bsnm{Leroy}, \binits{A.K.}},
\bauthor{\bsnm{Egorov}, \binits{O.V.}},
\bauthor{\bsnm{Williams}, \binits{T.G.}},
\bauthor{\bsnm{Bolatto}, \binits{A.D.}},
\bauthor{\bsnm{Boquien}, \binits{M.}},
\bauthor{\bsnm{Cao}, \binits{Y.}},
\bauthor{\bsnm{Dale}, \binits{D.A.}},
\bauthor{\bsnm{Lee}, \binits{J.C.}},
\bauthor{\bsnm{Rosolowsky}, \binits{E.}},
\bauthor{\bsnm{Schinnerer}, \binits{E.}},
\bauthor{\bsnm{Barnes}, \binits{{\relax Ashley}.T.}},
\bauthor{\bsnm{Belfiore}, \binits{F.}},
\bauthor{\bsnm{Bigiel}, \binits{F.}},
\bauthor{\bsnm{Chevance}, \binits{M.}},
\bauthor{\bsnm{Grasha}, \binits{K.}},
\bauthor{\bsnm{Groves}, \binits{B.}},
\bauthor{\bsnm{Hassani}, \binits{H.}},
\bauthor{\bsnm{Hughes}, \binits{A.}},
\bauthor{\bsnm{Klessen}, \binits{R.S.}},
\bauthor{\bsnm{Kruijssen}, \binits{J.M.D.}},
\bauthor{\bsnm{Larson}, \binits{K.L.}},
\bauthor{\bsnm{Liu}, \binits{D.}},
\bauthor{\bsnm{Lopez}, \binits{L.A.}},
\bauthor{\bsnm{Meidt}, \binits{S.E.}},
\bauthor{\bsnm{Murphy}, \binits{E.J.}},
\bauthor{\bsnm{Sormani}, \binits{M.C.}},
\bauthor{\bsnm{Thilker}, \binits{D.A.}},
\bauthor{\bsnm{Watkins}, \binits{E.J.}}:
\batitle{{{PHANGS-JWST First Results}}: {{Mapping}} the 3.3
  {\textbackslash}ensuremath{\textbackslash}mum {{Polycyclic Aromatic
  Hydrocarbon Vibrational Band}} in {{Nearby Galaxies}} with {{NIRCam Medium
  Bands}}}.
\bjtitle{Astrophys. J. Lett.}
\bvolume{944}(\bissue{2}),
\bfpage{7}
(\byear{2023})
\doiurl{10.3847/2041-8213/acb0cf}
\end{barticle}
\endbibitem

%%% 47
\bibitem[\protect\citeauthoryear{Kennicutt et~al.}{2008}]{Kennicutt2008}
\begin{barticle}
\bauthor{\bsnm{Kennicutt}, \binits{R.C.} \bsuffix{Jr.}},
\bauthor{\bsnm{Lee}, \binits{J.C.}},
\bauthor{\bsnm{Funes}, \binits{J.G.}},
\bauthor{\bsnm{J.}, \binits{S.}},
\bauthor{\bsnm{Sakai}, \binits{S.}},
\bauthor{\bsnm{Akiyama}, \binits{S.}}:
\batitle{An {{H$\alpha$ Imaging Survey}} of {{Galaxies}} in the {{Local}} 11
  {{Mpc Volume}}}.
\bjtitle{Astrophys. J. Suppl. Ser.}
\bvolume{178},
\bfpage{247}--\blpage{279}
(\byear{2008})
\doiurl{10.1086/590058}
\end{barticle}
\endbibitem

%%% 48
\bibitem[\protect\citeauthoryear{Lorenzo et~al.}{2022}]{Lorenzo2022}
\begin{barticle}
\bauthor{\bsnm{Lorenzo}, \binits{M.}},
\bauthor{\bsnm{Garcia}, \binits{M.}},
\bauthor{\bsnm{Najarro}, \binits{F.}},
\bauthor{\bsnm{Herrero}, \binits{A.}},
\bauthor{\bsnm{Cervi{\~n}o}, \binits{M.}},
\bauthor{\bsnm{Castro}, \binits{N.}}:
\batitle{A new reference catalogue for the very metal-poor {{Universe}}: +150
  {{OB}} stars in {{Sextans A}}}.
\bjtitle{Mon. Not. R. Astron. Soc.}
\bvolume{516}(\bissue{3}),
\bfpage{4164}--\blpage{4179}
(\byear{2022})
\doiurl{10.1093/mnras/stac2050}
\end{barticle}
\endbibitem

%%% 49
\bibitem[\protect\citeauthoryear{Sandstrom et~al.}{2023}]{Sandstrom2023PAHISM}
\begin{barticle}
\bauthor{\bsnm{Sandstrom}, \binits{K.M.}},
\bauthor{\bsnm{Koch}, \binits{E.W.}},
\bauthor{\bsnm{Leroy}, \binits{A.K.}},
\bauthor{\bsnm{Rosolowsky}, \binits{E.}},
\bauthor{\bsnm{Emsellem}, \binits{E.}},
\bauthor{\bsnm{Smith}, \binits{R.J.}},
\bauthor{\bsnm{Egorov}, \binits{O.V.}},
\bauthor{\bsnm{Williams}, \binits{T.G.}},
\bauthor{\bsnm{Larson}, \binits{K.L.}},
\bauthor{\bsnm{Lee}, \binits{J.C.}},
\bauthor{\bsnm{Schinnerer}, \binits{E.}},
\bauthor{\bsnm{Thilker}, \binits{D.A.}},
\bauthor{\bsnm{Barnes}, \binits{A.T.}},
\bauthor{\bsnm{Belfiore}, \binits{F.}},
\bauthor{\bsnm{Bigiel}, \binits{F.}},
\bauthor{\bsnm{Blanc}, \binits{G.A.}},
\bauthor{\bsnm{Bolatto}, \binits{A.D.}},
\bauthor{\bsnm{Boquien}, \binits{M.}},
\bauthor{\bsnm{Cao}, \binits{Y.}},
\bauthor{\bsnm{Chastenet}, \binits{J.}},
\bauthor{\bsnm{Chevance}, \binits{M.}},
\bauthor{\bsnm{Chiang}, \binits{I.-D.}},
\bauthor{\bsnm{Dale}, \binits{D.A.}},
\bauthor{\bsnm{Faesi}, \binits{C.M.}},
\bauthor{\bsnm{Glover}, \binits{S.C.O.}},
\bauthor{\bsnm{Grasha}, \binits{K.}},
\bauthor{\bsnm{Groves}, \binits{B.}},
\bauthor{\bsnm{Hassani}, \binits{H.}},
\bauthor{\bsnm{Henshaw}, \binits{J.D.}},
\bauthor{\bsnm{Hughes}, \binits{A.}},
\bauthor{\bsnm{Kim}, \binits{J.}},
\bauthor{\bsnm{Klessen}, \binits{R.S.}},
\bauthor{\bsnm{Kreckel}, \binits{K.}},
\bauthor{\bsnm{Kruijssen}, \binits{J.M.D.}},
\bauthor{\bsnm{Lopez}, \binits{L.A.}},
\bauthor{\bsnm{Liu}, \binits{D.}},
\bauthor{\bsnm{Meidt}, \binits{S.E.}},
\bauthor{\bsnm{Murphy}, \binits{E.J.}},
\bauthor{\bsnm{Pan}, \binits{H.-A.}},
\bauthor{\bsnm{Querejeta}, \binits{M.}},
\bauthor{\bsnm{Saito}, \binits{T.}},
\bauthor{\bsnm{Sardone}, \binits{A.}},
\bauthor{\bsnm{Sormani}, \binits{M.C.}},
\bauthor{\bsnm{Sutter}, \binits{J.}},
\bauthor{\bsnm{Usero}, \binits{A.}},
\bauthor{\bsnm{Watkins}, \binits{E.J.}}:
\batitle{{{PHANGS-JWST First Results}}: {{Tracing}} the {{Diffuse Interstellar
  Medium}} with {{JWST Imaging}} of {{Polycyclic Aromatic Hydrocarbon
  Emission}} in {{Nearby Galaxies}}}.
\bjtitle{Astrophys. J. Lett.}
\bvolume{944}(\bissue{2}),
\bfpage{8}
(\byear{2023})
\doiurl{10.3847/2041-8213/aca972}
\end{barticle}
\endbibitem

%%% 50
\bibitem[\protect\citeauthoryear{Sutter et~al.}{2024}]{Sutter2024}
\begin{barticle}
\bauthor{\bsnm{Sutter}, \binits{J.}},
\bauthor{\bsnm{Sandstrom}, \binits{K.}},
\bauthor{\bsnm{Chastenet}, \binits{J.}},
\bauthor{\bsnm{Leroy}, \binits{A.K.}},
\bauthor{\bsnm{Koch}, \binits{E.W.}},
\bauthor{\bsnm{Williams}, \binits{T.G.}},
\bauthor{\bsnm{Chown}, \binits{R.}},
\bauthor{\bsnm{Belfiore}, \binits{F.}},
\bauthor{\bsnm{Bigiel}, \binits{F.}},
\bauthor{\bsnm{Boquien}, \binits{M.}},
\bauthor{\bsnm{Cao}, \binits{Y.}},
\bauthor{\bsnm{Chevance}, \binits{M.}},
\bauthor{\bsnm{Dale}, \binits{D.A.}},
\bauthor{\bsnm{Egorov}, \binits{O.V.}},
\bauthor{\bsnm{Glover}, \binits{S.C.O.}},
\bauthor{\bsnm{Groves}, \binits{B.}},
\bauthor{\bsnm{Klessen}, \binits{R.S.}},
\bauthor{\bsnm{Kreckel}, \binits{K.}},
\bauthor{\bsnm{Larson}, \binits{K.L.}},
\bauthor{\bsnm{Oakes}, \binits{E.K.}},
\bauthor{\bsnm{Pathak}, \binits{D.}},
\bauthor{\bsnm{Ramambason}, \binits{L.}},
\bauthor{\bsnm{Rosolowsky}, \binits{E.}},
\bauthor{\bsnm{Watkins}, \binits{E.J.}}:
\batitle{The {{Fraction}} of {{Dust Mass}} in the {{Form}} of {{Polycyclic
  Aromatic Hydrocarbons}} on 10--50 pc {{Scales}} in {{Nearby Galaxies}}}.
\bjtitle{Astrophys. J.}
\bvolume{971}(\bissue{2}),
\bfpage{178}
(\byear{2024})
\doiurl{10.3847/1538-4357/ad54bd}
\end{barticle}
\endbibitem

%%% 51
\bibitem[\protect\citeauthoryear{Clark et~al.}{2025}]{Clark2025}
\begin{barticle}
\bauthor{\bsnm{Clark}, \binits{I.Y.}},
\bauthor{\bsnm{Sandstrom}, \binits{K.}},
\bauthor{\bsnm{Wolfire}, \binits{M.}},
\bauthor{\bsnm{Bolatto}, \binits{A.D.}},
\bauthor{\bsnm{Chastenet}, \binits{J.}},
\bauthor{\bsnm{Dale}, \binits{D.A.}},
\bauthor{\bsnm{Gaches}, \binits{B.A.L.}},
\bauthor{\bsnm{Glover}, \binits{S.C.O.}},
\bauthor{\bsnm{Goicoechea}, \binits{J.R.}},
\bauthor{\bsnm{Gordon}, \binits{K.D.}},
\bauthor{\bsnm{Groves}, \binits{B.}},
\bauthor{\bsnm{Hands}, \binits{L.}},
\bauthor{\bsnm{Klessen}, \binits{R.}},
\bauthor{\bsnm{De~Looze}, \binits{I.}},
\bauthor{\bsnm{Smith}, \binits{J.D.T.}},
\bauthor{\bsnm{Van De~Putte}, \binits{D.}},
\bauthor{\bsnm{Walch}, \binits{S.K.}}:
\batitle{The {{Resolved Structure}} of a {{Low-metallicity Photodissociation
  Region}}}.
\bjtitle{Astrophys. J.}
\bvolume{990},
\bfpage{209}
(\byear{2025})
\doiurl{10.3847/1538-4357/adef38}
\end{barticle}
\endbibitem

%%% 52
\bibitem[\protect\citeauthoryear{Rubio et~al.}{2015}]{Rubio2015}
\begin{barticle}
\bauthor{\bsnm{Rubio}, \binits{M.}},
\bauthor{\bsnm{Elmegreen}, \binits{B.G.}},
\bauthor{\bsnm{Hunter}, \binits{D.A.}},
\bauthor{\bsnm{Brinks}, \binits{E.}},
\bauthor{\bsnm{Cort{\'e}s}, \binits{J.R.}},
\bauthor{\bsnm{Cigan}, \binits{P.}}:
\batitle{Dense cloud cores revealed by {{CO}} in the low metallicity dwarf
  galaxy {{WLM}}}.
\bjtitle{Nature}
\bvolume{525}(\bissue{7568}),
\bfpage{218}--\blpage{221}
(\byear{2015})
\doiurl{10.1038/nature14901}
\end{barticle}
\endbibitem

%%% 53
\bibitem[\protect\citeauthoryear{Shi et~al.}{2020}]{Shi2020}
\begin{barticle}
\bauthor{\bsnm{Shi}, \binits{Y.}},
\bauthor{\bsnm{Wang}, \binits{J.}},
\bauthor{\bsnm{Zhang}, \binits{Z.-Y.}},
\bauthor{\bsnm{Zhang}, \binits{Q.}},
\bauthor{\bsnm{Gao}, \binits{Y.}},
\bauthor{\bsnm{Zhou}, \binits{L.}},
\bauthor{\bsnm{Gu}, \binits{Q.}},
\bauthor{\bsnm{Qiu}, \binits{K.}},
\bauthor{\bsnm{Xia}, \binits{X.-Y.}},
\bauthor{\bsnm{Hao}, \binits{C.-N.}},
\bauthor{\bsnm{Chen}, \binits{Y.}}:
\batitle{Oversized {{Gas Clumps}} in an {{Extremely Metal-poor Molecular Cloud
  Revealed}} by {{ALMA}}'s {{Parsec-scale Maps}}}.
\bjtitle{Astrophys. J.}
\bvolume{892}(\bissue{2}),
\bfpage{147}
(\byear{2020})
\doiurl{10.3847/1538-4357/ab7a12}
\end{barticle}
\endbibitem

%%% 54
\bibitem[\protect\citeauthoryear{Bolatto et~al.}{1999}]{Bolatto1999}
\begin{barticle}
\bauthor{\bsnm{Bolatto}, \binits{A.D.}},
\bauthor{\bsnm{Jackson}, \binits{J.M.}},
\bauthor{\bsnm{Ingalls}, \binits{J.G.}}:
\batitle{A {{Semianalytical Model}} for the {{Observational Properties}} of the
  {{Dominant Carbon Species}} at {{Different Metallicities}}}.
\bjtitle{Astrophys. J.}
\bvolume{513},
\bfpage{275}--\blpage{286}
(\byear{1999})
\doiurl{10.1086/306849}
\end{barticle}
\endbibitem

%%% 55
\bibitem[\protect\citeauthoryear{Wolfire et~al.}{2010}]{Wolfire2010}
\begin{barticle}
\bauthor{\bsnm{Wolfire}, \binits{M.G.}},
\bauthor{\bsnm{Hollenbach}, \binits{D.}},
\bauthor{\bsnm{McKee}, \binits{C.F.}}:
\batitle{The {{Dark Molecular Gas}}}.
\bjtitle{Astrophys. J.}
\bvolume{716}(\bissue{2}),
\bfpage{1191}--\blpage{1207}
(\byear{2010})
\doiurl{10.1088/0004-637X/716/2/1191}
\end{barticle}
\endbibitem

%%% 56
\bibitem[\protect\citeauthoryear{Chastenet et~al.}{2023}]{Chastenet2023_RPAH}
\begin{barticle}
\bauthor{\bsnm{Chastenet}, \binits{J.}},
\bauthor{\bsnm{Sutter}, \binits{J.}},
\bauthor{\bsnm{Sandstrom}, \binits{K.}},
\bauthor{\bsnm{Belfiore}, \binits{F.}},
\bauthor{\bsnm{Egorov}, \binits{O.V.}},
\bauthor{\bsnm{Larson}, \binits{K.L.}},
\bauthor{\bsnm{Leroy}, \binits{A.K.}},
\bauthor{\bsnm{Liu}, \binits{D.}},
\bauthor{\bsnm{Rosolowsky}, \binits{E.}},
\bauthor{\bsnm{Thilker}, \binits{D.A.}},
\bauthor{\bsnm{Watkins}, \binits{E.J.}},
\bauthor{\bsnm{Williams}, \binits{T.G.}},
\bauthor{\bsnm{Barnes}, \binits{{\relax Ashley}.T.}},
\bauthor{\bsnm{Bigiel}, \binits{F.}},
\bauthor{\bsnm{Boquien}, \binits{M.}},
\bauthor{\bsnm{Chevance}, \binits{M.}},
\bauthor{\bsnm{Chiang}, \binits{I.-D.}},
\bauthor{\bsnm{Dale}, \binits{D.A.}},
\bauthor{\bsnm{Kruijssen}, \binits{J.M.D.}},
\bauthor{\bsnm{Emsellem}, \binits{E.}},
\bauthor{\bsnm{Grasha}, \binits{K.}},
\bauthor{\bsnm{Groves}, \binits{B.}},
\bauthor{\bsnm{Hassani}, \binits{H.}},
\bauthor{\bsnm{Hughes}, \binits{A.}},
\bauthor{\bsnm{Kreckel}, \binits{K.}},
\bauthor{\bsnm{Meidt}, \binits{S.E.}},
\bauthor{\bsnm{Rickards~Vaught}, \binits{R.J.}},
\bauthor{\bsnm{Sardone}, \binits{A.}},
\bauthor{\bsnm{Schinnerer}, \binits{E.}}:
\batitle{{{PHANGS-JWST First Results}}: {{Variations}} in {{PAH Fraction}} as a
  {{Function}} of {{ISM Phase}} and {{Metallicity}}}.
\bjtitle{Astrophys. J. Lett.}
\bvolume{944}(\bissue{2}),
\bfpage{11}
(\byear{2023})
\doiurl{10.3847/2041-8213/acadd7}
\end{barticle}
\endbibitem

%%% 57
\bibitem[\protect\citeauthoryear{Allamandola et~al.}{1999}]{Allamandola1999}
\begin{barticle}
\bauthor{\bsnm{Allamandola}, \binits{L.J.}},
\bauthor{\bsnm{Hudgins}, \binits{D.M.}},
\bauthor{\bsnm{Sandford}, \binits{S.A.}}:
\batitle{Modeling the {{Unidentified Infrared Emission}} with {{Combinations}}
  of {{Polycyclic Aromatic Hydrocarbons}}}.
\bjtitle{Astrophys. J.}
\bvolume{511},
\bfpage{115}--\blpage{119}
(\byear{1999})
\doiurl{10.1086/311843}
\end{barticle}
\endbibitem

%%% 58
\bibitem[\protect\citeauthoryear{Maragkoudakis
  et~al.}{2020}]{Maragkoudakis2020}
\begin{barticle}
\bauthor{\bsnm{Maragkoudakis}, \binits{A.}},
\bauthor{\bsnm{Peeters}, \binits{E.}},
\bauthor{\bsnm{Ricca}, \binits{A.}}:
\batitle{Probing the size and charge of polycyclic aromatic hydrocarbons}.
\bjtitle{Mon. Not. R. Astron. Soc.}
\bvolume{494},
\bfpage{642}--\blpage{664}
(\byear{2020})
\doiurl{10.1093/mnras/staa681}
\end{barticle}
\endbibitem

%%% 59
\bibitem[\protect\citeauthoryear{Schutte et~al.}{1993}]{Schutte1993}
\begin{barticle}
\bauthor{\bsnm{Schutte}, \binits{W.A.}},
\bauthor{\bsnm{Tielens}, \binits{A.G.G.M.}},
\bauthor{\bsnm{Allamandola}, \binits{L.J.}}:
\batitle{Theoretical {{Modeling}} of the {{Infrared Fluorescence}} from
  {{Interstellar Polycyclic Aromatic Hydrocarbons}}}.
\bjtitle{Astrophys. J.}
\bvolume{415},
\bfpage{397}
(\byear{1993})
\doiurl{10.1086/173173}
\end{barticle}
\endbibitem

%%% 60
\bibitem[\protect\citeauthoryear{Dale et~al.}{2009}]{Dale2009}
\begin{barticle}
\bauthor{\bsnm{Dale}, \binits{D.A.}},
\bauthor{\bsnm{Cohen}, \binits{S.A.}},
\bauthor{\bsnm{Johnson}, \binits{L.C.}},
\bauthor{\bsnm{Schuster}, \binits{M.D.}},
\bauthor{\bsnm{Calzetti}, \binits{D.}},
\bauthor{\bsnm{Engelbracht}, \binits{C.W.}},
\bauthor{\bsnm{{Gil de Paz}}, \binits{A.}},
\bauthor{\bsnm{Kennicutt}, \binits{R.C.}},
\bauthor{\bsnm{Lee}, \binits{J.C.}},
\bauthor{\bsnm{Begum}, \binits{A.}},
\bauthor{\bsnm{Block}, \binits{M.}},
\bauthor{\bsnm{Dalcanton}, \binits{J.J.}},
\bauthor{\bsnm{Funes}, \binits{J.G.}},
\bauthor{\bsnm{Gordon}, \binits{K.D.}},
\bauthor{\bsnm{Johnson}, \binits{B.D.}},
\bauthor{\bsnm{Marble}, \binits{A.R.}},
\bauthor{\bsnm{Sakai}, \binits{S.}},
\bauthor{\bsnm{Skillman}, \binits{E.D.}},
\bauthor{\bsnm{{van Zee}}, \binits{L.}},
\bauthor{\bsnm{Walter}, \binits{F.}},
\bauthor{\bsnm{Weisz}, \binits{D.R.}},
\bauthor{\bsnm{Williams}, \binits{B.}},
\bauthor{\bsnm{Wu}, \binits{S.-Y.}},
\bauthor{\bsnm{Wu}, \binits{Y.}}:
\batitle{The {{Spitzer Local Volume Legacy}}: {{Survey Description}} and
  {{Infrared Photometry}}}.
\bjtitle{Astrophys. J.}
\bvolume{703},
\bfpage{517}--\blpage{556}
(\byear{2009})
\doiurl{10.1088/0004-637X/703/1/517}
\end{barticle}
\endbibitem

%%% 61
\bibitem[\protect\citeauthoryear{Bruzual and Charlot}{2003}]{Bruzual2003}
\begin{barticle}
\bauthor{\bsnm{Bruzual}, \binits{G.}},
\bauthor{\bsnm{Charlot}, \binits{S.}}:
\batitle{Stellar population synthesis at the resolution of 2003}.
\bjtitle{Mon. Not. R. Astron. Soc.}
\bvolume{344},
\bfpage{1000}--\blpage{1028}
(\byear{2003})
\doiurl{10.1046/j.1365-8711.2003.06897.x}
\end{barticle}
\endbibitem

%%% 62
\bibitem[\protect\citeauthoryear{Draine and Li}{2007}]{DraineLi2007}
\begin{barticle}
\bauthor{\bsnm{Draine}, \binits{B.T.}},
\bauthor{\bsnm{Li}, \binits{A.}}:
\batitle{Infrared {{Emission}} from {{Interstellar Dust}}. {{IV}}. {{The
  Silicate-Graphite-PAH Model}} in the {{Post-Spitzer Era}}}.
\bjtitle{Astrophys. J.}
\bvolume{657},
\bfpage{810}--\blpage{837}
(\byear{2007})
\doiurl{10.1086/511055}
\end{barticle}
\endbibitem

%%% 63
\bibitem[\protect\citeauthoryear{Chown et~al.}{2024}]{Chown2024}
\begin{barticle}
\bauthor{\bsnm{Chown}, \binits{R.}},
\bauthor{\bsnm{Sidhu}, \binits{A.}},
\bauthor{\bsnm{Peeters}, \binits{E.}},
\bauthor{\bsnm{Tielens}, \binits{A.G.G.M.}},
\bauthor{\bsnm{Cami}, \binits{J.}},
\bauthor{\bsnm{Bern{\'e}}, \binits{O.}},
\bauthor{\bsnm{Habart}, \binits{E.}},
\bauthor{\bsnm{Alarc{\'o}n}, \binits{F.}},
\bauthor{\bsnm{Canin}, \binits{A.}},
\bauthor{\bsnm{Schroetter}, \binits{I.}},
\bauthor{\bsnm{Trahin}, \binits{B.}},
\bauthor{\bsnm{Van De~Putte}, \binits{D.}},
\bauthor{\bsnm{Abergel}, \binits{A.}},
\bauthor{\bsnm{Bergin}, \binits{E.A.}},
\bauthor{\bsnm{{Bernard-Salas}}, \binits{J.}},
\bauthor{\bsnm{Boersma}, \binits{C.}},
\bauthor{\bsnm{Bron}, \binits{E.}},
\bauthor{\bsnm{Cuadrado}, \binits{S.}},
\bauthor{\bsnm{Dartois}, \binits{E.}},
\bauthor{\bsnm{Dicken}, \binits{D.}},
\bauthor{\bsnm{{El-Yajouri}}, \binits{M.}},
\bauthor{\bsnm{Fuente}, \binits{A.}},
\bauthor{\bsnm{Goicoechea}, \binits{J.R.}},
\bauthor{\bsnm{Gordon}, \binits{K.D.}},
\bauthor{\bsnm{Issa}, \binits{L.}},
\bauthor{\bsnm{Joblin}, \binits{C.}},
\bauthor{\bsnm{Kannavou}, \binits{O.}},
\bauthor{\bsnm{Khan}, \binits{B.}},
\bauthor{\bsnm{Lacinbala}, \binits{O.}},
\bauthor{\bsnm{Languignon}, \binits{D.}},
\bauthor{\bsnm{Le~Gal}, \binits{R.}},
\bauthor{\bsnm{Maragkoudakis}, \binits{A.}},
\bauthor{\bsnm{Meshaka}, \binits{R.}},
\bauthor{\bsnm{Okada}, \binits{Y.}},
\bauthor{\bsnm{Onaka}, \binits{T.}},
\bauthor{\bsnm{Pasquini}, \binits{S.}},
\bauthor{\bsnm{Pound}, \binits{M.W.}},
\bauthor{\bsnm{Robberto}, \binits{M.}},
\bauthor{\bsnm{R{\"o}llig}, \binits{M.}},
\bauthor{\bsnm{Schefter}, \binits{B.}},
\bauthor{\bsnm{Schirmer}, \binits{T.}},
\bauthor{\bsnm{Vicente}, \binits{S.}},
\bauthor{\bsnm{Wolfire}, \binits{M.G.}},
\bauthor{\bsnm{Zannese}, \binits{M.}},
\bauthor{\bsnm{Aleman}, \binits{I.}},
\bauthor{\bsnm{Allamandola}, \binits{L.}},
\bauthor{\bsnm{Auchettl}, \binits{R.}},
\bauthor{\bsnm{Baratta}, \binits{G.A.}},
\bauthor{\bsnm{Bejaoui}, \binits{S.}},
\bauthor{\bsnm{Bera}, \binits{P.P.}},
\bauthor{\bsnm{Black}, \binits{J.H.}},
\bauthor{\bsnm{Boulanger}, \binits{F.}},
\bauthor{\bsnm{Bouwman}, \binits{J.}},
\bauthor{\bsnm{Brandl}, \binits{B.}},
\bauthor{\bsnm{Brechignac}, \binits{P.}},
\bauthor{\bsnm{Br{\"u}nken}, \binits{S.}},
\bauthor{\bsnm{Buragohain}, \binits{M.}},
\bauthor{\bsnm{Burkhardt}, \binits{A.}},
\bauthor{\bsnm{Candian}, \binits{A.}},
\bauthor{\bsnm{Cazaux}, \binits{S.}},
\bauthor{\bsnm{Cernicharo}, \binits{J.}},
\bauthor{\bsnm{Chabot}, \binits{M.}},
\bauthor{\bsnm{Chakraborty}, \binits{S.}},
\bauthor{\bsnm{Champion}, \binits{J.}},
\bauthor{\bsnm{Colgan}, \binits{S.W.J.}},
\bauthor{\bsnm{Cooke}, \binits{I.R.}},
\bauthor{\bsnm{Coutens}, \binits{A.}},
\bauthor{\bsnm{Cox}, \binits{N.L.J.}},
\bauthor{\bsnm{Demyk}, \binits{K.}},
\bauthor{\bsnm{Meyer}, \binits{J.D.}},
\bauthor{\bsnm{Foschino}, \binits{S.}},
\bauthor{\bsnm{{Garc{\'i}a-Lario}}, \binits{P.}},
\bauthor{\bsnm{Gavilan}, \binits{L.}},
\bauthor{\bsnm{Gerin}, \binits{M.}},
\bauthor{\bsnm{Gottlieb}, \binits{C.A.}},
\bauthor{\bsnm{Guillard}, \binits{P.}},
\bauthor{\bsnm{Gusdorf}, \binits{A.}},
\bauthor{\bsnm{Hartigan}, \binits{P.}},
\bauthor{\bsnm{He}, \binits{J.}},
\bauthor{\bsnm{Herbst}, \binits{E.}},
\bauthor{\bsnm{Hornekaer}, \binits{L.}},
\bauthor{\bsnm{J{\"a}ger}, \binits{C.}},
\bauthor{\bsnm{{Janot-Pacheco}}, \binits{E.}},
\bauthor{\bsnm{Kaufman}, \binits{M.}},
\bauthor{\bsnm{Kemper}, \binits{F.}},
\bauthor{\bsnm{Kendrew}, \binits{S.}},
\bauthor{\bsnm{Kirsanova}, \binits{M.S.}},
\bauthor{\bsnm{Klaassen}, \binits{P.}},
\bauthor{\bsnm{Kwok}, \binits{S.}},
\bauthor{\bsnm{Labiano}, \binits{{\'A}.}},
\bauthor{\bsnm{Lai}, \binits{T.S.-Y.}},
\bauthor{\bsnm{Lee}, \binits{T.J.}},
\bauthor{\bsnm{Lefloch}, \binits{B.}},
\bauthor{\bsnm{Le~Petit}, \binits{F.}},
\bauthor{\bsnm{Li}, \binits{A.}},
\bauthor{\bsnm{Linz}, \binits{H.}},
\bauthor{\bsnm{Mackie}, \binits{C.J.}},
\bauthor{\bsnm{Madden}, \binits{S.C.}},
\bauthor{\bsnm{Mascetti}, \binits{J.}},
\bauthor{\bsnm{McGuire}, \binits{B.A.}},
\bauthor{\bsnm{Merino}, \binits{P.}},
\bauthor{\bsnm{Micelotta}, \binits{E.R.}},
\bauthor{\bsnm{Misselt}, \binits{K.}},
\bauthor{\bsnm{Morse}, \binits{J.A.}},
\bauthor{\bsnm{Mulas}, \binits{G.}},
\bauthor{\bsnm{Neelamkodan}, \binits{N.}},
\bauthor{\bsnm{Ohsawa}, \binits{R.}},
\bauthor{\bsnm{Omont}, \binits{A.}},
\bauthor{\bsnm{Paladini}, \binits{R.}},
\bauthor{\bsnm{Palumbo}, \binits{M.E.}},
\bauthor{\bsnm{Pathak}, \binits{A.}},
\bauthor{\bsnm{Pendleton}, \binits{Y.J.}},
\bauthor{\bsnm{Petrignani}, \binits{A.}},
\bauthor{\bsnm{Pino}, \binits{T.}},
\bauthor{\bsnm{Puga}, \binits{E.}},
\bauthor{\bsnm{Rangwala}, \binits{N.}},
\bauthor{\bsnm{Rapacioli}, \binits{M.}},
\bauthor{\bsnm{Ricca}, \binits{A.}},
\bauthor{\bsnm{{Roman-Duval}}, \binits{J.}},
\bauthor{\bsnm{Roser}, \binits{J.}},
\bauthor{\bsnm{Roueff}, \binits{E.}},
\bauthor{\bsnm{Rouill{\'e}}, \binits{G.}},
\bauthor{\bsnm{Salama}, \binits{F.}},
\bauthor{\bsnm{Sales}, \binits{D.A.}},
\bauthor{\bsnm{Sandstrom}, \binits{K.}},
\bauthor{\bsnm{Sarre}, \binits{P.}},
\bauthor{\bsnm{{Sciamma-O'Brien}}, \binits{E.}},
\bauthor{\bsnm{Sellgren}, \binits{K.}},
\bauthor{\bsnm{Shenoy}, \binits{S.S.}},
\bauthor{\bsnm{Teyssier}, \binits{D.}},
\bauthor{\bsnm{Thomas}, \binits{R.D.}},
\bauthor{\bsnm{Togi}, \binits{A.}},
\bauthor{\bsnm{Verstraete}, \binits{L.}},
\bauthor{\bsnm{Witt}, \binits{A.N.}},
\bauthor{\bsnm{Wootten}, \binits{A.}},
\bauthor{\bsnm{Zettergren}, \binits{H.}},
\bauthor{\bsnm{Zhang}, \binits{Y.}},
\bauthor{\bsnm{Zhang}, \binits{Z.E.}},
\bauthor{\bsnm{Zhen}, \binits{J.}}:
\batitle{{{PDRs4All}}. {{IV}}. {{An}} embarrassment of riches: {{Aromatic}}
  infrared bands in the {{Orion Bar}}}.
\bjtitle{Astron. Astrophys.}
\bvolume{685},
\bfpage{75}
(\byear{2024})
\doiurl{10.1051/0004-6361/202346662}
\end{barticle}
\endbibitem

%%% 64
\bibitem[\protect\citeauthoryear{Peeters et~al.}{2024}]{Peeters2024}
\begin{barticle}
\bauthor{\bsnm{Peeters}, \binits{E.}},
\bauthor{\bsnm{Habart}, \binits{E.}},
\bauthor{\bsnm{Bern{\'e}}, \binits{O.}},
\bauthor{\bsnm{Sidhu}, \binits{A.}},
\bauthor{\bsnm{Chown}, \binits{R.}},
\bauthor{\bsnm{Van De~Putte}, \binits{D.}},
\bauthor{\bsnm{Trahin}, \binits{B.}},
\bauthor{\bsnm{Schroetter}, \binits{I.}},
\bauthor{\bsnm{Canin}, \binits{A.}},
\bauthor{\bsnm{Alarc{\'o}n}, \binits{F.}},
\bauthor{\bsnm{Schefter}, \binits{B.}},
\bauthor{\bsnm{Khan}, \binits{B.}},
\bauthor{\bsnm{Pasquini}, \binits{S.}},
\bauthor{\bsnm{Tielens}, \binits{A.G.G.M.}},
\bauthor{\bsnm{Wolfire}, \binits{M.G.}},
\bauthor{\bsnm{Dartois}, \binits{E.}},
\bauthor{\bsnm{Goicoechea}, \binits{J.R.}},
\bauthor{\bsnm{Maragkoudakis}, \binits{A.}},
\bauthor{\bsnm{Onaka}, \binits{T.}},
\bauthor{\bsnm{Pound}, \binits{M.W.}},
\bauthor{\bsnm{Vicente}, \binits{S.}},
\bauthor{\bsnm{Abergel}, \binits{A.}},
\bauthor{\bsnm{Bergin}, \binits{E.A.}},
\bauthor{\bsnm{{Bernard-Salas}}, \binits{J.}},
\bauthor{\bsnm{Boersma}, \binits{C.}},
\bauthor{\bsnm{Bron}, \binits{E.}},
\bauthor{\bsnm{Cami}, \binits{J.}},
\bauthor{\bsnm{Cuadrado}, \binits{S.}},
\bauthor{\bsnm{Dicken}, \binits{D.}},
\bauthor{\bsnm{Elyajouri}, \binits{M.}},
\bauthor{\bsnm{Fuente}, \binits{A.}},
\bauthor{\bsnm{Gordon}, \binits{K.D.}},
\bauthor{\bsnm{Issa}, \binits{L.}},
\bauthor{\bsnm{Joblin}, \binits{C.}},
\bauthor{\bsnm{Kannavou}, \binits{O.}},
\bauthor{\bsnm{Lacinbala}, \binits{O.}},
\bauthor{\bsnm{Languignon}, \binits{D.}},
\bauthor{\bsnm{Le~Gal}, \binits{R.}},
\bauthor{\bsnm{Meshaka}, \binits{R.}},
\bauthor{\bsnm{Okada}, \binits{Y.}},
\bauthor{\bsnm{Robberto}, \binits{M.}},
\bauthor{\bsnm{R{\"o}llig}, \binits{M.}},
\bauthor{\bsnm{Schirmer}, \binits{T.}},
\bauthor{\bsnm{Tabone}, \binits{B.}},
\bauthor{\bsnm{Zannese}, \binits{M.}},
\bauthor{\bsnm{Aleman}, \binits{I.}},
\bauthor{\bsnm{Allamandola}, \binits{L.}},
\bauthor{\bsnm{Auchettl}, \binits{R.}},
\bauthor{\bsnm{Baratta}, \binits{G.A.}},
\bauthor{\bsnm{Bejaoui}, \binits{S.}},
\bauthor{\bsnm{Bera}, \binits{P.P.}},
\bauthor{\bsnm{Black}, \binits{J.H.}},
\bauthor{\bsnm{Boulanger}, \binits{F.}},
\bauthor{\bsnm{Bouwman}, \binits{J.}},
\bauthor{\bsnm{Brandl}, \binits{B.}},
\bauthor{\bsnm{Brechignac}, \binits{P.}},
\bauthor{\bsnm{Br{\"u}nken}, \binits{S.}},
\bauthor{\bsnm{Buragohain}, \binits{M.}},
\bauthor{\bsnm{Burkhardt}, \binits{A.}},
\bauthor{\bsnm{Candian}, \binits{A.}},
\bauthor{\bsnm{Cazaux}, \binits{S.}},
\bauthor{\bsnm{Cernicharo}, \binits{J.}},
\bauthor{\bsnm{Chabot}, \binits{M.}},
\bauthor{\bsnm{Chakraborty}, \binits{S.}},
\bauthor{\bsnm{Champion}, \binits{J.}},
\bauthor{\bsnm{Colgan}, \binits{S.W.J.}},
\bauthor{\bsnm{Cooke}, \binits{I.R.}},
\bauthor{\bsnm{Coutens}, \binits{A.}},
\bauthor{\bsnm{Cox}, \binits{N.L.J.}},
\bauthor{\bsnm{Demyk}, \binits{K.}},
\bauthor{\bsnm{Meyer}, \binits{J.D.}},
\bauthor{\bsnm{Foschino}, \binits{S.}},
\bauthor{\bsnm{{Garc{\'i}a-Lario}}, \binits{P.}},
\bauthor{\bsnm{Gerin}, \binits{M.}},
\bauthor{\bsnm{Gottlieb}, \binits{C.A.}},
\bauthor{\bsnm{Guillard}, \binits{P.}},
\bauthor{\bsnm{Gusdorf}, \binits{A.}},
\bauthor{\bsnm{Hartigan}, \binits{P.}},
\bauthor{\bsnm{He}, \binits{J.}},
\bauthor{\bsnm{Herbst}, \binits{E.}},
\bauthor{\bsnm{Hornekaer}, \binits{L.}},
\bauthor{\bsnm{J{\"a}ger}, \binits{C.}},
\bauthor{\bsnm{{Janot-Pacheco}}, \binits{E.}},
\bauthor{\bsnm{Kaufman}, \binits{M.}},
\bauthor{\bsnm{Kendrew}, \binits{S.}},
\bauthor{\bsnm{Kirsanova}, \binits{M.S.}},
\bauthor{\bsnm{Klaassen}, \binits{P.}},
\bauthor{\bsnm{Kwok}, \binits{S.}},
\bauthor{\bsnm{Labiano}, \binits{{\'A}.}},
\bauthor{\bsnm{Lai}, \binits{T.S.-Y.}},
\bauthor{\bsnm{Lee}, \binits{T.J.}},
\bauthor{\bsnm{Lefloch}, \binits{B.}},
\bauthor{\bsnm{Le~Petit}, \binits{F.}},
\bauthor{\bsnm{Li}, \binits{A.}},
\bauthor{\bsnm{Linz}, \binits{H.}},
\bauthor{\bsnm{Mackie}, \binits{C.J.}},
\bauthor{\bsnm{Madden}, \binits{S.C.}},
\bauthor{\bsnm{Mascetti}, \binits{J.}},
\bauthor{\bsnm{McGuire}, \binits{B.A.}},
\bauthor{\bsnm{Merino}, \binits{P.}},
\bauthor{\bsnm{Micelotta}, \binits{E.R.}},
\bauthor{\bsnm{Misselt}, \binits{K.}},
\bauthor{\bsnm{Morse}, \binits{J.A.}},
\bauthor{\bsnm{Mulas}, \binits{G.}},
\bauthor{\bsnm{Neelamkodan}, \binits{N.}},
\bauthor{\bsnm{Ohsawa}, \binits{R.}},
\bauthor{\bsnm{Paladini}, \binits{R.}},
\bauthor{\bsnm{Palumbo}, \binits{M.E.}},
\bauthor{\bsnm{Pathak}, \binits{A.}},
\bauthor{\bsnm{Pendleton}, \binits{Y.J.}},
\bauthor{\bsnm{Petrignani}, \binits{A.}},
\bauthor{\bsnm{Pino}, \binits{T.}},
\bauthor{\bsnm{Puga}, \binits{E.}},
\bauthor{\bsnm{Rangwala}, \binits{N.}},
\bauthor{\bsnm{Rapacioli}, \binits{M.}},
\bauthor{\bsnm{Ricca}, \binits{A.}},
\bauthor{\bsnm{{Roman-Duval}}, \binits{J.}},
\bauthor{\bsnm{Roser}, \binits{J.}},
\bauthor{\bsnm{Roueff}, \binits{E.}},
\bauthor{\bsnm{Rouill{\'e}}, \binits{G.}},
\bauthor{\bsnm{Salama}, \binits{F.}},
\bauthor{\bsnm{Sales}, \binits{D.A.}},
\bauthor{\bsnm{Sandstrom}, \binits{K.}},
\bauthor{\bsnm{Sarre}, \binits{P.}},
\bauthor{\bsnm{{Sciamma-O'Brien}}, \binits{E.}},
\bauthor{\bsnm{Sellgren}, \binits{K.}},
\bauthor{\bsnm{Shenoy}, \binits{S.S.}},
\bauthor{\bsnm{Teyssier}, \binits{D.}},
\bauthor{\bsnm{Thomas}, \binits{R.D.}},
\bauthor{\bsnm{Togi}, \binits{A.}},
\bauthor{\bsnm{Verstraete}, \binits{L.}},
\bauthor{\bsnm{Witt}, \binits{A.N.}},
\bauthor{\bsnm{Wootten}, \binits{A.}},
\bauthor{\bsnm{Ysard}, \binits{N.}},
\bauthor{\bsnm{Zettergren}, \binits{H.}},
\bauthor{\bsnm{Zhang}, \binits{Y.}},
\bauthor{\bsnm{Zhang}, \binits{Z.E.}},
\bauthor{\bsnm{Zhen}, \binits{J.}}:
\batitle{{{PDRs4All}}: {{III}}. {{JWST}}'s {{NIR}} spectroscopic view of the
  {{Orion Bar}}}.
\bjtitle{Astron. Astrophys.}
\bvolume{685},
\bfpage{74}
(\byear{2024})
\doiurl{10.1051/0004-6361/202348244}
\end{barticle}
\endbibitem

%%% 65
\bibitem[\protect\citeauthoryear{Jones et~al.}{2017}]{Jones2017}
\begin{barticle}
\bauthor{\bsnm{Jones}, \binits{A.P.}},
\bauthor{\bsnm{K{\"o}hler}, \binits{M.}},
\bauthor{\bsnm{Ysard}, \binits{N.}},
\bauthor{\bsnm{Bocchio}, \binits{M.}},
\bauthor{\bsnm{Verstraete}, \binits{L.}}:
\batitle{The global dust modelling framework {{THEMIS}}}.
\bjtitle{Astron. Astrophys.}
\bvolume{602},
\bfpage{46}
(\byear{2017})
\doiurl{10.1051/0004-6361/201630225}
\end{barticle}
\endbibitem

%%% 66
\bibitem[\protect\citeauthoryear{Whitcomb et~al.}{2026}]{Whitcomb2026}
\begin{barticle}
\bauthor{\bsnm{Whitcomb}, \binits{C.M.}},
\bauthor{\bsnm{Smith}, \binits{J.-D.T.}},
\bauthor{\bsnm{Tarantino}, \binits{E.}},
\bauthor{\bsnm{Sandstrom}, \binits{K.}},
\bauthor{\bsnm{Lai}, \binits{T.S.-Y.}},
\bauthor{\bsnm{Armus}, \binits{L.}},
\bauthor{\bsnm{Bolatto}, \binits{A.}},
\bauthor{\bsnm{Boyer}, \binits{M.}},
\bauthor{\bsnm{Dale}, \binits{D.A.}},
\bauthor{\bsnm{Draine}, \binits{B.T.}},
\bauthor{\bsnm{Hensley}, \binits{B.S.}},
\bauthor{\bsnm{Narayanan}, \binits{D.}},
\bauthor{\bsnm{{Roman-Duval}}, \binits{J.}},
\bauthor{\bsnm{Skillman}, \binits{E.D.}}:
\batitle{The {{Metallicity Dependence}} of {{PAH Emission}} in {{Galaxies}}.
  {{II}}. {{Insights}} from {{JWST}}/{{NIRCam Imaging}} of the {{Smallest Dust
  Grains}} in {{M101}}}.
\bjtitle{Astrophys. J.}
\bvolume{1000},
\bfpage{211}
(\byear{2026})
\doiurl{10.3847/1538-4357/ae4902}
\end{barticle}
\endbibitem

%%% 67
\bibitem[\protect\citeauthoryear{Lai et~al.}{2025}]{Lai2025}
\begin{barticle}
\bauthor{\bsnm{Lai}, \binits{T.S.-Y.}},
\bauthor{\bsnm{Duval}, \binits{S.}},
\bauthor{\bsnm{Smith}, \binits{J.D.T.}},
\bauthor{\bsnm{Armus}, \binits{L.}},
\bauthor{\bsnm{Witt}, \binits{A.N.}},
\bauthor{\bsnm{Sandstrom}, \binits{K.}},
\bauthor{\bsnm{Tarantino}, \binits{E.}},
\bauthor{\bsnm{Baba}, \binits{S.}},
\bauthor{\bsnm{Bolatto}, \binits{A.}},
\bauthor{\bsnm{Donnelly}, \binits{G.P.}},
\bauthor{\bsnm{Hensley}, \binits{B.S.}},
\bauthor{\bsnm{Imanishi}, \binits{M.}},
\bauthor{\bsnm{Lenkic}, \binits{L.}},
\bauthor{\bsnm{Linden}, \binits{S.}},
\bauthor{\bsnm{Nakagawa}, \binits{T.}},
\bauthor{\bsnm{Spoon}, \binits{H.W.W.}},
\bauthor{\bsnm{Togi}, \binits{A.}},
\bauthor{\bsnm{Whitcomb}, \binits{C.M.}}:
\batitle{Resolving {{Emission}} from {{Small Dust Grains}} in the {{Blue
  Compact Dwarf II Zw}} 40 with {{JWST}}}.
\bjtitle{Astrophys. J.}
\bvolume{991},
\bfpage{56}
(\byear{2025})
\doiurl{10.3847/2041-8213/ae0467}
\end{barticle}
\endbibitem

%%% 68
\bibitem[\protect\citeauthoryear{Zhang et~al.}{2025}]{Zhang2025}
\begin{barticle}
\bauthor{\bsnm{Zhang}, \binits{C.}},
\bauthor{\bsnm{Hales}, \binits{J.}},
\bauthor{\bsnm{Peeters}, \binits{E.}},
\bauthor{\bsnm{Cami}, \binits{J.}},
\bauthor{\bsnm{Sidhu}, \binits{A.}},
\bauthor{\bsnm{Zhen}, \binits{J.}}:
\batitle{A {{JWST Study}} of {{Polycyclic Aromatic Hydrocarbon Emission}} in a
  {{Region}} of 30 {{Doradus}}}.
\bjtitle{Astrophys. J. Suppl. Ser.}
\bvolume{280},
\bfpage{4}
(\byear{2025})
\doiurl{10.3847/1538-4365/adea6b}
\end{barticle}
\endbibitem

%%% 69
\bibitem[\protect\citeauthoryear{McGuire et~al.}{2021}]{McGuire2021}
\begin{barticle}
\bauthor{\bsnm{McGuire}, \binits{B.A.}},
\bauthor{\bsnm{Loomis}, \binits{R.A.}},
\bauthor{\bsnm{Burkhardt}, \binits{A.M.}},
\bauthor{\bsnm{Lee}, \binits{K.L.K.}},
\bauthor{\bsnm{Shingledecker}, \binits{C.N.}},
\bauthor{\bsnm{Charnley}, \binits{S.B.}},
\bauthor{\bsnm{Cooke}, \binits{I.R.}},
\bauthor{\bsnm{Cordiner}, \binits{M.A.}},
\bauthor{\bsnm{Herbst}, \binits{E.}},
\bauthor{\bsnm{Kalenskii}, \binits{S.}},
\bauthor{\bsnm{Siebert}, \binits{M.A.}},
\bauthor{\bsnm{Willis}, \binits{E.R.}},
\bauthor{\bsnm{Xue}, \binits{C.}},
\bauthor{\bsnm{Remijan}, \binits{A.J.}},
\bauthor{\bsnm{McCarthy}, \binits{M.C.}}:
\batitle{Detection of two interstellar polycyclic aromatic hydrocarbons via
  spectral matched filtering}.
\bjtitle{Science}
\bvolume{371}(\bissue{6535}),
\bfpage{1265}
(\byear{2021})
\doiurl{10.1126/science.abb7535}
\end{barticle}
\endbibitem

%%% 70
\bibitem[\protect\citeauthoryear{Wenzel et~al.}{2024}]{Wenzel2024}
\begin{barticle}
\bauthor{\bsnm{Wenzel}, \binits{G.}},
\bauthor{\bsnm{Cooke}, \binits{I.R.}},
\bauthor{\bsnm{Changala}, \binits{P.B.}},
\bauthor{\bsnm{Bergin}, \binits{E.A.}},
\bauthor{\bsnm{Zhang}, \binits{S.}},
\bauthor{\bsnm{Burkhardt}, \binits{A.M.}},
\bauthor{\bsnm{Byrne}, \binits{A.N.}},
\bauthor{\bsnm{Charnley}, \binits{S.B.}},
\bauthor{\bsnm{Cordiner}, \binits{M.A.}},
\bauthor{\bsnm{Duffy}, \binits{M.}},
\bauthor{\bsnm{Fried}, \binits{Z.T.P.}},
\bauthor{\bsnm{Gupta}, \binits{H.}},
\bauthor{\bsnm{Holdren}, \binits{M.S.}},
\bauthor{\bsnm{Lipnicky}, \binits{A.}},
\bauthor{\bsnm{Loomis}, \binits{R.A.}},
\bauthor{\bsnm{Shay}, \binits{H.T.}},
\bauthor{\bsnm{Shingledecker}, \binits{C.N.}},
\bauthor{\bsnm{Siebert}, \binits{M.A.}},
\bauthor{\bsnm{Stewart}, \binits{D.A.}},
\bauthor{\bsnm{Willis}, \binits{R.H.J.}},
\bauthor{\bsnm{Xue}, \binits{C.}},
\bauthor{\bsnm{Remijan}, \binits{A.J.}},
\bauthor{\bsnm{Wendlandt}, \binits{A.E.}},
\bauthor{\bsnm{McCarthy}, \binits{M.C.}},
\bauthor{\bsnm{McGuire}, \binits{B.A.}}:
\batitle{Detection of interstellar 1-cyanopyrene: {{A}} four-ring polycyclic
  aromatic hydrocarbon}.
\bjtitle{Science}
\bvolume{386},
\bfpage{810}--\blpage{813}
(\byear{2024})
\doiurl{10.1126/science.adq6391}
\end{barticle}
\endbibitem

%%% 71
\bibitem[\protect\citeauthoryear{Zhang et~al.}{2025}]{ZhangHensleyGreen2025}
\begin{barticle}
\bauthor{\bsnm{Zhang}, \binits{X.}},
\bauthor{\bsnm{Hensley}, \binits{B.S.}},
\bauthor{\bsnm{Green}, \binits{G.M.}}:
\batitle{Dust-extinction-curve {{Variation}} in the {{Translucent Interstellar
  Medium Is Driven}} by {{Polycyclic Aromatic Hydrocarbon Growth}}}.
\bjtitle{Astrophys. J.}
\bvolume{979},
\bfpage{17}
(\byear{2025})
\doiurl{10.3847/2041-8213/ada28f}
\end{barticle}
\endbibitem

%%% 72
\bibitem[\protect\citeauthoryear{Allain et~al.}{1996}]{Allain1996}
\begin{barticle}
\bauthor{\bsnm{Allain}, \binits{T.}},
\bauthor{\bsnm{Leach}, \binits{S.}},
\bauthor{\bsnm{Sedlmayr}, \binits{E.}}:
\batitle{Photodestruction of {{PAHs}} in the interstellar medium. {{II}}.
  {{Influence}} of the states of ionization and hydrogenation.}
\bjtitle{Astron. Astrophys.}
\bvolume{305},
\bfpage{616}
(\byear{1996})
\end{barticle}
\endbibitem

%%% 73
\bibitem[\protect\citeauthoryear{Leslie et~al.}{2020}]{Leslie2020}
\begin{barticle}
\bauthor{\bsnm{Leslie}, \binits{S.K.}},
\bauthor{\bsnm{Schinnerer}, \binits{E.}},
\bauthor{\bsnm{Liu}, \binits{D.}},
\bauthor{\bsnm{Magnelli}, \binits{B.}},
\bauthor{\bsnm{Algera}, \binits{H.}},
\bauthor{\bsnm{Karim}, \binits{A.}},
\bauthor{\bsnm{Davidzon}, \binits{I.}},
\bauthor{\bsnm{Gozaliasl}, \binits{G.}},
\bauthor{\bsnm{{Jim{\'e}nez-Andrade}}, \binits{E.F.}},
\bauthor{\bsnm{Lang}, \binits{P.}},
\bauthor{\bsnm{Sargent}, \binits{M.T.}},
\bauthor{\bsnm{Novak}, \binits{M.}},
\bauthor{\bsnm{Groves}, \binits{B.}},
\bauthor{\bsnm{Smol{\v c}i{\'c}}, \binits{V.}},
\bauthor{\bsnm{Zamorani}, \binits{G.}},
\bauthor{\bsnm{Vaccari}, \binits{M.}},
\bauthor{\bsnm{Battisti}, \binits{A.}},
\bauthor{\bsnm{Vardoulaki}, \binits{E.}},
\bauthor{\bsnm{Peng}, \binits{Y.}},
\bauthor{\bsnm{Kartaltepe}, \binits{J.}}:
\batitle{The {{VLA-COSMOS}} 3 {{GHz Large Project}}: {{Evolution}} of
  {{Specific Star Formation Rates}} out to z {$\sim$} 5}.
\bjtitle{Astrophys. J.}
\bvolume{899},
\bfpage{58}
(\byear{2020})
\doiurl{10.3847/1538-4357/aba044}
\end{barticle}
\endbibitem

%%% 74
\bibitem[\protect\citeauthoryear{Gardner et~al.}{2006}]{Gardner2006}
\begin{barticle}
\bauthor{\bsnm{Gardner}, \binits{J.P.}},
\bauthor{\bsnm{Mather}, \binits{J.C.}},
\bauthor{\bsnm{Clampin}, \binits{M.}},
\bauthor{\bsnm{Doyon}, \binits{R.}},
\bauthor{\bsnm{Greenhouse}, \binits{M.A.}},
\bauthor{\bsnm{Hammel}, \binits{H.B.}},
\bauthor{\bsnm{Hutchings}, \binits{J.B.}},
\bauthor{\bsnm{Jakobsen}, \binits{P.}},
\bauthor{\bsnm{Lilly}, \binits{S.J.}},
\bauthor{\bsnm{Long}, \binits{K.S.}},
\bauthor{\bsnm{Lunine}, \binits{J.I.}},
\bauthor{\bsnm{McCaughrean}, \binits{M.J.}},
\bauthor{\bsnm{Mountain}, \binits{M.}},
\bauthor{\bsnm{Nella}, \binits{J.}},
\bauthor{\bsnm{Rieke}, \binits{G.H.}},
\bauthor{\bsnm{Rieke}, \binits{M.J.}},
\bauthor{\bsnm{Rix}, \binits{H.-W.}},
\bauthor{\bsnm{Smith}, \binits{E.P.}},
\bauthor{\bsnm{Sonneborn}, \binits{G.}},
\bauthor{\bsnm{Stiavelli}, \binits{M.}},
\bauthor{\bsnm{Stockman}, \binits{H.S.}},
\bauthor{\bsnm{Windhorst}, \binits{R.A.}},
\bauthor{\bsnm{Wright}, \binits{G.S.}}:
\batitle{The {{James Webb Space Telescope}}}.
\bjtitle{Space Sci. Rev.}
\bvolume{123},
\bfpage{485}--\blpage{606}
(\byear{2006})
\doiurl{10.1007/s11214-006-8315-7}
\end{barticle}
\endbibitem

%%% 75
\bibitem[\protect\citeauthoryear{Rigby et~al.}{2023}]{Rigby2023}
\begin{barticle}
\bauthor{\bsnm{Rigby}, \binits{J.}},
\bauthor{\bsnm{Perrin}, \binits{M.}},
\bauthor{\bsnm{McElwain}, \binits{M.}},
\bauthor{\bsnm{Kimble}, \binits{R.}},
\bauthor{\bsnm{Friedman}, \binits{S.}},
\bauthor{\bsnm{Lallo}, \binits{M.}},
\bauthor{\bsnm{Doyon}, \binits{R.}},
\bauthor{\bsnm{Feinberg}, \binits{L.}},
\bauthor{\bsnm{Ferruit}, \binits{P.}},
\bauthor{\bsnm{Glasse}, \binits{A.}},
\bauthor{\bsnm{Rieke}, \binits{M.}},
\bauthor{\bsnm{Rieke}, \binits{G.}},
\bauthor{\bsnm{Wright}, \binits{G.}},
\bauthor{\bsnm{Willott}, \binits{C.}},
\bauthor{\bsnm{Colon}, \binits{K.}},
\bauthor{\bsnm{Milam}, \binits{S.}},
\bauthor{\bsnm{Neff}, \binits{S.}},
\bauthor{\bsnm{Stark}, \binits{C.}},
\bauthor{\bsnm{Valenti}, \binits{J.}},
\bauthor{\bsnm{Abell}, \binits{J.}},
\bauthor{\bsnm{Abney}, \binits{F.}},
\bauthor{\bsnm{{Abul-Huda}}, \binits{Y.}},
\bauthor{\bsnm{Acton}, \binits{D.S.}},
\bauthor{\bsnm{Adams}, \binits{E.}},
\bauthor{\bsnm{Adler}, \binits{D.}},
\bauthor{\bsnm{Aguilar}, \binits{J.}},
\bauthor{\bsnm{Ahmed}, \binits{N.}},
\bauthor{\bsnm{Albert}, \binits{L.}},
\bauthor{\bsnm{Alberts}, \binits{S.}},
\bauthor{\bsnm{Aldridge}, \binits{D.}},
\bauthor{\bsnm{Allen}, \binits{M.}},
\bauthor{\bsnm{Altenburg}, \binits{M.}},
\bauthor{\bsnm{{\'A}lvarez-M{\'a}rquez}, \binits{J.}},
\bauthor{\bsnm{{Alves de Oliveira}}, \binits{C.}},
\bauthor{\bsnm{Andersen}, \binits{G.}},
\bauthor{\bsnm{Anderson}, \binits{H.}},
\bauthor{\bsnm{Anderson}, \binits{S.}},
\bauthor{\bsnm{Argyriou}, \binits{I.}},
\bauthor{\bsnm{Armstrong}, \binits{A.}},
\bauthor{\bsnm{Arribas}, \binits{S.}},
\bauthor{\bsnm{Artigau}, \binits{E.}},
\bauthor{\bsnm{Arvai}, \binits{A.}},
\bauthor{\bsnm{Atkinson}, \binits{C.}},
\bauthor{\bsnm{Bacon}, \binits{G.}},
\bauthor{\bsnm{Bair}, \binits{T.}},
\bauthor{\bsnm{Banks}, \binits{K.}},
\bauthor{\bsnm{Barrientes}, \binits{J.}},
\bauthor{\bsnm{Barringer}, \binits{B.}},
\bauthor{\bsnm{Bartosik}, \binits{P.}},
\bauthor{\bsnm{Bast}, \binits{W.}},
\bauthor{\bsnm{Baudoz}, \binits{P.}},
\bauthor{\bsnm{Beatty}, \binits{T.}},
\bauthor{\bsnm{Bechtold}, \binits{K.}},
\bauthor{\bsnm{Beck}, \binits{T.}},
\bauthor{\bsnm{Bergeron}, \binits{E.}},
\bauthor{\bsnm{Bergkoetter}, \binits{M.}},
\bauthor{\bsnm{Bhatawdekar}, \binits{R.}},
\bauthor{\bsnm{Birkmann}, \binits{S.}},
\bauthor{\bsnm{Blazek}, \binits{R.}},
\bauthor{\bsnm{Blome}, \binits{C.}},
\bauthor{\bsnm{Boccaletti}, \binits{A.}},
\bauthor{\bsnm{B{\"o}ker}, \binits{T.}},
\bauthor{\bsnm{Boia}, \binits{J.}},
\bauthor{\bsnm{Bonaventura}, \binits{N.}},
\bauthor{\bsnm{Bond}, \binits{N.}},
\bauthor{\bsnm{Bosley}, \binits{K.}},
\bauthor{\bsnm{Boucarut}, \binits{R.}},
\bauthor{\bsnm{Bourque}, \binits{M.}},
\bauthor{\bsnm{Bouwman}, \binits{J.}},
\bauthor{\bsnm{Bower}, \binits{G.}},
\bauthor{\bsnm{Bowers}, \binits{C.}},
\bauthor{\bsnm{Boyer}, \binits{M.}},
\bauthor{\bsnm{Bradley}, \binits{L.}},
\bauthor{\bsnm{Brady}, \binits{G.}},
\bauthor{\bsnm{Braun}, \binits{H.}},
\bauthor{\bsnm{Breda}, \binits{D.}},
\bauthor{\bsnm{Bresnahan}, \binits{P.}},
\bauthor{\bsnm{Bright}, \binits{S.}},
\bauthor{\bsnm{Britt}, \binits{C.}},
\bauthor{\bsnm{Bromenschenkel}, \binits{A.}},
\bauthor{\bsnm{Brooks}, \binits{B.}},
\bauthor{\bsnm{Brooks}, \binits{K.}},
\bauthor{\bsnm{Brown}, \binits{B.}},
\bauthor{\bsnm{Brown}, \binits{M.}},
\bauthor{\bsnm{Brown}, \binits{P.}},
\bauthor{\bsnm{Bunker}, \binits{A.}},
\bauthor{\bsnm{Burger}, \binits{M.}},
\bauthor{\bsnm{Bushouse}, \binits{H.}},
\bauthor{\bsnm{Cale}, \binits{S.}},
\bauthor{\bsnm{Cameron}, \binits{A.}},
\bauthor{\bsnm{Cameron}, \binits{P.}},
\bauthor{\bsnm{Canipe}, \binits{A.}},
\bauthor{\bsnm{Caplinger}, \binits{J.}},
\bauthor{\bsnm{Caputo}, \binits{F.}},
\bauthor{\bsnm{Cara}, \binits{M.}},
\bauthor{\bsnm{Carey}, \binits{L.}},
\bauthor{\bsnm{Carniani}, \binits{S.}},
\bauthor{\bsnm{Carrasquilla}, \binits{M.}},
\bauthor{\bsnm{Carruthers}, \binits{M.}},
\bauthor{\bsnm{Case}, \binits{M.}},
\bauthor{\bsnm{Catherine}, \binits{R.}},
\bauthor{\bsnm{Chance}, \binits{D.}},
\bauthor{\bsnm{Chapman}, \binits{G.}},
\bauthor{\bsnm{Charlot}, \binits{S.}},
\bauthor{\bsnm{Charlow}, \binits{B.}},
\bauthor{\bsnm{Chayer}, \binits{P.}},
\bauthor{\bsnm{Chen}, \binits{B.}},
\bauthor{\bsnm{Cherinka}, \binits{B.}},
\bauthor{\bsnm{Chichester}, \binits{S.}},
\bauthor{\bsnm{Chilton}, \binits{Z.}},
\bauthor{\bsnm{Chonis}, \binits{T.}},
\bauthor{\bsnm{Clampin}, \binits{M.}},
\bauthor{\bsnm{Clark}, \binits{C.}},
\bauthor{\bsnm{Clark}, \binits{K.}},
\bauthor{\bsnm{Coe}, \binits{D.}},
\bauthor{\bsnm{Coleman}, \binits{B.}},
\bauthor{\bsnm{Comber}, \binits{B.}},
\bauthor{\bsnm{Comeau}, \binits{T.}},
\bauthor{\bsnm{Connolly}, \binits{D.}},
\bauthor{\bsnm{Cooper}, \binits{J.}},
\bauthor{\bsnm{Cooper}, \binits{R.}},
\bauthor{\bsnm{Coppock}, \binits{E.}},
\bauthor{\bsnm{Correnti}, \binits{M.}},
\bauthor{\bsnm{Cossou}, \binits{C.}},
\bauthor{\bsnm{Coulais}, \binits{A.}},
\bauthor{\bsnm{Coyle}, \binits{L.}},
\bauthor{\bsnm{Cracraft}, \binits{M.}},
\bauthor{\bsnm{Curti}, \binits{M.}},
\bauthor{\bsnm{Cuturic}, \binits{S.}},
\bauthor{\bsnm{Davis}, \binits{K.}},
\bauthor{\bsnm{Davis}, \binits{M.}},
\bauthor{\bsnm{Dean}, \binits{B.}},
\bauthor{\bsnm{DeLisa}, \binits{A.}},
\bauthor{\bsnm{{deMeester}}, \binits{W.}},
\bauthor{\bsnm{Dencheva}, \binits{N.}},
\bauthor{\bsnm{Dencheva}, \binits{N.}},
\bauthor{\bsnm{DePasquale}, \binits{J.}},
\bauthor{\bsnm{Deschenes}, \binits{J.}},
\bauthor{\bsnm{Hunor~Detre}, \binits{{\"O}.}},
\bauthor{\bsnm{Diaz}, \binits{R.}},
\bauthor{\bsnm{Dicken}, \binits{D.}},
\bauthor{\bsnm{DiFelice}, \binits{A.}},
\bauthor{\bsnm{Dillman}, \binits{M.}},
\bauthor{\bsnm{Dixon}, \binits{W.}},
\bauthor{\bsnm{Doggett}, \binits{J.}},
\bauthor{\bsnm{Donaldson}, \binits{T.}},
\bauthor{\bsnm{Douglas}, \binits{R.}},
\bauthor{\bsnm{DuPrie}, \binits{K.}},
\bauthor{\bsnm{Dupuis}, \binits{J.}},
\bauthor{\bsnm{Durning}, \binits{J.}},
\bauthor{\bsnm{Easmin}, \binits{N.}},
\bauthor{\bsnm{Eck}, \binits{W.}},
\bauthor{\bsnm{Edeani}, \binits{C.}},
\bauthor{\bsnm{Egami}, \binits{E.}},
\bauthor{\bsnm{Ehrenwinkler}, \binits{R.}},
\bauthor{\bsnm{Eisenhamer}, \binits{J.}},
\bauthor{\bsnm{Eisenhower}, \binits{M.}},
\bauthor{\bsnm{Elie}, \binits{M.}},
\bauthor{\bsnm{Elliott}, \binits{J.}},
\bauthor{\bsnm{Elliott}, \binits{K.}},
\bauthor{\bsnm{Ellis}, \binits{T.}},
\bauthor{\bsnm{Engesser}, \binits{M.}},
\bauthor{\bsnm{Espinoza}, \binits{N.}},
\bauthor{\bsnm{Etienne}, \binits{O.}},
\bauthor{\bsnm{Etxaluze}, \binits{M.}},
\bauthor{\bsnm{Falini}, \binits{P.}},
\bauthor{\bsnm{Feeney}, \binits{M.}},
\bauthor{\bsnm{Ferry}, \binits{M.}},
\bauthor{\bsnm{Filippazzo}, \binits{J.}},
\bauthor{\bsnm{Fincham}, \binits{B.}},
\bauthor{\bsnm{Fix}, \binits{M.}},
\bauthor{\bsnm{Flagey}, \binits{N.}},
\bauthor{\bsnm{Florian}, \binits{M.}},
\bauthor{\bsnm{Flynn}, \binits{J.}},
\bauthor{\bsnm{Fontanella}, \binits{E.}},
\bauthor{\bsnm{Ford}, \binits{T.}},
\bauthor{\bsnm{Forshay}, \binits{P.}},
\bauthor{\bsnm{Fox}, \binits{O.}},
\bauthor{\bsnm{Franz}, \binits{D.}},
\bauthor{\bsnm{Fu}, \binits{H.}},
\bauthor{\bsnm{Fullerton}, \binits{A.}},
\bauthor{\bsnm{Galkin}, \binits{S.}},
\bauthor{\bsnm{Galyer}, \binits{A.}},
\bauthor{\bsnm{Garc{\'i}a~Mar{\'i}n}, \binits{M.}},
\bauthor{\bsnm{Gardner}, \binits{J.P.}},
\bauthor{\bsnm{Gardner}, \binits{L.}},
\bauthor{\bsnm{Garland}, \binits{D.}},
\bauthor{\bsnm{Garrett}, \binits{B.}},
\bauthor{\bsnm{Gasman}, \binits{D.}},
\bauthor{\bsnm{Gaspar}, \binits{A.}},
\bauthor{\bsnm{Gaudreau}, \binits{D.}},
\bauthor{\bsnm{Gauthier}, \binits{P.}},
\bauthor{\bsnm{Geers}, \binits{V.}},
\bauthor{\bsnm{Geithner}, \binits{P.}},
\bauthor{\bsnm{Gennaro}, \binits{M.}},
\bauthor{\bsnm{Giardino}, \binits{G.}},
\bauthor{\bsnm{Girard}, \binits{J.}},
\bauthor{\bsnm{Giuliano}, \binits{M.}},
\bauthor{\bsnm{Glassmire}, \binits{K.}},
\bauthor{\bsnm{Glauser}, \binits{A.}}:
\batitle{The {{Science Performance}} of {{JWST}} as {{Characterized}} in
  {{Commissioning}}}.
\bjtitle{Publ. Astron. Soc. Pac.}
\bvolume{135},
\bfpage{048001}
(\byear{2023})
\doiurl{10.1088/1538-3873/acb293}
\end{barticle}
\endbibitem

%%% 76
\bibitem[\protect\citeauthoryear{Rieke et~al.}{2005}]{Rieke2005}
\begin{bchapter}
\bauthor{\bsnm{Rieke}, \binits{M.J.}},
\bauthor{\bsnm{Kelly}, \binits{D.}},
\bauthor{\bsnm{Horner}, \binits{S.}}:
\bctitle{Overview of {{James Webb Space Telescope}} and {{NIRCam}}'s {{Role}}}.
In: \bbtitle{Cryogenic {{Optical Systems}} and {{Instruments XI}}},
vol. \bseriesno{5904},
pp. \bfpage{1}--\blpage{8}
(\byear{2005}).
\doiurl{10.1117/12.615554} .
\burl{https://ui.adsabs.harvard.edu/abs/2005SPIE.5904....1R}
\end{bchapter}
\endbibitem

%%% 77
\bibitem[\protect\citeauthoryear{Rieke et~al.}{2023}]{Rieke2023}
\begin{barticle}
\bauthor{\bsnm{Rieke}, \binits{M.J.}},
\bauthor{\bsnm{Kelly}, \binits{D.M.}},
\bauthor{\bsnm{Misselt}, \binits{K.}},
\bauthor{\bsnm{Stansberry}, \binits{J.}},
\bauthor{\bsnm{Boyer}, \binits{M.}},
\bauthor{\bsnm{Beatty}, \binits{T.}},
\bauthor{\bsnm{Egami}, \binits{E.}},
\bauthor{\bsnm{Florian}, \binits{M.}},
\bauthor{\bsnm{Greene}, \binits{T.P.}},
\bauthor{\bsnm{Hainline}, \binits{K.}},
\bauthor{\bsnm{Leisenring}, \binits{J.}},
\bauthor{\bsnm{Roellig}, \binits{T.}},
\bauthor{\bsnm{Schlawin}, \binits{E.}},
\bauthor{\bsnm{Sun}, \binits{F.}},
\bauthor{\bsnm{Tinnin}, \binits{L.}},
\bauthor{\bsnm{Williams}, \binits{C.C.}},
\bauthor{\bsnm{Willmer}, \binits{C.N.A.}},
\bauthor{\bsnm{Wilson}, \binits{D.}},
\bauthor{\bsnm{Clark}, \binits{C.R.}},
\bauthor{\bsnm{Rohrbach}, \binits{S.}},
\bauthor{\bsnm{Brooks}, \binits{B.}},
\bauthor{\bsnm{Canipe}, \binits{A.}},
\bauthor{\bsnm{Correnti}, \binits{M.}},
\bauthor{\bsnm{DiFelice}, \binits{A.}},
\bauthor{\bsnm{Gennaro}, \binits{M.}},
\bauthor{\bsnm{Girard}, \binits{J.H.}},
\bauthor{\bsnm{Hartig}, \binits{G.}},
\bauthor{\bsnm{Hilbert}, \binits{B.}},
\bauthor{\bsnm{Koekemoer}, \binits{A.M.}},
\bauthor{\bsnm{Nikolov}, \binits{N.K.}},
\bauthor{\bsnm{Pirzkal}, \binits{N.}},
\bauthor{\bsnm{Rest}, \binits{A.}},
\bauthor{\bsnm{Robberto}, \binits{M.}},
\bauthor{\bsnm{Sunnquist}, \binits{B.}},
\bauthor{\bsnm{Telfer}, \binits{R.}},
\bauthor{\bsnm{Wu}, \binits{C.R.}},
\bauthor{\bsnm{Ferry}, \binits{M.}},
\bauthor{\bsnm{Lewis}, \binits{D.}},
\bauthor{\bsnm{Baum}, \binits{S.}},
\bauthor{\bsnm{Beichman}, \binits{C.}},
\bauthor{\bsnm{Doyon}, \binits{R.}},
\bauthor{\bsnm{Dressler}, \binits{A.}},
\bauthor{\bsnm{Eisenstein}, \binits{D.J.}},
\bauthor{\bsnm{Ferrarese}, \binits{L.}},
\bauthor{\bsnm{Hodapp}, \binits{K.}},
\bauthor{\bsnm{Horner}, \binits{S.}},
\bauthor{\bsnm{Jaffe}, \binits{D.T.}},
\bauthor{\bsnm{Johnstone}, \binits{D.}},
\bauthor{\bsnm{Krist}, \binits{J.}},
\bauthor{\bsnm{Martin}, \binits{P.}},
\bauthor{\bsnm{McCarthy}, \binits{D.W.}},
\bauthor{\bsnm{Meyer}, \binits{M.}},
\bauthor{\bsnm{Rieke}, \binits{G.H.}},
\bauthor{\bsnm{Trauger}, \binits{J.}},
\bauthor{\bsnm{Young}, \binits{E.T.}}:
\batitle{Performance of {{NIRCam}} on {{JWST}} in {{Flight}}}.
\bjtitle{Publ. Astron. Soc. Pac.}
\bvolume{135},
\bfpage{028001}
(\byear{2023})
\doiurl{10.1088/1538-3873/acac53}
\end{barticle}
\endbibitem

%%% 78
\bibitem[\protect\citeauthoryear{Rieke et~al.}{2015}]{Rieke2015}
\begin{barticle}
\bauthor{\bsnm{Rieke}, \binits{G.H.}},
\bauthor{\bsnm{Wright}, \binits{G.S.}},
\bauthor{\bsnm{B{\"o}ker}, \binits{T.}},
\bauthor{\bsnm{Bouwman}, \binits{J.}},
\bauthor{\bsnm{Colina}, \binits{L.}},
\bauthor{\bsnm{Glasse}, \binits{A.}},
\bauthor{\bsnm{Gordon}, \binits{K.D.}},
\bauthor{\bsnm{Greene}, \binits{T.P.}},
\bauthor{\bsnm{G{\"u}del}, \binits{M.}},
\bauthor{\bsnm{Henning}, \binits{{\relax Th}.}},
\bauthor{\bsnm{Justtanont}, \binits{K.}},
\bauthor{\bsnm{Lagage}, \binits{P.-O.}},
\bauthor{\bsnm{Meixner}, \binits{M.E.}},
\bauthor{\bsnm{{N{\o}rgaard-Nielsen}}, \binits{H.-U.}},
\bauthor{\bsnm{Ray}, \binits{T.P.}},
\bauthor{\bsnm{Ressler}, \binits{M.E.}},
\bauthor{\bsnm{{van Dishoeck}}, \binits{E.F.}},
\bauthor{\bsnm{Waelkens}, \binits{C.}}:
\batitle{The {{Mid-Infrared Instrument}} for the {{James Webb Space
  Telescope}}, {{I}}: {{Introduction}}}.
\bjtitle{Publ. Astron. Soc. Pac.}
\bvolume{127}(\bissue{953}),
\bfpage{584}
(\byear{2015})
\doiurl{10.1086/682252}
\end{barticle}
\endbibitem

%%% 79
\bibitem[\protect\citeauthoryear{Wright et~al.}{2023}]{Wright2023}
\begin{barticle}
\bauthor{\bsnm{Wright}, \binits{G.S.}},
\bauthor{\bsnm{Rieke}, \binits{G.H.}},
\bauthor{\bsnm{Glasse}, \binits{A.}},
\bauthor{\bsnm{Ressler}, \binits{M.}},
\bauthor{\bsnm{Garc{\'i}a~Mar{\'i}n}, \binits{M.}},
\bauthor{\bsnm{Aguilar}, \binits{J.}},
\bauthor{\bsnm{Alberts}, \binits{S.}},
\bauthor{\bsnm{{\'A}lvarez-M{\'a}rquez}, \binits{J.}},
\bauthor{\bsnm{Argyriou}, \binits{I.}},
\bauthor{\bsnm{Banks}, \binits{K.}},
\bauthor{\bsnm{Baudoz}, \binits{P.}},
\bauthor{\bsnm{Boccaletti}, \binits{A.}},
\bauthor{\bsnm{Bouchet}, \binits{P.}},
\bauthor{\bsnm{Bouwman}, \binits{J.}},
\bauthor{\bsnm{Brandl}, \binits{B.R.}},
\bauthor{\bsnm{Breda}, \binits{D.}},
\bauthor{\bsnm{Bright}, \binits{S.}},
\bauthor{\bsnm{Cale}, \binits{S.}},
\bauthor{\bsnm{Colina}, \binits{L.}},
\bauthor{\bsnm{Cossou}, \binits{C.}},
\bauthor{\bsnm{Coulais}, \binits{A.}},
\bauthor{\bsnm{Cracraft}, \binits{M.}},
\bauthor{\bsnm{De~Meester}, \binits{W.}},
\bauthor{\bsnm{Dicken}, \binits{D.}},
\bauthor{\bsnm{Engesser}, \binits{M.}},
\bauthor{\bsnm{Etxaluze}, \binits{M.}},
\bauthor{\bsnm{Fox}, \binits{O.D.}},
\bauthor{\bsnm{Friedman}, \binits{S.}},
\bauthor{\bsnm{Fu}, \binits{H.}},
\bauthor{\bsnm{Gasman}, \binits{D.}},
\bauthor{\bsnm{G{\'a}sp{\'a}r}, \binits{A.}},
\bauthor{\bsnm{Gastaud}, \binits{R.}},
\bauthor{\bsnm{Geers}, \binits{V.}},
\bauthor{\bsnm{Glauser}, \binits{A.M.}},
\bauthor{\bsnm{Gordon}, \binits{K.D.}},
\bauthor{\bsnm{Greene}, \binits{T.}},
\bauthor{\bsnm{Greve}, \binits{T.R.}},
\bauthor{\bsnm{Grundy}, \binits{T.}},
\bauthor{\bsnm{G{\"u}del}, \binits{M.}},
\bauthor{\bsnm{Guillard}, \binits{P.}},
\bauthor{\bsnm{Haderlein}, \binits{P.}},
\bauthor{\bsnm{Hashimoto}, \binits{R.}},
\bauthor{\bsnm{Henning}, \binits{T.}},
\bauthor{\bsnm{Hines}, \binits{D.}},
\bauthor{\bsnm{Holler}, \binits{B.}},
\bauthor{\bsnm{Detre}, \binits{{\"O}.H.}},
\bauthor{\bsnm{Jahromi}, \binits{A.}},
\bauthor{\bsnm{James}, \binits{B.}},
\bauthor{\bsnm{Jones}, \binits{O.C.}},
\bauthor{\bsnm{Justtanont}, \binits{K.}},
\bauthor{\bsnm{Kavanagh}, \binits{P.}},
\bauthor{\bsnm{Kendrew}, \binits{S.}},
\bauthor{\bsnm{Klaassen}, \binits{P.}},
\bauthor{\bsnm{Krause}, \binits{O.}},
\bauthor{\bsnm{Labiano}, \binits{A.}},
\bauthor{\bsnm{Lagage}, \binits{P.-O.}},
\bauthor{\bsnm{Lambros}, \binits{S.}},
\bauthor{\bsnm{Larson}, \binits{K.}},
\bauthor{\bsnm{Law}, \binits{D.}},
\bauthor{\bsnm{Lee}, \binits{D.}},
\bauthor{\bsnm{Libralato}, \binits{M.}},
\bauthor{\bsnm{Alverez}, \binits{J.L.}},
\bauthor{\bsnm{Meixner}, \binits{M.}},
\bauthor{\bsnm{Morrison}, \binits{J.}},
\bauthor{\bsnm{Mueller}, \binits{M.}},
\bauthor{\bsnm{Murray}, \binits{K.}},
\bauthor{\bsnm{Mycroft}, \binits{M.}},
\bauthor{\bsnm{Myers}, \binits{R.}},
\bauthor{\bsnm{Nayak}, \binits{O.}},
\bauthor{\bsnm{Naylor}, \binits{B.}},
\bauthor{\bsnm{Nickson}, \binits{B.}},
\bauthor{\bsnm{{Noriega-Crespo}}, \binits{A.}},
\bauthor{\bsnm{{\"O}stlin}, \binits{G.}},
\bauthor{\bsnm{O'Sullivan}, \binits{B.}},
\bauthor{\bsnm{Ottens}, \binits{R.}},
\bauthor{\bsnm{Patapis}, \binits{P.}},
\bauthor{\bsnm{Penanen}, \binits{K.}},
\bauthor{\bsnm{Pietraszkiewicz}, \binits{M.}},
\bauthor{\bsnm{Ray}, \binits{T.}},
\bauthor{\bsnm{Regan}, \binits{M.}},
\bauthor{\bsnm{Roteliuk}, \binits{A.}},
\bauthor{\bsnm{Royer}, \binits{P.}},
\bauthor{\bsnm{{Samara-Ratna}}, \binits{P.}},
\bauthor{\bsnm{Samuelson}, \binits{B.}},
\bauthor{\bsnm{Sargent}, \binits{B.A.}},
\bauthor{\bsnm{Scheithauer}, \binits{S.}},
\bauthor{\bsnm{Schneider}, \binits{A.}},
\bauthor{\bsnm{Schreiber}, \binits{J.}},
\bauthor{\bsnm{Shaughnessy}, \binits{B.}},
\bauthor{\bsnm{Sheehan}, \binits{E.}},
\bauthor{\bsnm{Shivaei}, \binits{I.}},
\bauthor{\bsnm{Sloan}, \binits{G.C.}},
\bauthor{\bsnm{Tamas}, \binits{L.}},
\bauthor{\bsnm{Teague}, \binits{K.}},
\bauthor{\bsnm{Temim}, \binits{T.}},
\bauthor{\bsnm{Tikkanen}, \binits{T.}},
\bauthor{\bsnm{Tustain}, \binits{S.}},
\bauthor{\bsnm{{van Dishoeck}}, \binits{E.F.}},
\bauthor{\bsnm{Vandenbussche}, \binits{B.}},
\bauthor{\bsnm{Weilert}, \binits{M.}},
\bauthor{\bsnm{Whitehouse}, \binits{P.}},
\bauthor{\bsnm{Wolff}, \binits{S.}}:
\batitle{The {{Mid-infrared Instrument}} for {{JWST}} and {{Its In-flight
  Performance}}}.
\bjtitle{Publ. Astron. Soc. Pac.}
\bvolume{135}(\bissue{1046}),
\bfpage{048003}
(\byear{2023})
\doiurl{10.1088/1538-3873/acbe66}
\end{barticle}
\endbibitem

%%% 80
\bibitem[\protect\citeauthoryear{Pontoppidan et~al.}{2016}]{Pontoppidan2016}
\begin{bchapter}
\bauthor{\bsnm{Pontoppidan}, \binits{K.M.}},
\bauthor{\bsnm{Pickering}, \binits{T.E.}},
\bauthor{\bsnm{Laidler}, \binits{V.G.}},
\bauthor{\bsnm{Gilbert}, \binits{K.}},
\bauthor{\bsnm{Sontag}, \binits{C.D.}},
\bauthor{\bsnm{Slocum}, \binits{C.}},
\bauthor{\bsnm{Sienkiewicz}, \binits{M.J.}},
\bauthor{\bsnm{Hanley}, \binits{C.}},
\bauthor{\bsnm{Earl}, \binits{N.M.}},
\bauthor{\bsnm{Pueyo}, \binits{L.}},
\bauthor{\bsnm{Ravindranath}, \binits{S.}},
\bauthor{\bsnm{Karakla}, \binits{D.M.}},
\bauthor{\bsnm{Robberto}, \binits{M.}},
\bauthor{\bsnm{{Noriega-Crespo}}, \binits{A.}},
\bauthor{\bsnm{Barker}, \binits{E.A.}}:
\bctitle{Pandeia: A multi-mission exposure time calculator for {{JWST}} and
  {{WFIRST}}}.
In: \bbtitle{Observatory {{Operations}}: {{Strategies}}, {{Processes}}, And
  {{Systems VI}}},
vol. \bseriesno{9910}.
\bconflocation{eprint: arXiv:1707.02202},
p. \bfpage{991016}
(\byear{2016}).
\doiurl{10.1117/12.2231768} .
\burl{https://ui.adsabs.harvard.edu/abs/2016SPIE.9910E..16P}
\end{bchapter}
\endbibitem

%%% 81
\bibitem[\protect\citeauthoryear{}{}]{Bushouse2025}
\begin{botherref}
{{JWST Calibration Pipeline}}.
\url{https://zenodo.org/records/17101851}
\end{botherref}
\endbibitem

%%% 82
\bibitem[\protect\citeauthoryear{Greenfield and Miller}{2016}]{Greenfield2016}
\begin{barticle}
\bauthor{\bsnm{Greenfield}, \binits{P.}},
\bauthor{\bsnm{Miller}, \binits{T.}}:
\batitle{The {{Calibration Reference Data System}}}.
\bjtitle{Astron. Comput.}
\bvolume{16},
\bfpage{41}--\blpage{53}
(\byear{2016})
\doiurl{10.1016/j.ascom.2016.04.001}
\end{barticle}
\endbibitem

%%% 83
\bibitem[\protect\citeauthoryear{Willott et~al.}{2022}]{Willott2022}
\begin{barticle}
\bauthor{\bsnm{Willott}, \binits{C.J.}},
\bauthor{\bsnm{Doyon}, \binits{R.}},
\bauthor{\bsnm{Albert}, \binits{L.}},
\bauthor{\bsnm{Brammer}, \binits{G.B.}},
\bauthor{\bsnm{Dixon}, \binits{W.V.}},
\bauthor{\bsnm{Muzic}, \binits{K.}},
\bauthor{\bsnm{Ravindranath}, \binits{S.}},
\bauthor{\bsnm{Scholz}, \binits{A.}},
\bauthor{\bsnm{Abraham}, \binits{R.}},
\bauthor{\bsnm{Artigau}, \binits{{\'E}.}},
\bauthor{\bsnm{Brada{\v c}}, \binits{M.}},
\bauthor{\bsnm{Goudfrooij}, \binits{P.}},
\bauthor{\bsnm{Hutchings}, \binits{J.B.}},
\bauthor{\bsnm{Iyer}, \binits{K.G.}},
\bauthor{\bsnm{Jayawardhana}, \binits{R.}},
\bauthor{\bsnm{LaMassa}, \binits{S.}},
\bauthor{\bsnm{Martis}, \binits{N.}},
\bauthor{\bsnm{Meyer}, \binits{M.R.}},
\bauthor{\bsnm{Morishita}, \binits{T.}},
\bauthor{\bsnm{Mowla}, \binits{L.}},
\bauthor{\bsnm{Muzzin}, \binits{A.}},
\bauthor{\bsnm{Noirot}, \binits{G.}},
\bauthor{\bsnm{Pacifici}, \binits{C.}},
\bauthor{\bsnm{Rowlands}, \binits{N.}},
\bauthor{\bsnm{Sarrouh}, \binits{G.}},
\bauthor{\bsnm{Sawicki}, \binits{M.}},
\bauthor{\bsnm{Taylor}, \binits{J.M.}},
\bauthor{\bsnm{Volk}, \binits{K.}},
\bauthor{\bsnm{Zabl}, \binits{J.}}:
\batitle{The {{Near-infrared Imager}} and {{Slitless Spectrograph}} for the
  {{James Webb Space Telescope}}. {{II}}. {{Wide Field Slitless
  Spectroscopy}}}.
\bjtitle{Publ. Astron. Soc. Pac.}
\bvolume{134}(\bissue{1032}),
\bfpage{025002}
(\byear{2022})
\doiurl{10.1088/1538-3873/ac5158}
\end{barticle}
\endbibitem

%%% 84
\bibitem[\protect\citeauthoryear{Rigby et~al.}{2023}]{Rigby2023b}
\begin{barticle}
\bauthor{\bsnm{Rigby}, \binits{J.R.}},
\bauthor{\bsnm{Lightsey}, \binits{P.A.}},
\bauthor{\bsnm{Garc{\'i}a~Mar{\'i}n}, \binits{M.}},
\bauthor{\bsnm{Bowers}, \binits{C.W.}},
\bauthor{\bsnm{Smith}, \binits{E.C.}},
\bauthor{\bsnm{Glasse}, \binits{A.}},
\bauthor{\bsnm{McElwain}, \binits{M.W.}},
\bauthor{\bsnm{Rieke}, \binits{G.H.}},
\bauthor{\bsnm{Chary}, \binits{R.-R.}},
\bauthor{\bsnm{Liu}, \binits{X.C.}},
\bauthor{\bsnm{Clampin}, \binits{M.}},
\bauthor{\bsnm{Kimble}, \binits{R.A.}},
\bauthor{\bsnm{Kinzel}, \binits{W.}},
\bauthor{\bsnm{Laidler}, \binits{V.}},
\bauthor{\bsnm{Mehalick}, \binits{K.I.}},
\bauthor{\bsnm{{Noriega-Crespo}}, \binits{A.}},
\bauthor{\bsnm{Shivaei}, \binits{I.}},
\bauthor{\bsnm{Skelton}, \binits{D.}},
\bauthor{\bsnm{Stark}, \binits{C.}},
\bauthor{\bsnm{Temim}, \binits{T.}},
\bauthor{\bsnm{Wei}, \binits{Z.}},
\bauthor{\bsnm{Willott}, \binits{C.J.}}:
\batitle{How {{Dark}} the {{Sky}}: {{The JWST Backgrounds}}}.
\bjtitle{Publ. Astron. Soc. Pac.}
\bvolume{135}(\bissue{1046}),
\bfpage{048002}
(\byear{2023})
\doiurl{10.1088/1538-3873/acbcf4}
\end{barticle}
\endbibitem

%%% 85
\bibitem[\protect\citeauthoryear{Aniano et~al.}{2011}]{Aniano2011}
\begin{barticle}
\bauthor{\bsnm{Aniano}, \binits{G.}},
\bauthor{\bsnm{Draine}, \binits{B.T.}},
\bauthor{\bsnm{Gordon}, \binits{K.D.}},
\bauthor{\bsnm{Sandstrom}, \binits{K.}}:
\batitle{Common-{{Resolution Convolution Kernels}} for {{Space-}} and
  {{Ground-Based Telescopes}}}.
\bjtitle{Publ. Astron. Soc. Pac.}
\bvolume{123}(\bissue{908}),
\bfpage{1218}
(\byear{2011})
\doiurl{10.1086/662219}
\end{barticle}
\endbibitem

%%% 86
\bibitem[\protect\citeauthoryear{Perrin et~al.}{2014}]{Perrin2014}
\begin{bchapter}
\bauthor{\bsnm{Perrin}, \binits{M.D.}},
\bauthor{\bsnm{Sivaramakrishnan}, \binits{A.}},
\bauthor{\bsnm{Lajoie}, \binits{C.-P.}},
\bauthor{\bsnm{Elliott}, \binits{E.}},
\bauthor{\bsnm{Pueyo}, \binits{L.}},
\bauthor{\bsnm{Ravindranath}, \binits{S.}},
\bauthor{\bsnm{Albert}, \binits{{\relax Lo{\"i}c}.}}:
\bctitle{Updated point spread function simulations for {{JWST}} with
  {{WebbPSF}}}.
In: \beditor{\bsnm{Oschmann}, \binits{{\relax Jr}.} \bsuffix{Jacobus~M.}},
\beditor{\bsnm{Clampin}, \binits{M.}},
\beditor{\bsnm{Fazio}, \binits{G.G.}},
\beditor{\bsnm{MacEwen}, \binits{H.A.}} (eds.)
\bbtitle{Space {{Telesc}}. {{Instrum}}. 2014 {{Opt}}. {{Infrared Millim}}.
  {{Wave}}}.
\bsertitle{Society of {{Photo-Optical Instrumentation Engineers}} ({{SPIE}})
  {{Conference Series}}},
vol. \bseriesno{9143},
p. \bfpage{91433}
(\byear{2014}).
\doiurl{10.1117/12.2056689}
\end{bchapter}
\endbibitem

%%% 87
\bibitem[\protect\citeauthoryear{Libralato et~al.}{2024}]{Libralato2024}
\begin{barticle}
\bauthor{\bsnm{Libralato}, \binits{M.}},
\bauthor{\bsnm{Argyriou}, \binits{I.}},
\bauthor{\bsnm{Dicken}, \binits{D.}},
\bauthor{\bsnm{Garc{\'i}a~Mar{\'i}n}, \binits{M.}},
\bauthor{\bsnm{Guillard}, \binits{P.}},
\bauthor{\bsnm{Hines}, \binits{D.C.}},
\bauthor{\bsnm{Kavanagh}, \binits{P.J.}},
\bauthor{\bsnm{Kendrew}, \binits{S.}},
\bauthor{\bsnm{Law}, \binits{D.R.}},
\bauthor{\bsnm{{Noriega-Crespo}}, \binits{A.}},
\bauthor{\bsnm{{\'A}lvarez-M{\'a}rquez}, \binits{J.}}:
\batitle{High-precision {{Astrometry}} and {{Photometry}} with the
  {{JWST}}/{{MIRI Imager}}}.
\bjtitle{Publ. Astron. Soc. Pac.}
\bvolume{136},
\bfpage{034502}
(\byear{2024})
\doiurl{10.1088/1538-3873/ad2551}
\end{barticle}
\endbibitem

%%% 88
\bibitem[\protect\citeauthoryear{Lai et~al.}{2020}]{Lai2020}
\begin{barticle}
\bauthor{\bsnm{Lai}, \binits{T.S.-Y.}},
\bauthor{\bsnm{Smith}, \binits{J.D.T.}},
\bauthor{\bsnm{Baba}, \binits{S.}},
\bauthor{\bsnm{Spoon}, \binits{H.W.W.}},
\bauthor{\bsnm{Imanishi}, \binits{M.}}:
\batitle{All the {{PAHs}}: {{An AKARI-Spitzer Cross-archival Spectroscopic
  Survey}} of {{Aromatic Emission}} in {{Galaxies}}}.
\bjtitle{Astrophys. J.}
\bvolume{905}(\bissue{1}),
\bfpage{55}
(\byear{2020})
\doiurl{10.3847/1538-4357/abc002}
\end{barticle}
\endbibitem

%%% 89
\bibitem[\protect\citeauthoryear{Bolatto et~al.}{2024}]{Bolatto2024}
\begin{barticle}
\bauthor{\bsnm{Bolatto}, \binits{A.D.}},
\bauthor{\bsnm{Levy}, \binits{R.C.}},
\bauthor{\bsnm{Tarantino}, \binits{E.}},
\bauthor{\bsnm{Boyer}, \binits{M.L.}},
\bauthor{\bsnm{Fisher}, \binits{D.B.}},
\bauthor{\bsnm{Cronin}, \binits{S.A.}},
\bauthor{\bsnm{Leroy}, \binits{A.K.}},
\bauthor{\bsnm{Klessen}, \binits{R.S.}},
\bauthor{\bsnm{Smith}, \binits{J.D.}},
\bauthor{\bsnm{Berg}, \binits{D.A.}},
\bauthor{\bsnm{B{\"o}ker}, \binits{T.}},
\bauthor{\bsnm{Boogaard}, \binits{L.A.}},
\bauthor{\bsnm{Ostriker}, \binits{E.C.}},
\bauthor{\bsnm{Thompson}, \binits{T.A.}},
\bauthor{\bsnm{Ott}, \binits{J.}},
\bauthor{\bsnm{Lenki{\'c}}, \binits{L.}},
\bauthor{\bsnm{Lopez}, \binits{L.A.}},
\bauthor{\bsnm{Dale}, \binits{D.A.}},
\bauthor{\bsnm{Veilleux}, \binits{S.}},
\bauthor{\bsnm{{van der Werf}}, \binits{P.P.}},
\bauthor{\bsnm{Glover}, \binits{S.C.O.}},
\bauthor{\bsnm{Sandstrom}, \binits{K.M.}},
\bauthor{\bsnm{Skillman}, \binits{E.D.}},
\bauthor{\bsnm{Chisholm}, \binits{J.}},
\bauthor{\bsnm{Villanueva}, \binits{V.}},
\bauthor{\bsnm{Lai}, \binits{T.S.-Y.}},
\bauthor{\bsnm{Lopez}, \binits{S.}},
\bauthor{\bsnm{Mills}, \binits{E.A.C.}},
\bauthor{\bsnm{Emig}, \binits{K.L.}},
\bauthor{\bsnm{Armus}, \binits{L.}},
\bauthor{\bsnm{Mayya}, \binits{D.}},
\bauthor{\bsnm{Meier}, \binits{D.S.}},
\bauthor{\bsnm{De~Looze}, \binits{I.}},
\bauthor{\bsnm{{Herrera-Camus}}, \binits{R.}},
\bauthor{\bsnm{Walter}, \binits{F.}},
\bauthor{\bsnm{Rela{\~n}o}, \binits{M.}},
\bauthor{\bsnm{Koziol}, \binits{H.B.}},
\bauthor{\bsnm{Marvil}, \binits{J.}},
\bauthor{\bsnm{{Jim{\'e}nez-Donaire}}, \binits{M.J.}},
\bauthor{\bsnm{Martini}, \binits{P.}}:
\batitle{{{JWST Observations}} of {{Starbursts}}: {{Polycyclic Aromatic
  Hydrocarbon Emission}} at the {{Base}} of the {{M82 Galactic Wind}}}.
\bjtitle{Astrophys. J.}
\bvolume{967}(\bissue{1}),
\bfpage{63}
(\byear{2024})
\doiurl{10.3847/1538-4357/ad33c8}
\end{barticle}
\endbibitem

%%% 90
\bibitem[\protect\citeauthoryear{Chown et~al.}{2025}]{Chown2025_pdrs}
\begin{barticle}
\bauthor{\bsnm{Chown}, \binits{R.}},
\bauthor{\bsnm{Okada}, \binits{Y.}},
\bauthor{\bsnm{Peeters}, \binits{E.}},
\bauthor{\bsnm{Sidhu}, \binits{A.}},
\bauthor{\bsnm{Khan}, \binits{B.}},
\bauthor{\bsnm{Schefter}, \binits{B.}},
\bauthor{\bsnm{Trahin}, \binits{B.}},
\bauthor{\bsnm{Canin}, \binits{A.}},
\bauthor{\bsnm{Van De~Putte}, \binits{D.}},
\bauthor{\bsnm{Alarc{\'o}n}, \binits{F.}},
\bauthor{\bsnm{Schroetter}, \binits{I.}},
\bauthor{\bsnm{Kannavou}, \binits{O.}},
\bauthor{\bsnm{Habart}, \binits{E.}},
\bauthor{\bsnm{Bern{\'e}}, \binits{O.}},
\bauthor{\bsnm{Boersma}, \binits{C.}},
\bauthor{\bsnm{Cami}, \binits{J.}},
\bauthor{\bsnm{Dartois}, \binits{E.}},
\bauthor{\bsnm{Goicoechea}, \binits{J.}},
\bauthor{\bsnm{Gordon}, \binits{K.}},
\bauthor{\bsnm{Onaka}, \binits{T.}}:
\batitle{{{PDRs4All}}: {{XIII}}. {{Empirical}} prescriptions for the
  interpretation of {{JWST}} imaging observations of star-forming regions}.
\bjtitle{Astron. Astrophys.}
\bvolume{698},
\bfpage{86}
(\byear{2025})
\doiurl{10.1051/0004-6361/202452940}
\end{barticle}
\endbibitem

%%% 91
\bibitem[\protect\citeauthoryear{Donnelly et~al.}{2025}]{Donnelly2025}
\begin{barticle}
\bauthor{\bsnm{Donnelly}, \binits{G.P.}},
\bauthor{\bsnm{Lai}, \binits{T.S.-Y.}},
\bauthor{\bsnm{Armus}, \binits{L.}},
\bauthor{\bsnm{{D{\'i}az-Santos}}, \binits{T.}},
\bauthor{\bsnm{Larson}, \binits{K.L.}},
\bauthor{\bsnm{{Barcos-Mu{\~n}oz}}, \binits{L.}},
\bauthor{\bsnm{Bianchin}, \binits{M.}},
\bauthor{\bsnm{Bohn}, \binits{T.}},
\bauthor{\bsnm{B{\"o}ker}, \binits{T.}},
\bauthor{\bsnm{Buiten}, \binits{V.A.}},
\bauthor{\bsnm{Charmandaris}, \binits{V.}},
\bauthor{\bsnm{Evans}, \binits{A.S.}},
\bauthor{\bsnm{Howell}, \binits{J.}},
\bauthor{\bsnm{Inami}, \binits{H.}},
\bauthor{\bsnm{Kakkad}, \binits{D.}},
\bauthor{\bsnm{Lenki{\'c}}, \binits{L.}},
\bauthor{\bsnm{Linden}, \binits{S.T.}},
\bauthor{\bsnm{Lofaro}, \binits{C.M.}},
\bauthor{\bsnm{Malkan}, \binits{M.A.}},
\bauthor{\bsnm{Medling}, \binits{A.M.}},
\bauthor{\bsnm{Privon}, \binits{G.C.}},
\bauthor{\bsnm{Ricci}, \binits{C.}},
\bauthor{\bsnm{Smith}, \binits{J.D.T.}},
\bauthor{\bsnm{Song}, \binits{Y.}},
\bauthor{\bsnm{Stierwalt}, \binits{S.}},
\bauthor{\bsnm{{van der Werf}}, \binits{P.P.}},
\bauthor{\bsnm{U}, \binits{V.}}:
\batitle{A {{Spectroscopically Calibrated Prescription}} for {{Extracting
  Polycyclic Aromatic Hydrocarbon Flux}} from {{JWST MIRI Imaging}}}.
\bjtitle{Astrophys. J.}
\bvolume{983},
\bfpage{79}
(\byear{2025})
\doiurl{10.3847/1538-4357/adb97f}
\end{barticle}
\endbibitem

%%% 92
\bibitem[\protect\citeauthoryear{Joblin et~al.}{1996}]{Joblin1996}
\begin{barticle}
\bauthor{\bsnm{Joblin}, \binits{C.}},
\bauthor{\bsnm{Tielens}, \binits{A.G.G.M.}},
\bauthor{\bsnm{Allamandola}, \binits{L.J.}},
\bauthor{\bsnm{Geballe}, \binits{T.R.}}:
\batitle{Spatial {{Variation}} of the 3.29 and 3.40 {{Micron Emission Bands}}
  within {{Reflection Nebulae}} and the {{Photochemical Evolution}} of
  {{Methylated Polycyclic Aromatic Hydrocarbons}}}.
\bjtitle{Astrophys. J.}
\bvolume{458},
\bfpage{610}
(\byear{1996})
\doiurl{10.1086/176843}
\end{barticle}
\endbibitem

%%% 93
\bibitem[\protect\citeauthoryear{Hammonds et~al.}{2015}]{Hammonds2015}
\begin{barticle}
\bauthor{\bsnm{Hammonds}, \binits{M.}},
\bauthor{\bsnm{Mori}, \binits{T.}},
\bauthor{\bsnm{Usui}, \binits{F.}},
\bauthor{\bsnm{Onaka}, \binits{T.}}:
\batitle{Variations in the 3.3 {\textbackslash}ensuremath{\textbackslash}mum
  feature and carbonaceous dust in {{AKARI}} data}.
\bjtitle{Planet. Space Sci.}
\bvolume{116},
\bfpage{73}--\blpage{83}
(\byear{2015})
\doiurl{10.1016/j.pss.2015.05.010}
\end{barticle}
\endbibitem

%%% 94
\bibitem[\protect\citeauthoryear{Van De~Putte et~al.}{2024}]{VanDePutte2024}
\begin{barticle}
\bauthor{\bsnm{Van De~Putte}, \binits{D.}},
\bauthor{\bsnm{Meshaka}, \binits{R.}},
\bauthor{\bsnm{Trahin}, \binits{B.}},
\bauthor{\bsnm{Habart}, \binits{E.}},
\bauthor{\bsnm{Peeters}, \binits{E.}},
\bauthor{\bsnm{Bern{\'e}}, \binits{O.}},
\bauthor{\bsnm{Alarc{\'o}n}, \binits{F.}},
\bauthor{\bsnm{Canin}, \binits{A.}},
\bauthor{\bsnm{Chown}, \binits{R.}},
\bauthor{\bsnm{Schroetter}, \binits{I.}},
\bauthor{\bsnm{Sidhu}, \binits{A.}},
\bauthor{\bsnm{Boersma}, \binits{C.}},
\bauthor{\bsnm{Bron}, \binits{E.}},
\bauthor{\bsnm{Dartois}, \binits{E.}},
\bauthor{\bsnm{Goicoechea}, \binits{J.R.}},
\bauthor{\bsnm{Gordon}, \binits{K.D.}},
\bauthor{\bsnm{Onaka}, \binits{T.}},
\bauthor{\bsnm{Tielens}, \binits{A.G.G.M.}},
\bauthor{\bsnm{Verstraete}, \binits{L.}},
\bauthor{\bsnm{Wolfire}, \binits{M.G.}},
\bauthor{\bsnm{Abergel}, \binits{A.}},
\bauthor{\bsnm{Bergin}, \binits{E.A.}},
\bauthor{\bsnm{{Bernard-Salas}}, \binits{J.}},
\bauthor{\bsnm{Cami}, \binits{J.}},
\bauthor{\bsnm{Cuadrado}, \binits{S.}},
\bauthor{\bsnm{Dicken}, \binits{D.}},
\bauthor{\bsnm{Elyajouri}, \binits{M.}},
\bauthor{\bsnm{Fuente}, \binits{A.}},
\bauthor{\bsnm{Joblin}, \binits{C.}},
\bauthor{\bsnm{Khan}, \binits{B.}},
\bauthor{\bsnm{Lacinbala}, \binits{O.}},
\bauthor{\bsnm{Languignon}, \binits{D.}},
\bauthor{\bsnm{Le~Gal}, \binits{R.}},
\bauthor{\bsnm{Maragkoudakis}, \binits{A.}},
\bauthor{\bsnm{Okada}, \binits{Y.}},
\bauthor{\bsnm{Pasquini}, \binits{S.}},
\bauthor{\bsnm{Pound}, \binits{M.W.}},
\bauthor{\bsnm{Robberto}, \binits{M.}},
\bauthor{\bsnm{R{\"o}llig}, \binits{M.}},
\bauthor{\bsnm{Schefter}, \binits{B.}},
\bauthor{\bsnm{Schirmer}, \binits{T.}},
\bauthor{\bsnm{Tabone}, \binits{B.}},
\bauthor{\bsnm{Vicente}, \binits{S.}},
\bauthor{\bsnm{Zannese}, \binits{M.}},
\bauthor{\bsnm{Colgan}, \binits{S.W.J.}},
\bauthor{\bsnm{He}, \binits{J.}},
\bauthor{\bsnm{Rouill{\'e}}, \binits{G.}},
\bauthor{\bsnm{Togi}, \binits{A.}},
\bauthor{\bsnm{Aleman}, \binits{I.}},
\bauthor{\bsnm{Auchettl}, \binits{R.}},
\bauthor{\bsnm{Baratta}, \binits{G.A.}},
\bauthor{\bsnm{Bejaoui}, \binits{S.}},
\bauthor{\bsnm{Bera}, \binits{P.P.}},
\bauthor{\bsnm{Black}, \binits{J.H.}},
\bauthor{\bsnm{Boulanger}, \binits{F.}},
\bauthor{\bsnm{Bouwman}, \binits{J.}},
\bauthor{\bsnm{Brandl}, \binits{B.}},
\bauthor{\bsnm{Brechignac}, \binits{P.}},
\bauthor{\bsnm{Br{\"u}nken}, \binits{S.}},
\bauthor{\bsnm{Buragohain}, \binits{M.}},
\bauthor{\bsnm{Burkhardt}, \binits{A.}},
\bauthor{\bsnm{Candian}, \binits{A.}},
\bauthor{\bsnm{Cazaux}, \binits{S.}},
\bauthor{\bsnm{Cernicharo}, \binits{J.}},
\bauthor{\bsnm{Chabot}, \binits{M.}},
\bauthor{\bsnm{Chakraborty}, \binits{S.}},
\bauthor{\bsnm{Champion}, \binits{J.}},
\bauthor{\bsnm{Cooke}, \binits{I.R.}},
\bauthor{\bsnm{Coutens}, \binits{A.}},
\bauthor{\bsnm{Cox}, \binits{N.L.J.}},
\bauthor{\bsnm{Demyk}, \binits{K.}},
\bauthor{\bsnm{Meyer}, \binits{J.D.}},
\bauthor{\bsnm{Foschino}, \binits{S.}},
\bauthor{\bsnm{{Garc{\'i}a-Lario}}, \binits{P.}},
\bauthor{\bsnm{Gerin}, \binits{M.}},
\bauthor{\bsnm{Gottlieb}, \binits{C.A.}},
\bauthor{\bsnm{Guillard}, \binits{P.}},
\bauthor{\bsnm{Gusdorf}, \binits{A.}},
\bauthor{\bsnm{Hartigan}, \binits{P.}},
\bauthor{\bsnm{Herbst}, \binits{E.}},
\bauthor{\bsnm{Hornekaer}, \binits{L.}},
\bauthor{\bsnm{Issa}, \binits{L.}},
\bauthor{\bsnm{J{\"a}ger}, \binits{C.}},
\bauthor{\bsnm{{Janot-Pacheco}}, \binits{E.}},
\bauthor{\bsnm{Kannavou}, \binits{O.}},
\bauthor{\bsnm{Kaufman}, \binits{M.}},
\bauthor{\bsnm{Kemper}, \binits{F.}},
\bauthor{\bsnm{Kendrew}, \binits{S.}},
\bauthor{\bsnm{Kirsanova}, \binits{M.S.}},
\bauthor{\bsnm{Klaassen}, \binits{P.}},
\bauthor{\bsnm{Kwok}, \binits{S.}},
\bauthor{\bsnm{Labiano}, \binits{{\'A}.}},
\bauthor{\bsnm{Lai}, \binits{T.S.-Y.}},
\bauthor{\bsnm{Le~Floch}, \binits{B.}},
\bauthor{\bsnm{Le~Petit}, \binits{F.}},
\bauthor{\bsnm{Li}, \binits{A.}},
\bauthor{\bsnm{Linz}, \binits{H.}},
\bauthor{\bsnm{Mackie}, \binits{C.J.}},
\bauthor{\bsnm{Madden}, \binits{S.C.}},
\bauthor{\bsnm{Mascetti}, \binits{J.}},
\bauthor{\bsnm{McGuire}, \binits{B.A.}},
\bauthor{\bsnm{Merino}, \binits{P.}},
\bauthor{\bsnm{Micelotta}, \binits{E.R.}},
\bauthor{\bsnm{Morse}, \binits{J.A.}},
\bauthor{\bsnm{Mulas}, \binits{G.}},
\bauthor{\bsnm{Neelamkodan}, \binits{N.}},
\bauthor{\bsnm{Ohsawa}, \binits{R.}},
\bauthor{\bsnm{Omont}, \binits{A.}},
\bauthor{\bsnm{Paladini}, \binits{R.}},
\bauthor{\bsnm{Palumbo}, \binits{M.E.}},
\bauthor{\bsnm{Pathak}, \binits{A.}},
\bauthor{\bsnm{Pendleton}, \binits{Y.J.}},
\bauthor{\bsnm{Petrignani}, \binits{A.}},
\bauthor{\bsnm{Pino}, \binits{T.}},
\bauthor{\bsnm{Puga}, \binits{E.}},
\bauthor{\bsnm{Rangwala}, \binits{N.}},
\bauthor{\bsnm{Rapacioli}, \binits{M.}},
\bauthor{\bsnm{Rho}, \binits{J.}},
\bauthor{\bsnm{Ricca}, \binits{A.}},
\bauthor{\bsnm{{Roman-Duval}}, \binits{J.}},
\bauthor{\bsnm{Roser}, \binits{J.}},
\bauthor{\bsnm{Roueff}, \binits{E.}},
\bauthor{\bsnm{Salama}, \binits{F.}},
\bauthor{\bsnm{Sales}, \binits{D.A.}},
\bauthor{\bsnm{Sandstrom}, \binits{K.}},
\bauthor{\bsnm{Sarre}, \binits{P.}},
\bauthor{\bsnm{{Sciamma-O'Brien}}, \binits{E.}},
\bauthor{\bsnm{Sellgren}, \binits{K.}},
\bauthor{\bsnm{Shenoy}, \binits{S.S.}},
\bauthor{\bsnm{Teyssier}, \binits{D.}},
\bauthor{\bsnm{Thomas}, \binits{R.D.}},
\bauthor{\bsnm{Witt}, \binits{A.N.}},
\bauthor{\bsnm{Wootten}, \binits{A.}},
\bauthor{\bsnm{Ysard}, \binits{N.}},
\bauthor{\bsnm{Zettergren}, \binits{H.}},
\bauthor{\bsnm{Zhang}, \binits{Y.}},
\bauthor{\bsnm{Zhang}, \binits{Z.E.}},
\bauthor{\bsnm{Zhen}, \binits{J.}}:
\batitle{{{PDRs4All}}: {{VIII}}. {{Mid-infrared}} emission line inventory of
  the {{Orion Bar}}}.
\bjtitle{Astron. Astrophys.}
\bvolume{687},
\bfpage{86}
(\byear{2024})
\doiurl{10.1051/0004-6361/202449295}
\end{barticle}
\endbibitem

%%% 95
\bibitem[\protect\citeauthoryear{Van De~Putte et~al.}{2025}]{VanDePutte2025}
\begin{barticle}
\bauthor{\bsnm{Van De~Putte}, \binits{D.}},
\bauthor{\bsnm{Peeters}, \binits{E.}},
\bauthor{\bsnm{Gordon}, \binits{K.D.}},
\bauthor{\bsnm{Smith}, \binits{J.-D.T.}},
\bauthor{\bsnm{Lai}, \binits{T.S.-Y.}},
\bauthor{\bsnm{Maragkoudakis}, \binits{A.}},
\bauthor{\bsnm{Schefter}, \binits{B.}},
\bauthor{\bsnm{Sidhu}, \binits{A.}},
\bauthor{\bsnm{Doshi}, \binits{D.}},
\bauthor{\bsnm{Bern{\'e}}, \binits{O.}},
\bauthor{\bsnm{Cami}, \binits{J.}},
\bauthor{\bsnm{Boersma}, \binits{C.}},
\bauthor{\bsnm{Dartois}, \binits{E.}},
\bauthor{\bsnm{Habart}, \binits{E.}},
\bauthor{\bsnm{Onaka}, \binits{T.}},
\bauthor{\bsnm{Tielens}, \binits{A.G.G.M.}}:
\batitle{{{PDRs4All}}: {{XVI}}. {{Tracing}} aromatic infrared band
  characteristics in photodissociation region spectra with {{PAHFIT}} in the
  {{JWST}} era}.
\bjtitle{Astron. Astrophys.}
\bvolume{701},
\bfpage{111}
(\byear{2025})
\doiurl{10.1051/0004-6361/202554991}
\end{barticle}
\endbibitem

%%% 96
\bibitem[\protect\citeauthoryear{Boersma et~al.}{2023}]{Boersma2023}
\begin{barticle}
\bauthor{\bsnm{Boersma}, \binits{C.}},
\bauthor{\bsnm{Allamandola}, \binits{L.J.}},
\bauthor{\bsnm{Esposito}, \binits{V.J.}},
\bauthor{\bsnm{Maragkoudakis}, \binits{A.}},
\bauthor{\bsnm{Bregman}, \binits{J.D.}},
\bauthor{\bsnm{Temi}, \binits{P.}},
\bauthor{\bsnm{Lee}, \binits{T.J.}},
\bauthor{\bsnm{Fortenberry}, \binits{R.C.}},
\bauthor{\bsnm{Peeters}, \binits{E.}}:
\batitle{{{JWST}}: {{Deuterated PAHs}}, {{PAH Nitriles}}, and {{PAH Overtone}}
  and {{Combination Bands}}. {{I}}. {{Program Description}} and {{First
  Look}}}.
\bjtitle{Astrophys. J.}
\bvolume{959},
\bfpage{74}
(\byear{2023})
\doiurl{10.3847/1538-4357/ad022b}
\end{barticle}
\endbibitem

%%% 97
\bibitem[\protect\citeauthoryear{Allamandola et~al.}{1989}]{Allamandola1989}
\begin{barticle}
\bauthor{\bsnm{Allamandola}, \binits{L.J.}},
\bauthor{\bsnm{Tielens}, \binits{A.G.G.M.}},
\bauthor{\bsnm{Barker}, \binits{J.R.}}:
\batitle{Interstellar {{Polycyclic Aromatic Hydrocarbons}}: {{The Infrared
  Emission Bands}}, the {{Excitation}}/{{Emission Mechanism}}, and the
  {{Astrophysical Implications}}}.
\bjtitle{Astrophys. J. Suppl. Ser.}
\bvolume{71},
\bfpage{733}
(\byear{1989})
\doiurl{10.1086/191396}
\end{barticle}
\endbibitem

%%% 98
\bibitem[\protect\citeauthoryear{Allamandola et~al.}{2021}]{Allamandola2021}
\begin{barticle}
\bauthor{\bsnm{Allamandola}, \binits{L.J.}},
\bauthor{\bsnm{Boersma}, \binits{C.}},
\bauthor{\bsnm{Lee}, \binits{T.J.}},
\bauthor{\bsnm{Bregman}, \binits{J.D.}},
\bauthor{\bsnm{Temi}, \binits{P.}}:
\batitle{{{PAH Spectroscopy}} from 1 to 5 {$M$}m}.
\bjtitle{Astrophys. J.}
\bvolume{917},
\bfpage{35}
(\byear{2021})
\doiurl{10.3847/2041-8213/ac17f0}
\end{barticle}
\endbibitem

%%% 99
\bibitem[\protect\citeauthoryear{Esposito et~al.}{2024}]{Esposito2024}
\begin{barticle}
\bauthor{\bsnm{Esposito}, \binits{V.J.}},
\bauthor{\bsnm{Allamandola}, \binits{L.J.}},
\bauthor{\bsnm{Boersma}, \binits{C.}},
\bauthor{\bsnm{Bregman}, \binits{J.D.}},
\bauthor{\bsnm{Fortenberry}, \binits{R.C.}},
\bauthor{\bsnm{Maragkoudakis}, \binits{A.}},
\bauthor{\bsnm{Temi}, \binits{P.}}:
\batitle{Anharmonic {{IR}} absorption spectra of the prototypical interstellar
  {{PAHs}} phenanthrene, pyrene, and pentacene in their neutral and cation
  states}.
\bjtitle{Mol. Phys.}
\bvolume{122}(\bissue{7-8}),
\bfpage{2252936}
(\byear{2024})
\doiurl{10.1080/00268976.2023.2252936}
\end{barticle}
\endbibitem

%%% 100
\bibitem[\protect\citeauthoryear{Dale et~al.}{2025}]{Dale2025}
\begin{barticle}
\bauthor{\bsnm{Dale}, \binits{D.A.}},
\bauthor{\bsnm{Graham}, \binits{G.B.}},
\bauthor{\bsnm{Barnes}, \binits{A.T.}},
\bauthor{\bsnm{Baron}, \binits{D.}},
\bauthor{\bsnm{Bigiel}, \binits{F.}},
\bauthor{\bsnm{Boquien}, \binits{M.}},
\bauthor{\bsnm{Chandar}, \binits{R.}},
\bauthor{\bsnm{Chastenet}, \binits{J.}},
\bauthor{\bsnm{Chown}, \binits{R.}},
\bauthor{\bsnm{Egorov}, \binits{O.V.}},
\bauthor{\bsnm{Glover}, \binits{S.C.O.}},
\bauthor{\bsnm{Hands}, \binits{L.}},
\bauthor{\bsnm{Henny}, \binits{K.F.}},
\bauthor{\bsnm{Indebetouw}, \binits{R.}},
\bauthor{\bsnm{Klessen}, \binits{R.S.}},
\bauthor{\bsnm{Larson}, \binits{K.L.}},
\bauthor{\bsnm{Lee}, \binits{J.C.}},
\bauthor{\bsnm{Leroy}, \binits{A.K.}},
\bauthor{\bsnm{Maschmann}, \binits{D.}},
\bauthor{\bsnm{Pathak}, \binits{D.}},
\bauthor{\bsnm{Rodr{\'i}guez}, \binits{M.J.}},
\bauthor{\bsnm{Rosolowsky}, \binits{E.}},
\bauthor{\bsnm{Sandstrom}, \binits{K.}},
\bauthor{\bsnm{Schinnerer}, \binits{E.}},
\bauthor{\bsnm{Sutter}, \binits{J.}},
\bauthor{\bsnm{Thilker}, \binits{D.A.}},
\bauthor{\bsnm{Weinbeck}, \binits{T.D.}},
\bauthor{\bsnm{Whitmore}, \binits{B.C.}},
\bauthor{\bsnm{Williams}, \binits{T.G.}},
\bauthor{\bsnm{Wofford}, \binits{A.}}:
\batitle{{{PAH Feature Ratios}} around {{Stellar Clusters}} and
  {{Associations}} in 19 {{Nearby Galaxies}}}.
\bjtitle{Astron. J.}
\bvolume{169},
\bfpage{133}
(\byear{2025})
\doiurl{10.3847/1538-3881/ada89f}
\end{barticle}
\endbibitem

%%% 101
\bibitem[\protect\citeauthoryear{Whitcomb et~al.}{2023}]{Whitcomb2023a}
\begin{barticle}
\bauthor{\bsnm{Whitcomb}, \binits{C.M.}},
\bauthor{\bsnm{Sandstrom}, \binits{K.}},
\bauthor{\bsnm{Smith}, \binits{J.-D.T.}}:
\batitle{{{JWST-MIRI Synthetic Photometry Composition}} using 460 {{Spitzer-IRS
  Spectra}} of {{Nearby Galaxies}}}.
\bjtitle{Res. Notes AAS}
\bvolume{7}(\bissue{3}),
\bfpage{38}
(\byear{2023})
\doiurl{10.3847/2515-5172/acc073}
\end{barticle}
\endbibitem

%%% 102
\bibitem[\protect\citeauthoryear{Misselt et~al.}{2025}]{Misselt2025}
\begin{barticle}
\bauthor{\bsnm{Misselt}, \binits{K.}},
\bauthor{\bsnm{Witt}, \binits{A.N.}},
\bauthor{\bsnm{Gordon}, \binits{K.D.}},
\bauthor{\bsnm{Van De~Putte}, \binits{D.}},
\bauthor{\bsnm{Trahin}, \binits{B.}},
\bauthor{\bsnm{Abergel}, \binits{A.}},
\bauthor{\bsnm{{Noriega-Crespo}}, \binits{A.}},
\bauthor{\bsnm{Guillard}, \binits{P.}},
\bauthor{\bsnm{Zannese}, \binits{M.}},
\bauthor{\bsnm{Dell'ova}, \binits{P.}},
\bauthor{\bsnm{Baes}, \binits{M.}},
\bauthor{\bsnm{Klaassen}, \binits{P.}},
\bauthor{\bsnm{Ysard}, \binits{N.}}:
\batitle{{{JWST}} observations of photodissociation regions: {{I}}.
  {{Aliphatic}} and aromatic carbonaceous dust, ices, and gas-phase spectral
  line inventory}.
\bjtitle{Astron. Astrophys.}
\bvolume{700},
\bfpage{158}
(\byear{2025})
\doiurl{10.1051/0004-6361/202554851}
\end{barticle}
\endbibitem

%%% 103
\bibitem[\protect\citeauthoryear{Williams et~al.}{2024}]{Williams2024}
\begin{barticle}
\bauthor{\bsnm{Williams}, \binits{T.G.}},
\bauthor{\bsnm{Lee}, \binits{J.C.}},
\bauthor{\bsnm{Larson}, \binits{K.L.}},
\bauthor{\bsnm{Leroy}, \binits{A.K.}},
\bauthor{\bsnm{Sandstrom}, \binits{K.}},
\bauthor{\bsnm{Schinnerer}, \binits{E.}},
\bauthor{\bsnm{Thilker}, \binits{D.A.}},
\bauthor{\bsnm{Belfiore}, \binits{F.}},
\bauthor{\bsnm{Egorov}, \binits{O.V.}},
\bauthor{\bsnm{Rosolowsky}, \binits{E.}},
\bauthor{\bsnm{Sutter}, \binits{J.}},
\bauthor{\bsnm{DePasquale}, \binits{J.}},
\bauthor{\bsnm{Pagan}, \binits{A.}},
\bauthor{\bsnm{Berger}, \binits{T.A.}},
\bauthor{\bsnm{Anand}, \binits{G.S.}},
\bauthor{\bsnm{Barnes}, \binits{A.T.}},
\bauthor{\bsnm{Bigiel}, \binits{F.}},
\bauthor{\bsnm{Boquien}, \binits{M.}},
\bauthor{\bsnm{Cao}, \binits{Y.}},
\bauthor{\bsnm{Chastenet}, \binits{J.}},
\bauthor{\bsnm{Chevance}, \binits{M.}},
\bauthor{\bsnm{Chown}, \binits{R.}},
\bauthor{\bsnm{Dale}, \binits{D.A.}},
\bauthor{\bsnm{Deger}, \binits{S.}},
\bauthor{\bsnm{Eibensteiner}, \binits{C.}},
\bauthor{\bsnm{Emsellem}, \binits{E.}},
\bauthor{\bsnm{Faesi}, \binits{C.M.}},
\bauthor{\bsnm{Glover}, \binits{S.C.O.}},
\bauthor{\bsnm{Grasha}, \binits{K.}},
\bauthor{\bsnm{Hannon}, \binits{S.}},
\bauthor{\bsnm{Hassani}, \binits{H.}},
\bauthor{\bsnm{Henshaw}, \binits{J.D.}},
\bauthor{\bsnm{{Jim{\'e}nez-Donaire}}, \binits{M.J.}},
\bauthor{\bsnm{Kim}, \binits{J.}},
\bauthor{\bsnm{Klessen}, \binits{R.S.}},
\bauthor{\bsnm{Koch}, \binits{E.W.}},
\bauthor{\bsnm{Li}, \binits{J.}},
\bauthor{\bsnm{Liu}, \binits{D.}},
\bauthor{\bsnm{Meidt}, \binits{S.E.}},
\bauthor{\bsnm{{M{\'e}ndez-Delgado}}, \binits{J.E.}},
\bauthor{\bsnm{Murphy}, \binits{E.J.}},
\bauthor{\bsnm{Neumann}, \binits{J.}},
\bauthor{\bsnm{Neumann}, \binits{L.}},
\bauthor{\bsnm{Neumayer}, \binits{N.}},
\bauthor{\bsnm{Oakes}, \binits{E.K.}},
\bauthor{\bsnm{Pathak}, \binits{D.}},
\bauthor{\bsnm{Pety}, \binits{J.}},
\bauthor{\bsnm{Pinna}, \binits{F.}},
\bauthor{\bsnm{Querejeta}, \binits{M.}},
\bauthor{\bsnm{Ramambason}, \binits{L.}},
\bauthor{\bsnm{Romanelli}, \binits{A.}},
\bauthor{\bsnm{Sormani}, \binits{M.C.}},
\bauthor{\bsnm{Stuber}, \binits{S.K.}},
\bauthor{\bsnm{Sun}, \binits{J.}},
\bauthor{\bsnm{Teng}, \binits{Y.-H.}},
\bauthor{\bsnm{Usero}, \binits{A.}},
\bauthor{\bsnm{Watkins}, \binits{E.J.}},
\bauthor{\bsnm{Weinbeck}, \binits{T.D.}}:
\batitle{{{PHANGS-JWST}}: {{Data-processing Pipeline}} and {{First Full Public
  Data Release}}}.
\bjtitle{Astrophys. J. Suppl. Ser.}
\bvolume{273},
\bfpage{13}
(\byear{2024})
\doiurl{10.3847/1538-4365/ad4be5}
\end{barticle}
\endbibitem

%%% 104
\bibitem[\protect\citeauthoryear{Hagen et~al.}{2017}]{Hagen2017}
\begin{barticle}
\bauthor{\bsnm{Hagen}, \binits{L.M.Z.}},
\bauthor{\bsnm{Siegel}, \binits{M.H.}},
\bauthor{\bsnm{Hoversten}, \binits{E.A.}},
\bauthor{\bsnm{Gronwall}, \binits{C.}},
\bauthor{\bsnm{Immler}, \binits{S.}},
\bauthor{\bsnm{Hagen}, \binits{A.}}:
\batitle{Swift {{Ultraviolet Survey}} of the {{Magellanic Clouds}} ({{SUMaC}})
  - {{I}}. {{Shape}} of the ultraviolet dust extinction law and recent star
  formation history of the {{Small Magellanic Cloud}}}.
\bjtitle{Mon. Not. R. Astron. Soc.}
\bvolume{466},
\bfpage{4540}--\blpage{4557}
(\byear{2017})
\doiurl{10.1093/mnras/stw2954}
\end{barticle}
\endbibitem

%%% 105
\bibitem[\protect\citeauthoryear{Rosolowsky et~al.}{2008}]{Rosolowsky2008}
\begin{barticle}
\bauthor{\bsnm{Rosolowsky}, \binits{E.W.}},
\bauthor{\bsnm{Pineda}, \binits{J.E.}},
\bauthor{\bsnm{Kauffmann}, \binits{J.}},
\bauthor{\bsnm{Goodman}, \binits{A.A.}}:
\batitle{Structural {{Analysis}} of {{Molecular Clouds}}: {{Dendrograms}}}.
\bjtitle{Astrophys. J.}
\bvolume{679},
\bfpage{1338}--\blpage{1351}
(\byear{2008})
\doiurl{10.1086/587685}
\end{barticle}
\endbibitem

%%% 106
\bibitem[\protect\citeauthoryear{Goodman et~al.}{2009}]{Goodman2009}
\begin{barticle}
\bauthor{\bsnm{Goodman}, \binits{A.A.}},
\bauthor{\bsnm{Rosolowsky}, \binits{E.W.}},
\bauthor{\bsnm{Borkin}, \binits{M.A.}},
\bauthor{\bsnm{Foster}, \binits{J.B.}},
\bauthor{\bsnm{Halle}, \binits{M.}},
\bauthor{\bsnm{Kauffmann}, \binits{J.}},
\bauthor{\bsnm{Pineda}, \binits{J.E.}}:
\batitle{A role for self-gravity at multiple length scales in the process of
  star formation}.
\bjtitle{Nature}
\bvolume{457},
\bfpage{63}--\blpage{66}
(\byear{2009})
\doiurl{10.1038/nature07609}
\end{barticle}
\endbibitem

%%% 107
\bibitem[\protect\citeauthoryear{Kirk et~al.}{2013}]{Kirk2013}
\begin{barticle}
\bauthor{\bsnm{Kirk}, \binits{J.M.}},
\bauthor{\bsnm{{Ward-Thompson}}, \binits{D.}},
\bauthor{\bsnm{Palmeirim}, \binits{P.}},
\bauthor{\bsnm{Andr{\'e}}, \binits{{\relax Ph}.}},
\bauthor{\bsnm{Griffin}, \binits{M.J.}},
\bauthor{\bsnm{Hargrave}, \binits{P.J.}},
\bauthor{\bsnm{K{\"o}nyves}, \binits{V.}},
\bauthor{\bsnm{Bernard}, \binits{J.-P.}},
\bauthor{\bsnm{Nutter}, \binits{D.J.}},
\bauthor{\bsnm{Sibthorpe}, \binits{B.}},
\bauthor{\bsnm{Di~Francesco}, \binits{J.}},
\bauthor{\bsnm{Abergel}, \binits{A.}},
\bauthor{\bsnm{Arzoumanian}, \binits{D.}},
\bauthor{\bsnm{Benedettini}, \binits{M.}},
\bauthor{\bsnm{Bontemps}, \binits{S.}},
\bauthor{\bsnm{Elia}, \binits{D.}},
\bauthor{\bsnm{Hennemann}, \binits{M.}},
\bauthor{\bsnm{Hill}, \binits{T.}},
\bauthor{\bsnm{Men'shchikov}, \binits{A.}},
\bauthor{\bsnm{Motte}, \binits{F.}},
\bauthor{\bsnm{{Nguyen-Luong}}, \binits{Q.}},
\bauthor{\bsnm{Peretto}, \binits{N.}},
\bauthor{\bsnm{Pezzuto}, \binits{S.}},
\bauthor{\bsnm{Rygl}, \binits{K.L.J.}},
\bauthor{\bsnm{Sadavoy}, \binits{S.I.}},
\bauthor{\bsnm{Schisano}, \binits{E.}},
\bauthor{\bsnm{Schneider}, \binits{N.}},
\bauthor{\bsnm{Testi}, \binits{L.}},
\bauthor{\bsnm{White}, \binits{G.}}:
\batitle{First results from the {{Herschel Gould Belt Survey}} in {{Taurus}}}.
\bjtitle{Mon. Not. R. Astron. Soc.}
\bvolume{432},
\bfpage{1424}--\blpage{1433}
(\byear{2013})
\doiurl{10.1093/mnras/stt561}
\end{barticle}
\endbibitem

%%% 108
\bibitem[\protect\citeauthoryear{{Lind-Thomsen}
  et~al.}{2025}]{Lind-Thomsen2025}
\begin{barticle}
\bauthor{\bsnm{{Lind-Thomsen}}, \binits{C.}},
\bauthor{\bsnm{Sneppen}, \binits{A.}},
\bauthor{\bsnm{Steinhardt}, \binits{C.L.}}:
\batitle{A {{Power Spectral Study}} of {{PHANGS Galaxies}} with {{JWST MIRI}}:
  {{On}} the {{Spatial Scales}} of {{Dust}} and {{Polycyclic Aromatic
  Hydrocarbons}}}.
\bjtitle{Astrophys. J.}
\bvolume{985},
\bfpage{144}
(\byear{2025})
\doiurl{10.3847/1538-4357/adc808}
\end{barticle}
\endbibitem

%%% 109
\bibitem[\protect\citeauthoryear{Gordon et~al.}{2022}]{Gordon2022}
\begin{barticle}
\bauthor{\bsnm{Gordon}, \binits{K.D.}},
\bauthor{\bsnm{Bohlin}, \binits{R.}},
\bauthor{\bsnm{Sloan}, \binits{G.C.}},
\bauthor{\bsnm{Rieke}, \binits{G.}},
\bauthor{\bsnm{Volk}, \binits{K.}},
\bauthor{\bsnm{Boyer}, \binits{M.}},
\bauthor{\bsnm{Muzerolle}, \binits{J.}},
\bauthor{\bsnm{Schlawin}, \binits{E.}},
\bauthor{\bsnm{Deustua}, \binits{S.E.}},
\bauthor{\bsnm{Hines}, \binits{D.C.}},
\bauthor{\bsnm{Kraemer}, \binits{K.E.}},
\bauthor{\bsnm{Mullally}, \binits{S.E.}},
\bauthor{\bsnm{Su}, \binits{K.Y.L.}}:
\batitle{The {{James Webb Space Telescope Absolute Flux Calibration}}. {{I}}.
  {{Program Design}} and {{Calibrator Stars}}}.
\bjtitle{Astron. J.}
\bvolume{163},
\bfpage{267}
(\byear{2022})
\doiurl{10.3847/1538-3881/ac66dc}
\end{barticle}
\endbibitem

%%% 110
\bibitem[\protect\citeauthoryear{Hensley and Draine}{2023}]{Hensley2023}
\begin{barticle}
\bauthor{\bsnm{Hensley}, \binits{B.S.}},
\bauthor{\bsnm{Draine}, \binits{B.T.}}:
\batitle{The {{Astrodust}}+{{PAH Model}}: {{A Unified Description}} of the
  {{Extinction}}, {{Emission}}, and {{Polarization}} from {{Dust}} in the
  {{Diffuse Interstellar Medium}}}.
\bjtitle{Astrophys. J.}
\bvolume{948}(\bissue{1}),
\bfpage{55}
(\byear{2023})
\doiurl{10.3847/1538-4357/acc4c2}
\end{barticle}
\endbibitem

%%% 111
\bibitem[\protect\citeauthoryear{Kennicutt et~al.}{2003}]{Kennicutt2003}
\begin{barticle}
\bauthor{\bsnm{Kennicutt}, \binits{R.C.} \bsuffix{Jr.}},
\bauthor{\bsnm{Armus}, \binits{L.}},
\bauthor{\bsnm{Bendo}, \binits{G.}},
\bauthor{\bsnm{Calzetti}, \binits{D.}},
\bauthor{\bsnm{Dale}, \binits{D.A.}},
\bauthor{\bsnm{Draine}, \binits{B.T.}},
\bauthor{\bsnm{Engelbracht}, \binits{C.W.}},
\bauthor{\bsnm{Gordon}, \binits{K.D.}},
\bauthor{\bsnm{Grauer}, \binits{A.D.}},
\bauthor{\bsnm{Helou}, \binits{G.}},
\bauthor{\bsnm{Hollenbach}, \binits{D.J.}},
\bauthor{\bsnm{Jarrett}, \binits{T.H.}},
\bauthor{\bsnm{Kewley}, \binits{L.J.}},
\bauthor{\bsnm{Leitherer}, \binits{C.}},
\bauthor{\bsnm{Li}, \binits{A.}},
\bauthor{\bsnm{Malhotra}, \binits{S.}},
\bauthor{\bsnm{Regan}, \binits{M.W.}},
\bauthor{\bsnm{Rieke}, \binits{G.H.}},
\bauthor{\bsnm{Rieke}, \binits{M.J.}},
\bauthor{\bsnm{Roussel}, \binits{H.}},
\bauthor{\bsnm{Smith}, \binits{J.-D.T.}},
\bauthor{\bsnm{Thornley}, \binits{M.D.}},
\bauthor{\bsnm{Walter}, \binits{F.}}:
\batitle{{{SINGS}}: {{The SIRTF Nearby Galaxies Survey}}}.
\bjtitle{Publ. Astron. Soc. Pac.}
\bvolume{115},
\bfpage{928}--\blpage{952}
(\byear{2003})
\doiurl{10.1086/376941}
\end{barticle}
\endbibitem

%%% 112
\bibitem[\protect\citeauthoryear{Whitcomb et~al.}{2023}]{Whitcomb2023}
\begin{barticle}
\bauthor{\bsnm{Whitcomb}, \binits{C.M.}},
\bauthor{\bsnm{Sandstrom}, \binits{K.}},
\bauthor{\bsnm{Leroy}, \binits{A.}},
\bauthor{\bsnm{Smith}, \binits{J.-D.T.}}:
\batitle{Star {{Formation}} and {{Molecular Gas Diagnostics}} with {{Mid-}} and
  {{Far-infrared Emission}}}.
\bjtitle{Astrophys. J.}
\bvolume{948}(\bissue{2}),
\bfpage{88}
(\byear{2023})
\doiurl{10.3847/1538-4357/acc316}
\end{barticle}
\endbibitem

%%% 113
\bibitem[\protect\citeauthoryear{Dalcanton et~al.}{2015}]{Dalcanton2015}
\begin{barticle}
\bauthor{\bsnm{Dalcanton}, \binits{J.J.}},
\bauthor{\bsnm{Fouesneau}, \binits{M.}},
\bauthor{\bsnm{Hogg}, \binits{D.W.}},
\bauthor{\bsnm{Lang}, \binits{D.}},
\bauthor{\bsnm{Leroy}, \binits{A.K.}},
\bauthor{\bsnm{Gordon}, \binits{K.D.}},
\bauthor{\bsnm{Sandstrom}, \binits{K.}},
\bauthor{\bsnm{Weisz}, \binits{D.R.}},
\bauthor{\bsnm{Williams}, \binits{B.F.}},
\bauthor{\bsnm{Bell}, \binits{E.F.}},
\bauthor{\bsnm{Dong}, \binits{H.}},
\bauthor{\bsnm{Gilbert}, \binits{K.M.}},
\bauthor{\bsnm{Gouliermis}, \binits{D.A.}},
\bauthor{\bsnm{Guhathakurta}, \binits{P.}},
\bauthor{\bsnm{Lauer}, \binits{T.R.}},
\bauthor{\bsnm{Schruba}, \binits{A.}},
\bauthor{\bsnm{Seth}, \binits{A.C.}},
\bauthor{\bsnm{Skillman}, \binits{E.D.}}:
\batitle{The {{Panchromatic Hubble Andromeda Treasury}}. {{VIII}}. {{A
  Wide-area}}, {{High-resolution Map}} of {{Dust Extinction}} in {{M31}}}.
\bjtitle{Astrophys. J.}
\bvolume{814},
\bfpage{3}
(\byear{2015})
\doiurl{10.1088/0004-637X/814/1/3}
\end{barticle}
\endbibitem

%%% 114
\bibitem[\protect\citeauthoryear{Gordon et~al.}{2016}]{Gordon2016}
\begin{barticle}
\bauthor{\bsnm{Gordon}, \binits{K.D.}},
\bauthor{\bsnm{Fouesneau}, \binits{M.}},
\bauthor{\bsnm{Arab}, \binits{H.}},
\bauthor{\bsnm{Tchernyshyov}, \binits{K.}},
\bauthor{\bsnm{Weisz}, \binits{D.R.}},
\bauthor{\bsnm{Dalcanton}, \binits{J.J.}},
\bauthor{\bsnm{Williams}, \binits{B.F.}},
\bauthor{\bsnm{Bell}, \binits{E.F.}},
\bauthor{\bsnm{Bianchi}, \binits{L.}},
\bauthor{\bsnm{Boyer}, \binits{M.}},
\bauthor{\bsnm{Choi}, \binits{Y.}},
\bauthor{\bsnm{Dolphin}, \binits{A.}},
\bauthor{\bsnm{Girardi}, \binits{L.}},
\bauthor{\bsnm{Hogg}, \binits{D.W.}},
\bauthor{\bsnm{Kalirai}, \binits{J.S.}},
\bauthor{\bsnm{Kapala}, \binits{M.}},
\bauthor{\bsnm{Lewis}, \binits{A.R.}},
\bauthor{\bsnm{Rix}, \binits{H.-W.}},
\bauthor{\bsnm{Sandstrom}, \binits{K.}},
\bauthor{\bsnm{Skillman}, \binits{E.D.}}:
\batitle{The {{Panchromatic Hubble Andromeda Treasury}}. {{XV}}. {{The BEAST}}:
  {{Bayesian Extinction}} and {{Stellar Tool}}}.
\bjtitle{Astrophys. J.}
\bvolume{826},
\bfpage{104}
(\byear{2016})
\doiurl{10.3847/0004-637X/826/2/104}
\end{barticle}
\endbibitem

%%% 115
\bibitem[\protect\citeauthoryear{Lindberg et~al.}{2025}]{Lindberg2025}
\begin{barticle}
\bauthor{\bsnm{Lindberg}, \binits{C.W.}},
\bauthor{\bsnm{Murray}, \binits{C.E.}},
\bauthor{\bsnm{{Yanchulova Merica-Jones}}, \binits{P.}},
\bauthor{\bsnm{Bot}, \binits{C.}},
\bauthor{\bsnm{Burhenne}, \binits{C.}},
\bauthor{\bsnm{Choi}, \binits{Y.}},
\bauthor{\bsnm{Clark}, \binits{C.J.R.}},
\bauthor{\bsnm{Cohen}, \binits{R.E.}},
\bauthor{\bsnm{Gilbert}, \binits{K.M.}},
\bauthor{\bsnm{Goldman}, \binits{S.R.}},
\bauthor{\bsnm{Gordon}, \binits{K.D.}},
\bauthor{\bsnm{Hirschauer}, \binits{A.S.}},
\bauthor{\bsnm{McQuinn}, \binits{K.B.W.}},
\bauthor{\bsnm{{Roman-Duval}}, \binits{J.C.}},
\bauthor{\bsnm{Sandstrom}, \binits{K.M.}},
\bauthor{\bsnm{Tarantino}, \binits{E.}},
\bauthor{\bsnm{Williams}, \binits{B.F.}}:
\batitle{Scylla. {{IV}}. {{Intrinsic Stellar Properties}} and {{Line-of-sight
  Dust Extinction Measurements}} toward 1.5 {{Million Stars}} in the {{SMC}}
  and {{LMC}}}.
\bjtitle{Astrophys. J.}
\bvolume{982},
\bfpage{33}
(\byear{2025})
\doiurl{10.3847/1538-4357/adb4e8}
\end{barticle}
\endbibitem

%%% 116
\bibitem[\protect\citeauthoryear{Hunter et~al.}{2011}]{Hunter2011}
\begin{barticle}
\bauthor{\bsnm{Hunter}, \binits{D.A.}},
\bauthor{\bsnm{Zahedy}, \binits{F.}},
\bauthor{\bsnm{Bowsher}, \binits{E.C.}},
\bauthor{\bsnm{Wilcots}, \binits{E.M.}},
\bauthor{\bsnm{Kepley}, \binits{A.A.}},
\bauthor{\bsnm{Gaal}, \binits{V.}}:
\batitle{Mapping the {{Extended H I Distribution}} of {{Three Dwarf
  Galaxies}}}.
\bjtitle{Astron. J.}
\bvolume{142},
\bfpage{173}
(\byear{2011})
\doiurl{10.1088/0004-6256/142/5/173}
\end{barticle}
\endbibitem

%%% 117
\bibitem[\protect\citeauthoryear{Elyajouri et~al.}{2024}]{Elyajouri2024}
\begin{barticle}
\bauthor{\bsnm{Elyajouri}, \binits{M.}},
\bauthor{\bsnm{Ysard}, \binits{N.}},
\bauthor{\bsnm{Abergel}, \binits{A.}},
\bauthor{\bsnm{Habart}, \binits{E.}},
\bauthor{\bsnm{Verstraete}, \binits{L.}},
\bauthor{\bsnm{Jones}, \binits{A.}},
\bauthor{\bsnm{Juvela}, \binits{M.}},
\bauthor{\bsnm{Schirmer}, \binits{T.}},
\bauthor{\bsnm{Meshaka}, \binits{R.}},
\bauthor{\bsnm{Dartois}, \binits{E.}},
\bauthor{\bsnm{Lebourlot}, \binits{J.}},
\bauthor{\bsnm{Rouill{\'e}}, \binits{G.}},
\bauthor{\bsnm{Onaka}, \binits{T.}},
\bauthor{\bsnm{Peeters}, \binits{E.}},
\bauthor{\bsnm{Bern{\'e}}, \binits{O.}},
\bauthor{\bsnm{Alarc{\'o}n}, \binits{F.}},
\bauthor{\bsnm{{Bernard-Salas}}, \binits{J.}},
\bauthor{\bsnm{Buragohain}, \binits{M.}},
\bauthor{\bsnm{Cami}, \binits{J.}},
\bauthor{\bsnm{Canin}, \binits{A.}},
\bauthor{\bsnm{Chown}, \binits{R.}},
\bauthor{\bsnm{Demyk}, \binits{K.}},
\bauthor{\bsnm{Gordon}, \binits{K.}},
\bauthor{\bsnm{Kannavou}, \binits{O.}},
\bauthor{\bsnm{Kirsanova}, \binits{M.}},
\bauthor{\bsnm{Madden}, \binits{S.}},
\bauthor{\bsnm{Paladini}, \binits{R.}},
\bauthor{\bsnm{Pendleton}, \binits{Y.}},
\bauthor{\bsnm{Salama}, \binits{F.}},
\bauthor{\bsnm{Schroetter}, \binits{I.}},
\bauthor{\bsnm{Sidhu}, \binits{A.}},
\bauthor{\bsnm{R{\"o}llig}, \binits{M.}},
\bauthor{\bsnm{Trahin}, \binits{B.}},
\bauthor{\bsnm{Van De~Putte}, \binits{D.}}:
\batitle{{{PDRs4All}}. {{V}}. {{Modelling}} the dust evolution across the
  illuminated edge of the {{Orion Bar}}}.
\bjtitle{Astron. Astrophys.}
\bvolume{685},
\bfpage{76}
(\byear{2024})
\doiurl{10.1051/0004-6361/202348728}
\end{barticle}
\endbibitem

%%% 118
\bibitem[\protect\citeauthoryear{Mathis et~al.}{1983}]{Mathis1983}
\begin{barticle}
\bauthor{\bsnm{Mathis}, \binits{J.S.}},
\bauthor{\bsnm{Mezger}, \binits{P.G.}},
\bauthor{\bsnm{Panagia}, \binits{N.}}:
\batitle{Interstellar radiation field and dust temperatures in the diffuse
  interstellar medium and in giant molecular clouds}.
\bjtitle{Astron. Astrophys.}
\bvolume{128},
\bfpage{212}--\blpage{229}
(\byear{1983})
\end{barticle}
\endbibitem

%%% 119
\bibitem[\protect\citeauthoryear{Compi{\`e}gne et~al.}{2010}]{Compiegne2010}
\begin{barticle}
\bauthor{\bsnm{Compi{\`e}gne}, \binits{M.}},
\bauthor{\bsnm{Flagey}, \binits{N.}},
\bauthor{\bsnm{{Noriega-Crespo}}, \binits{A.}},
\bauthor{\bsnm{Martin}, \binits{P.G.}},
\bauthor{\bsnm{Bernard}, \binits{J.-P.}},
\bauthor{\bsnm{Paladini}, \binits{R.}},
\bauthor{\bsnm{Molinari}, \binits{S.}}:
\batitle{Dust in the {{Diffuse Emission}} of the {{Galactic Plane}}: {{The
  Herschel}}/{{Spitzer Spectral Energy Distribution Fitting}}}.
\bjtitle{Astrophys. J.}
\bvolume{724},
\bfpage{44}--\blpage{47}
(\byear{2010})
\doiurl{10.1088/2041-8205/724/1/L44}
\end{barticle}
\endbibitem

%%% 120
\bibitem[\protect\citeauthoryear{Li and Draine}{2001}]{Li2001}
\begin{barticle}
\bauthor{\bsnm{Li}, \binits{A.}},
\bauthor{\bsnm{Draine}, \binits{B.T.}}:
\batitle{Infrared {{Emission}} from {{Interstellar Dust}}. {{II}}. {{The
  Diffuse Interstellar Medium}}}.
\bjtitle{Astrophys. J.}
\bvolume{554},
\bfpage{778}--\blpage{802}
(\byear{2001})
\doiurl{10.1086/323147}
\end{barticle}
\endbibitem

\end{thebibliography}
\end{document}